\documentclass[%
 reprint,
 amsmath,amssymb,
 aps,
]{revtex4-2}
\usepackage{graphicx}
\usepackage{dcolumn}
\usepackage{bm}
\usepackage{hyperref}
\usepackage[nice]{nicefrac}
\usepackage{adjustbox}
\usepackage{amsmath}
\usepackage[]{quantikz}
\usepackage[ruled,linesnumbered]{algorithm2e}
\usepackage[]{float}
\newcommand{\rf}[1]{Figure~\ref{#1}}
\newcommand{\rs}[1]{Section~\ref{#1}}

\begin{document}

\title{Mixed Quantum-Classical Dynamics for Near Term Quantum Computers}

\author{Daniel Bultrini}
 \email{daniel.bultrini@uni-heidelberg.de}
  \affiliation{Theoretische Chemie, Physikalisch-Chemisches Institut, Heidelberg}
\author{Oriol Vendrell}
 \email{oriol.vendrell@uni-heidelberg.de}
 \affiliation{Theoretische Chemie, Physikalisch-Chemisches Institut, Heidelberg University}%
\affiliation{Interdisciplinary Center for Scientific Computing, Heidelberg University
}

\begin{abstract}
Mixed quantum-classical dynamics is a set of methods often used to understand systems too complex to treat fully quantum mechanically. Many techniques exist for full quantum mechanical evolution on quantum computers, but mixed quantum-classical dynamics are less explored. We present a modular algorithm for general mixed quantum-classical dynamics where the quantum subsystem is coupled with the classical subsystem. We test it on a modified Shin-Metiu model in the first quantization through Ehrenfest propagation. We find that the Time-Dependent Variational Time Propagation algorithm performs well for short-time evolutions and retains qualitative results for longer-time evolutions.\end{abstract}

\maketitle

\section{Introduction}

Quantum computers have found great success in electronic structure theory through the variational quantum eigensolver \cite{peruzzoVariationalEigenvalueSolver2014} and subsequent algorithms known as variational quantum algorithms (VQAs) \cite{cerezoVariationalQuantumAlgorithms2020}. Adding additional electrons to a system greatly increases its complexity, this is something we hope quantum computers could handle. If one were to also consider the full nuclear dynamics, the problem becomes unmanageable much faster, potentially even for quantum computers \cite{ollitraultMolecularQuantumDynamics2021}. One way to reconcile this is to partition the system into interacting quantum and classical parts. This is the realm of mixed quantum-classical (MQC) approaches, which are a widely used set of tools for understanding chemical systems \cite{curchodInitioNonadiabaticQuantum2018,kirranderEhrenfestMethodsElectron2020}. In quantum computing, this area is less researched than the electronic structure problem, but it is actively being explored \cite{ollitraultNonadiabaticMolecularQuantum2020,ollitraultMolecularQuantumDynamics2021,sokolovMicrocanonicalFinitetemperatureInitio2021}. In this work we propose and explore a noisy intermediate-scale quantum (NISQ) friendly algorithm that can be used to study MQC dynamics. 

Using quantum computers alongside classical computers is the backbone of VQAs, but splitting a system into sections treated separately by each machine is not new. A DFT embedding scheme with a quantum computer expansion of the active space \cite{rossmannekQuantumHFDFTEmbedding2020} and has been found to outperform certain types of state-of-the-art approximate techniques such as CASSCF \cite{levineCASSCFExtremelyLarge2020} in finding ground state energies. Furthermore, ground state dynamics, geometry relaxation, and force measurements for MD applications have been explored with success in \cite{sokolovMicrocanonicalFinitetemperatureInitio2021}. In \cite{ollitraultMolecularQuantumDynamics2021}, dynamics are explored in both first and second quantization, but using a time-independent Hamiltonian. This is also the case for various other time propagation techniques \cite{linRealImaginaryTimeEvolution2021,barisonEfficientQuantumAlgorithm2021,berthusenQuantumDynamicsSimulations2022}; we draw inspiration from and use p-VQD \cite{barisonEfficientQuantumAlgorithm2021} in this work. 

Our contribution is the presentation of a general algorithmic structure to tackle non-adiabatic molecular dynamics (NAMD) by offloading the QM part to a quantum computer and evolving the classical system by the Ehrenfest method. Observables from the quantum mechanical (QM) subsystem are measured and used to update the classical system, which in turn will update the time-dependent Hamiltonian that is used to evolve the QM state in turn. This is all done in first quantization, which saves the algorithm from needing to measure nonadiabatic couplings, as these are treated directly within the wave function and its evolution in this setting. The algorithm is demonstrated in the Shin-Metiu model \cite{shinNonadiabaticEffectsCharge1995}, which is often used to test various non-adiabatic techniques \cite{albaredaUniversalStepsQuantum2016,erdmannCombinedElectronicNuclear2003,falgeQuantumWavePacketDynamics2012,gosselNumericalSolutionExact2019}. We modify this to be a NAMD-like problem by partitioning the system into a classical nucleus and quantum electron. The major contribution is the study of how the interaction between observable measurement and system updates play out as well as introducing a scheme that is suited to begin exploring other time-dependent phenomena on NISQ machines. 

The theoretical advantage of using quantum computers is that they have access to an exponentially growing computational space for each additional qubit in the system \cite{nielsenQuantumComputationQuantum2010}. Current machines have access to hundreds of qubits, which would ideally allow them to already outperform current supercomputers. This is not the case due to noise coming from interactions with the environment and imperfect gate implementations. As such, NISQ algorithms \cite{bhartiNoisyIntermediatescaleQuantum2021} have to contend with limits on the number of imperfect operations that can be made. But even if this were not the case and full quantum dynamics could be simulated, we probably will always want to tackle a problem bigger than current machines can handle, so these kinds of approximations will always be used. 

It should be noted that existing supercomputers by far outperform existing quantum computers in handling large quantum-chemistry problems, and applications to chemical problems will have to be deferred until there is a provable quantum advantage. Approximations like limiting the simulation to a selected active space \cite{rossmannekQuantumHFDFTEmbedding2020} can extend the reach of quantum computers, but analogues of these ideas apply to classical computers as well. For now, quantum algorithms in chemistry mostly study systems that are comfortably computable on current classical hardware.

\section{Results}

\subsection{Time-Dependent Hamiltonian Variational Quantum Propagation }

The time-dependent variational quantum propagation (TDVQP) algorithm builds on the circuit compression idea of "projected variational quantum dynamics" (p-VQD) \cite{barisonEfficientQuantumAlgorithm2021} by allowing the Hamiltonian to be time-dependent. For many large problems of interest to theoretical chemistry, especially in MD, it is impossible to fully simulate the system of interest quantum mechanically. As such, the system is subdivided into classical and quantum components. The evolution of both systems occurs in locked steps, with the classical system defining the Hamiltonian for the quantum evolution, and the quantum system then feeding back into the classical system in the way of some observable, usually the energy gradient (force). One mustn't limit themselves to molecular or even physical systems, as this algorithm would work with any set of observables that can be used to update the classical system of interest. 

The algorithm begins with a parameterized circuit initialized to some desired state. This is denoted as $\ket{\psi_0}$, which will have been generated according to a Hamiltonian based on an initial vector of classical parameters $\vec q_0$ of the classical coordinates, which we denote $H(\vec q_0):=H_0$. This is done by choosing some sufficiently expressive parameterized circuit ansatz $\hat C(\vec\theta)$ which takes the quantum computer's initial state, denoted \(\ket{0} \), to \(\ket{\psi _0} \). This can be done using a VQA to find a chosen state with respect to $H_0$, which returns the circuit parameters $\vec\theta_0$. Then $\ket{\psi_0}=\hat C(\vec\theta_0)\ket{0}$. A chosen set of observables $ \{\hat O^{(s)}(q_0)\}:=\{\hat O^{(s)}_0\}$ are measured from $\ket{\psi_0}$, which yield a set of expectation values $\{O^{(s)}_0\}$. These observables are used to evolve the classical state of the system, generating a new vector of classical parameters $\vec q_1$. These can then be used to generate $H_1$ and $\{\hat O^{(s)}_1\}$. Now, one evolves the state from $\ket{\psi_0}$ to $\ket{\psi_1}$ by applying the time evolution operator to the state, $\ket{\psi_1}=\exp{(-i\hat H_0\Delta t)}\ket{\psi_0}$. 

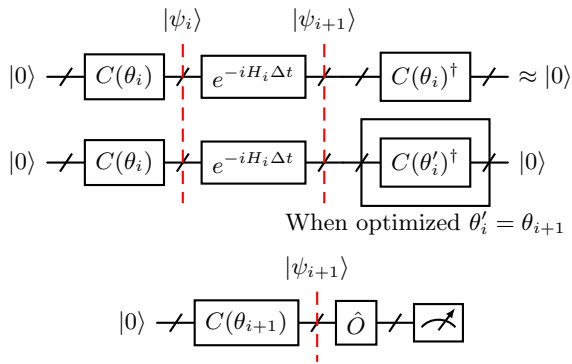
\begin{figure}[h]
     \centering
    \begin{quantikz}
        \lstick[]{$\ket{0}$}&\gate[]{C(\theta_i)}\slice{$\ket{\psi_{i}}$}\qwbundle[alternate]{}
        &
        \gate[]{e^{-iH_i\Delta t}}\slice{$\ket{\psi_{i+1}}$}\qwbundle[alternate]{} &\qwbundle[alternate]{}&
        \gate{C(\theta_i)^\dagger}\qwbundle[alternate]{}&
        \qwbundle[alternate]{} \rstick[]{$\approx\ket{0}$}\\
        \lstick[]{$\ket{0}$}&\gate[]{C(\theta_i)}\qwbundle[alternate]{}
        &
        \gate[]{e^{-iH_i\Delta t}}\qwbundle[alternate]{} &\qwbundle[alternate]{}&
        \gate{C(\theta_i^\prime)^\dagger}\gategroup[wires=1,steps
=1,label style={label position=below,yshift=-4,anchor=north}]{When optimized $\theta_i^\prime=\theta_{i+1}$}\qwbundle[alternate]{}&
        \qwbundle[alternate]{} \rstick[]{$\ket{0}$}
    \end{quantikz}  
    \begin{quantikz}
        \lstick[]{$\ket{0}$}&\gate[]{C(\theta_{i+1})}\slice{$\ket{\psi_{i+1}}$}\qwbundle[alternate]{}
        &\gate[]{\hat O}\qwbundle[alternate]{}&\meter{}\qwbundle[alternate]{}
    \end{quantikz}
     \caption{\textbf{Sketch of the TDVQP process with slices} showing the state after each gate at the initial condition and final condition. The initial guess for the parameter vector $\theta_i^\prime$ is $\theta_i$ and its final value is denoted $\theta_{i+1}$. Then observables $\hat O$ can be measured on $\ket{\psi_{i+1}}$ to update the Hamiltonian or the classical state of the system.}
     \label{fig:TDVQP}
 \end{figure}

The physical implementation of the time evolution can either be the Trotterized form of the operator or some other approximate time evolution. In this work, we simply use $\hat H_0$ to evolve the state with a first-order trotter expansion. The Hamiltonian used for the time evolution could be of higher order. For example, $(\hat H_0+\hat H_1)/2$ can be used with no extra cost in this scheme, but higher-order integrators will require an additional evolution and observable measurement for each timestep beyond $H_1$. In return, one gets higher-order symplectic integration. Now a single step of the p-VQD algorithm is applied which generates the new circuit parameters $\theta_1$ such that $\ket{\psi_1}\approx \hat C(\theta_1)\ket{0}$ to some desired threshold. This process is repeated until the desired timestep is reached. The entire process is more precisely described in Algorithm~\ref{alg:TDVQP}, and a depiction of the quantum circuit can be seen in Fig.~\ref{fig:TDVQP}. The overall number of circuit evaluations is linear with respect to the number of iterations, circuit parameters, timesteps and Pauli terms of the observables. This is treated in greater depth in \ref{app:resources}.

\begin{algorithm}\label{alg:TDVQP}
    \caption[short]{Time-Dependent Variational Quantum Propagation (TDVQP)}
    \SetKwFunction{ecost}{EvalCost}
    \SetKwFunction{eval}{ExpectationValue}
    \SetKwFunction{force}{MeasureObs}
    \SetKwFunction{update}{UpdateParameters}
    \SetKwFunction{upfun}{UpdateFunction}
    \SetKwFunction{upang}{UpdateAngles}
    \SetKwFunction{vqe}{VQE}

    \SetKwBlock{rep}{repeat}{}

    \SetKwProg{Fn}{Function}{:}{}
    \KwIn{$\hat H_{\text{gen}}(\vec q\ )$, \(C(\vec\theta)\ \), $ \{{\hat O}^{(s)}_{\text{gen}}(\vec q)\}$, \ecost, \(\vec q_{\text{0} } \)}
    \Fn{\vqe($\hat{H}_{\text{gen} }(\vec{q} ) $, $C(\vec\theta)$, $\vec q $) }{
        \(\vec\theta^\prime   = \mathop{\min}_{\theta } \) \eval(\((\hat{H} _\text{gen} (\vec q ),C(\vec\theta)\)))\; 
        \KwRet \(\vec\theta^\prime  \) 
    }
    \Fn{\update(\(\vec O(\vec q\ )\), $\vec q$)}{
        \(\vec {q^\prime } \)  = \upfun ($\vec{q} $, \(\vec O(\vec q\ )\))\;
        \KwRet \(\vec {q^\prime } \) 
    }
    \Fn{\upang(\(C(\vec\theta\ ),\ \hat{H} ,\ \Delta t\))}{
        \(\vec{\theta }\) = \(\mathop{\min} _{\vec{\theta_i^\prime } }\) \ecost(\( C(\vec{\theta_{\text{i}}^\prime })\exp(i\hat{H}\Delta t) C(\vec{\theta_{\text{i} } })^{\dagger}\)) \;
        \KwRet \(\vec{\theta }\)
    }

    \Fn{\force($\{{\hat O}^{(s)}_{\text{gen}}(\vec q)\}$, \(C(\vec\theta\ )\))}{
        for s in number of observables \rep{
        \( O^{(s)}\) =  \eval(\(\hat O_{\text{gen} }^{(s)}(\vec q\ ) \), \(C(\vec\theta\ )\)) \;}
        \KwRet \( {O_i^{(s)}}\) 
    }
    \Fn{TDVQP($H_{\text{gen} }(\vec q\ ) $, \(C(\vec\theta )\), \(\vec q_{0 }\))}{
        store all \(\vec \theta _i,\  \{O^{(s)}_i\},\ \vec{q} _i \) in arrays \(\pmb{\theta ,\ O,\ q} \) \; 
        \(\vec\theta_0\) = \vqe (\(\hat{H} _\text{gen} (\vec{q} _0)\), \(C(\vec{\theta} )\))  \;
        \(\{ O_0^{(s)}\}\)  =  \force ($\{\hat O^{(s)}_{\text{gen} }(\vec q_0)\}$, \(C(\vec \theta_0)\))\;
        for \(i = 1\) to \(n_t\) timesteps \rep{
            \(\vec q_i\) = \update ($\{ O_{i-1}^{(s)}\}$, $\vec q_{i-1}$) \;
            \(\vec \theta _i\) = \upang ($C(\theta _{i-1}), \hat{H}_{\text{gen} }(\vec{q}_{i-1}), \Delta t$) \;
            \(\{ O_i^{(s)}\}\)  =  \force ($\{\hat O_{\text{gen}}^{(s)}(\vec q_i)\}$, \(C(\vec \theta_i)\))\;
        }
        \KwRet \(\pmb{\theta ,\ O,\ q} \) 

    }

\end{algorithm}

 TDVQP should be thought of as a meta-algorithm that has replaceable components. The most directly replaceable part is the choice of ansatz \(\hat C(\vec \theta )\), which at the moment is generally a heuristic choice for most problems in NISQ devices. More advanced ansatze such as the family of adaptive ansatze, which changes the ansatz throughout the evolution would work, but could not use the previous step's $\theta$ parameters as effectively.  The very costly time evolution is currently a Trotterized form of the time evolution operator, as in this work, and in \cite{barisonEfficientQuantumAlgorithm2021, berthusenQuantumDynamicsSimulations2022}. This can be replaced by a plethora of more NISQ-friendly time evolutions as is done in \cite{lowHamiltonianSimulationQubitization2019, cirstoiuVariationalFastForwarding2020} if the form of the Hamiltonian allows this. The limit is the no-fast forwarding theorem \cite{atiaFastforwardingHamiltoniansExponentially2017,berryEfficientQuantumAlgorithms2007,childsLimitationsSimulationNonsparse}, which states that you cannot achieve a time evolution of time \(t\) in a sublinear gate count for a general Hamiltonian, but for shorter time evolutions, limited sizes and specific cases of Hamiltonians, including our sparse Hamiltonian using short time evolutions, this likely is not the case \cite{cirstoiuVariationalFastForwarding2020}.
 
In the classical evolution, the choice of integrator and the actual Hamiltonian used in the time evolution will depend on the type of problem and desired accuracy. Integrators like the Velocity Verlet algorithm require no additional resources. TDVQP becomes exact when \(\hat{C} (\vec\theta )\) can express the system perfectly for any configuration of classical parameters \(\vec q\), given that the exact parameters \(\theta \) can be found by optimization. This is only a statement of the best-case scenario. In reality, finding a good ansatz, VQA and shot-efficient optimizer is at the forefront of research in this area \cite{cerezoVariationalQuantumAlgorithms2020}, and it is out of scope for this work.

\subsubsection{Error propagation in TDVQP}

The TDVQP algorithm inherits all of the errors of its constituent parts. This includes the chosen circuit compression algorithm, time evolution approximation, and in the classical propagator. Nonetheless, it is important to have an intuition of the potential pitfalls of the algorithm. This section illustrates the sources of error in the wavefunction and observables and their interaction velocity Verlet integrator. A more thorough derivation and explanation can be found in the supplementary materials, \ref{app:errors}.

When running the algorithm, any coherent error on the wavefunction representation in the quantum computer $\ket{\tilde\psi}$ can be represented as a superposition of the desired state $\ket{\psi}$ and some combination of undesired orthogonal states $\ket{\phi}$, such that $\ket{\tilde\psi}=\sqrt{ 1-I^{2} }\ket{\psi}+I\ket{\phi}$, where $I$ is the infidelity. When we measure the expectation value of a   Hermitian observable ${\mathcal{O}}$ on $\ket{\tilde\psi}$ we will get 

\begin{align}
\bra{\tilde \psi} \mathcal{O} \ket {\tilde \psi}      = \left( 1-I^{2} \right)\bra{\psi} \mathcal{O} \ket {\psi} +I^{2}\bra{\phi} \mathcal{O} \ket {\phi}.
\end{align}

How this translates to the actual measured observable used here is completely system dependent. This will lead to an error in the observable, which in the case of the velocity Verlet integrator with a force error $F_i\epsilon\propto I^2$ in 1 dimension will give a new position $\tilde R_i$ of 
\begin{equation}
\tilde R_i = R_{i}+\frac{F_{i\epsilon}}{M}\Delta t^{2},\\
\end{equation}
shifted from the expected true position $R_i$. This is linear in the error of the force and quadratic with respect to the timestep $\Delta t$. The following time evolution Hamiltonian and observable operator will be based on this position with an error, which is again, system dependent. The effect is illustrated by the equations  
\begin{align}
\ket{\tilde \psi_{1}} &=\exp(-iH_{el}( R_{0})\Delta t)\ket{\tilde \psi_{0}},  \\ 
\ket{\tilde \psi_{i}} &=\exp(-iH_{el}( \tilde R_{i-1})\Delta t)\ket{\tilde \psi_{i-1}}.
\end{align}
Even in the one-dimensional model used in this work, this effect is not analytically computable, but it is small if the timesteps are sufficiently small. This then enters the velocity $(\dot R)$ update  as 
\begin{align}
\tilde{ \dot{R}}_{i+1}=  \dot{R}_{(i+1)}+\frac{F_{i\epsilon}+F_{(i+1)\epsilon}}{2M}\Delta t,
\end{align}
which is linear in the error and timestep. Assuming a constant error over all time of $F_\epsilon$, that is to say, that the force deviates from the correct one by a constant offset - this is equivalent to having an additional linear term on the potential. This has the overall effect on the position at iteration $i$ of 
\begin{equation}
    \tilde R_{i}=  R_{i}+\dot{R}_{i-1}\Delta t +\frac{(i^{2}+i){F_{\epsilon}}}{2M} \Delta t^{2}.
\end{equation}
This expression is quadratic in $i$ and quadratic in timestep. The effect on the fidelity of the TDVQP wavefunction compared to an exact propagation is nontrivial, but numerical examples are provided in the supplementary materials \ref{app:errors}. 

The other main source of error inherent to the p-VQD algorithm is that the optimizer never finds a perfect representation of the time-evolved wavefunction, but rather an approximation that meets some fidelity threshold $T<1$. If this threshold is met exactly at each p-VQD step, assuming all observable measurements are unaffected, then the decrease in the fidelity is modelled by 
\begin{equation}\label{eq:fidfall}
    \text{Fidelity(i)}=T^i.
\end{equation}

When the algorithm is run under limited quantum resources and thus subject to finite sampling noise, both the observable and p-VQD step fidelity measurements will have some Gaussian distribution, which will feed into the errors above on a simulation by simulation basis. The effect of this has been analysed numerically in the case of our modified Shin-Metiu Model.  

\subsection{\label{ssec:SM}The Shin Metiu Model}

The Shin-Metiu model is a numerically exactly solvable minimal model which captures essential nonadiabatic effects \cite{shinNonadiabaticEffectsCharge1995}. It is often used as a benchmark system for new techniques and is used to study the effects of different environments as has been done for polaritonic dynamics, coupling to cavities, and the effect of electromagnetic fields \cite{albaredaUniversalStepsQuantum2016,erdmannCombinedElectronicNuclear2003,falgeQuantumWavePacketDynamics2012,flickCavityBornOppenheimer2017}. It is simple to change its parameters for it to exhibit adiabatic to strongly non-adiabatic dynamics.

\begin{figure}[h]
    \centering
    \includegraphics[width=1\columnwidth]{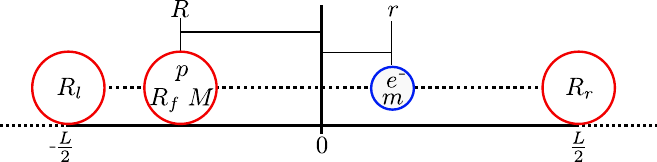}
    \caption{\textbf{Illustration of the Shin-Metiu model} with fixed ions at \(\frac{-L}{2},\ \frac{L}{2} \) as stationary boundaries, the mobile ion \(p\) of mass \(M\)  at distance \(R\) from the origin and the electron \(e^-\)  at distance \(r\) from the origin. \(R_{l} ,\ R_{r} \text{ and } R_{f} \) are constants for the regularized Coulomb potential in eq.~\ref{eq:shin_metiu}. }
    \label{fig:shin-metiu}
\end{figure}

In its simplest and original conception, the model shown in Fig. \ref{fig:shin-metiu} consists of two stationary ions separated by a distance of \(L\), specifically located at \(\frac{L}{2}\) and \(\frac{-L}{2}\). These  enclose a mobile ion \(p\)  of mass \(M\) at distance \(R\) from the origin and an electron \(e^-\)  at distance \(r\). The modified Coulomb potential is parameterized by the constants \(R_{l},\ R_{r} \text{ and } R_{f} \), as shown in Eq. \ref{eq:shin_metiu}. This is done to avoid singularities and make the system numerically simpler to simulate.

The full Hamiltonian of the system is 
\[
    H=-\frac{1}{2M}\frac{\partial }{\partial R^2} +H_e(r,R),
\]
with the electronic part being
\begin{align}\label{eq:shin_metiu}
    \begin{split}
    H_e=&-\frac{1}{2m}\frac{\partial ^2}{\partial r^2} + 
    \frac{1}{\left\vert \frac{L}{2} -R  \right\vert }+\frac{1}{\left\vert \frac{L}{2} + R  \right\vert  }-\\
    &-\frac{ \text{erf} (\nicefrac{\left\vert \frac{L}{2} -r \right\vert }{R_r})}{\left\vert \frac{L}{2} -r \right\vert  }-
    \frac{ \text{erf} (\nicefrac{\left\vert \frac{L}{2} +r \right\vert }{R_l})}{\left\vert \frac{L}{2} +r \right\vert  }-
    \frac{\text{erf} (\nicefrac{\left\vert R -r  \right\vert }{R_f})}{\left\vert R-r \right\vert }.
    \end{split}
\end{align}
The equation uses atomic units, setting \(e=Z=\hbar=1\), we also take \(m=1\) and \(M=1836\) in the simulation. The constants 
\(R_{l},\ R_{r} \text{ and } R_{f} \), as shown in \rf{fig:shin-metiu}
are chosen to create specific adiabatic surfaces with transitions we would like to observe, as in \rf{fig:shin-metiu_BOPES}.

\begin{figure}[h]
    \centering
    \includegraphics[width=1\columnwidth]{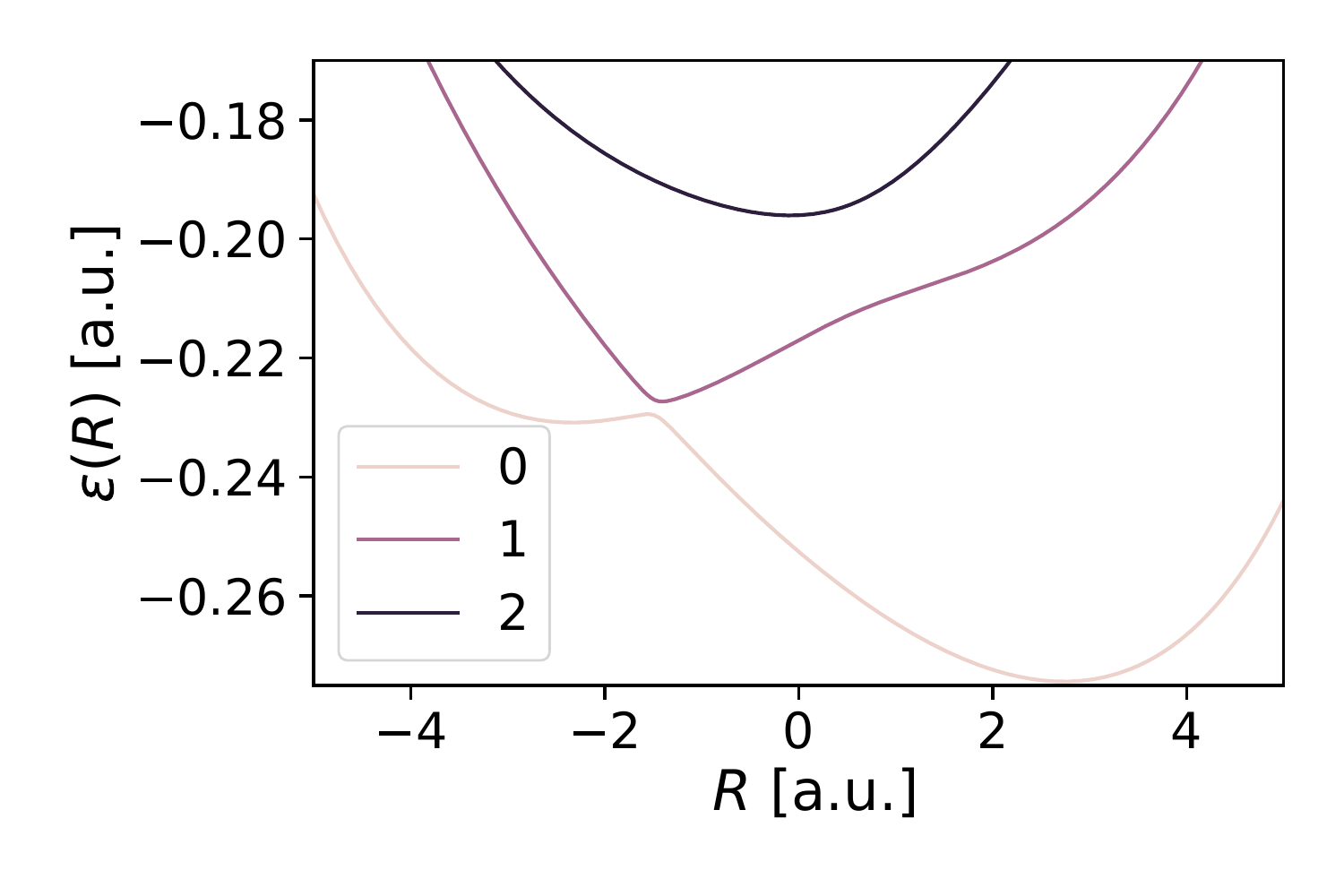}
    \caption{\textbf{Potential energy surfaces of the Shin-Metiu Model} for the coefficients \(R_f=5.0,\ R_l=4.0\) and \(R_r=3.2\) showing the avoided crossing around position \(R=-2\) with \(L=19\) .}
    \label{fig:shin-metiu_BOPES}
\end{figure}

We use the values \(R_f=5.0,\ R_l=4.0\) and \(R_r=3.2\), which resulting in avoided crossing around  \(R=-1.9\) when the distance between the ions is \(L=19\). These parameters were chosen to be similar to those used in several studies of the model \cite{gosselNumericalSolutionExact2019,albaredaUniversalStepsQuantum2016}. The shape of the Born-Oppenheimer potential energy surfaces (BOPES) can be seen in Fig.~\ref{fig:shin-metiu_BOPES} 

\subsubsection{Ehrenfest propagation of the model}

To perform Ehrenfest propagation of the Shin-Metiu model, we split the system in two. The  nucleus (\(p\) ) and the electron (\(e^-\) ). The electron subsystem is treated as a quantum particle described by Hamiltonian ~\ref{eq:shin_metiu}, where \(H_e\) is parameterized by the nuclear position (\(R\)). The nuclear subsystem is treated classically by tracking parameters of position (\(R\)) and velocity (\(\dot{R}\)). 

For initial coordinates \(R_0\text{ and }  \dot R_0\), we first prepare the electronic Hamiltonian \(H_e(R_0)\), which is used to compute the initial state of the electron \(\ket{\psi_0} \). Thanks to the simplicity of the model, we can use exact diagonalization to compute the eigenvectors and choose any arbitrary superposition of eigenvectors as the initial state.

The nucleus is evolved using the velocity Verlet method \cite{swopeComputerSimulationMethod1982} with the acceleration being computed from the Coulombic repulsion from the fixed ions and the force from the electronic state. The electronic state is evolved by unitary time evolution with the Hamiltonian at the nuclear position. We use a timestep \(\Delta t\), and the system state at timestep \(i\)  is denoted by under scripts \(i\), where the time is simply \(i\cdot \Delta t\). We set our initial conditions at timestep 0, and for the \(i^{\text{th} } \) step, we compute 

\begin{align}
\begin{split}\label{eq:verlet_start}
    F_e(R, \ket{\psi } ) &= - \bra{\psi} \frac{\partial H_e(R)}{\partial R}  
    \ket{\psi},  \\ 
\end{split}\\
\begin{split}
    R_i &= R_{i-1}+\dot R_{i-1} \Delta t +\frac{F_e(R_{i-1},\ket{\psi_{i-1}})}{M}\Delta t^2,\\\end{split}\\
\begin{split}
    \ket{\psi _i} &=  e^{-iH_{e}(R_{i-1})\Delta t}
                        \ket{\psi _{i-1}},\\
\end{split}\\
\begin{split}\label{eq:verlet_end}
  \dot R_i&=\dot R_{i-1} + \frac{F_e(R_{i-1},\ket{\psi_{i-1}} )+F_e(R_{i},\ket{\psi_{i}} )}{2M}\Delta t. \\ 
\end{split}
\end{align}

\subsection{Numerical results}

The results shown in this section are the result of two types of potential cases. The first, which is referred to as 'single' is the evolution of a single set of initial conditions meant to represent the precision of this algorithm to exactly reproduce a quantum-classical system. Although this is not the intended use case of TDVQP, it is nonetheless the most instructive to determine its behaviour. The second, referred to as 'MD' is the molecular dynamics-like use case, we use a single TDVQP evolution per trajectory. The trajectories are picked from random pairs of normal thermal distributions around the same initial state as the 'single' simulations, with the specific values written in \rs{sec:simparam}. Their average behaviour is taken as the approximation to the true system evolution. For NISQ devices there are some potential cases for different approaches to MD such as those mentioned in \cite{kuroiwaQuantumCarParrinelloMolecular2022}.

\begin{figure}[h]
    \centering
    \includegraphics[width=1\columnwidth]{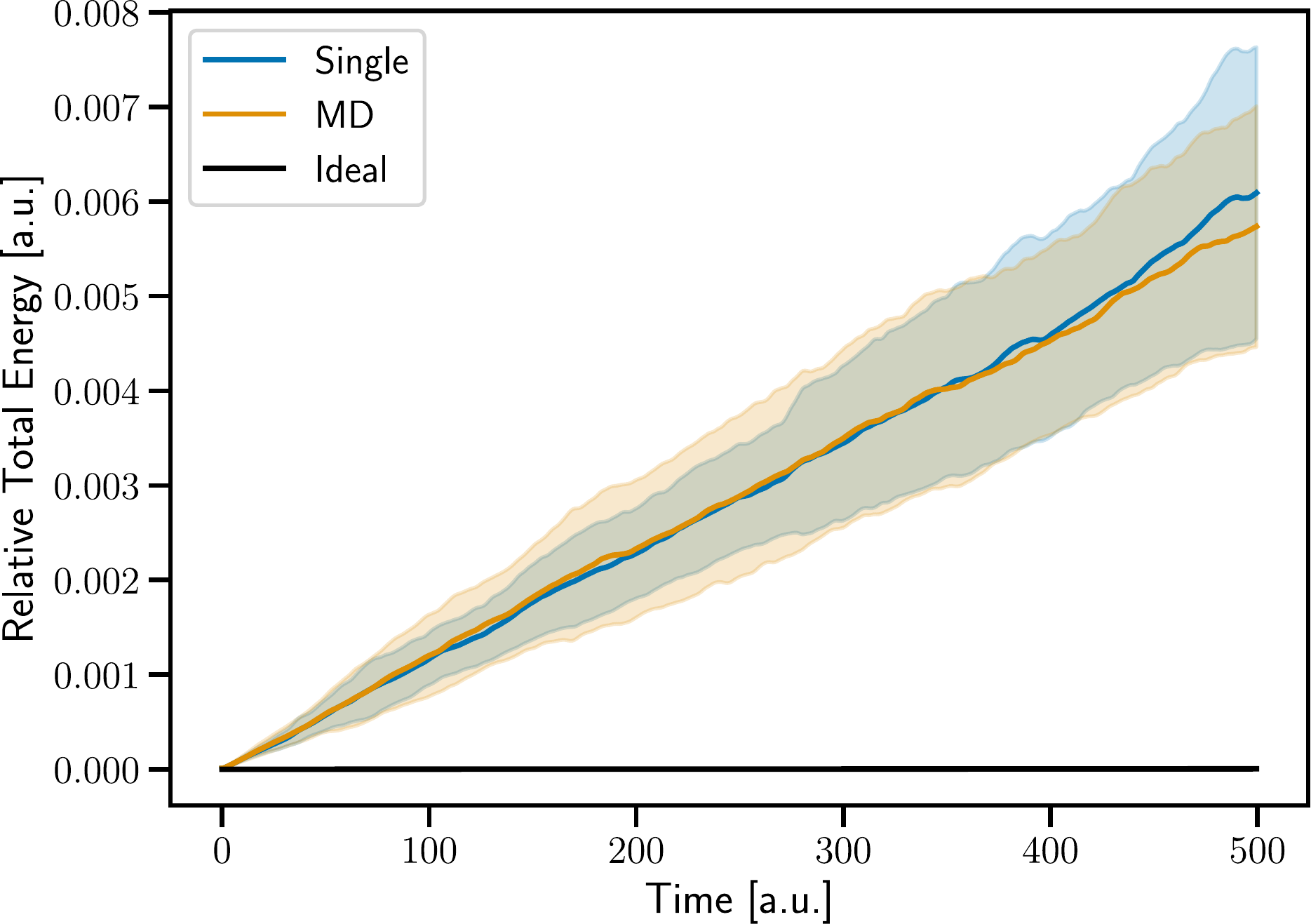}
    \caption{\textbf{Mean relative TDVQP energy} shown for "Single" initialization (blue) and for "MD" initialization (orange). The highlighted areas show the standard deviation of the distribution of 100 separate runs of 1000 timesteps of $0.5$ a.u. at the infinite shot limit. The other lines show the ideal initial state energy evolution. The energy continually increases in the TDVQP as higher energy levels are increasingly populated through leakage.}
    \label{fig:Senergy}
\end{figure}

\begin{figure}[h]
    \centering
    \includegraphics[width=1\columnwidth]{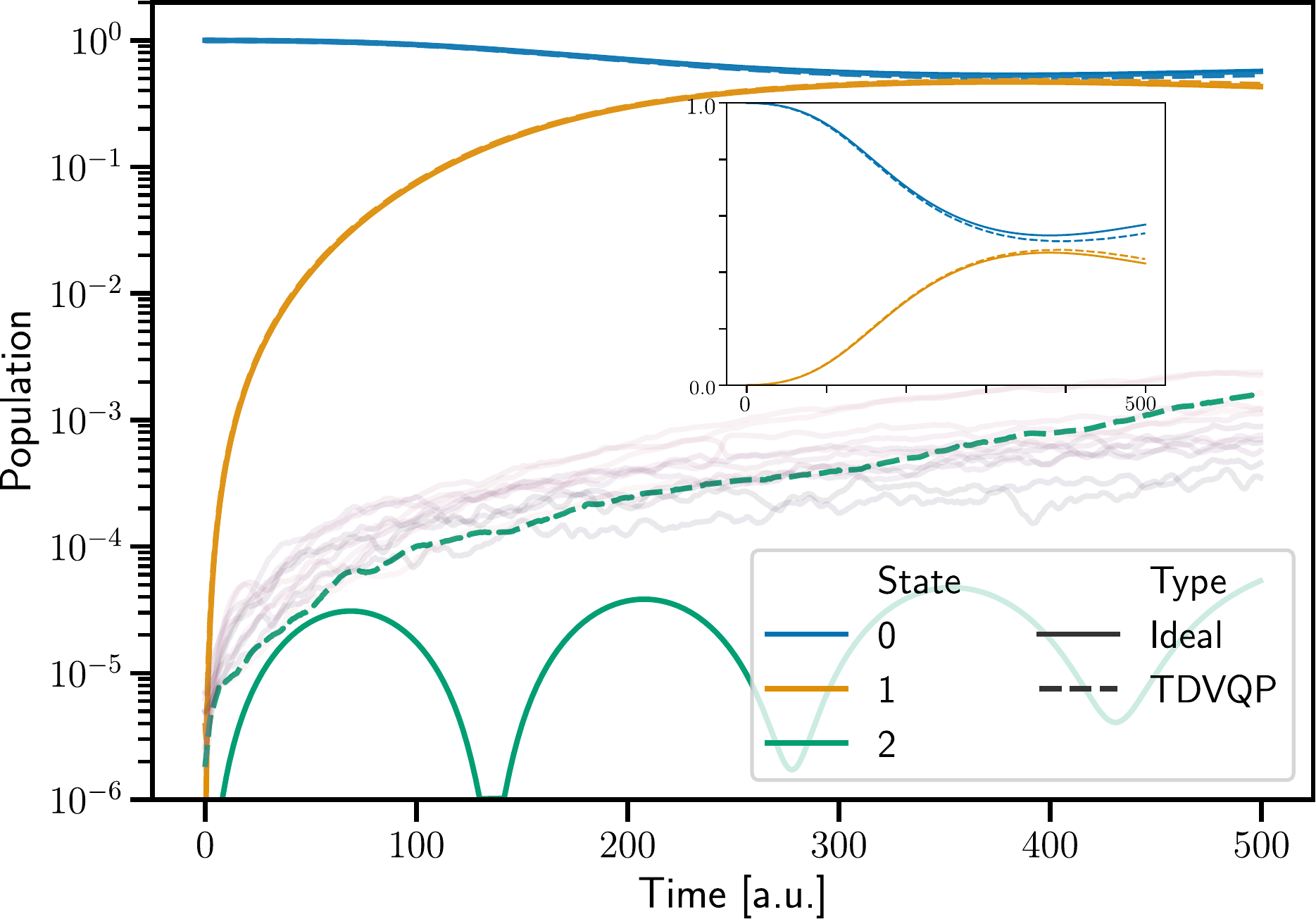}
    \caption{\textbf{Time evolution of state populations} between the ideal simulation (solid) and 100 instances of TDVQP evolution (dashed) over 1000 timesteps of $0.5$ a.u. at the infinite shot limit for the "single" simulation. The populated states are the same as those seen in Figure~\ref{fig:shin-metiu_BOPES}. The main graph is in logarithmic scale showing faint lines for higher energy levels populated by TDVQP, with the inset showing a linear scale of the two most populated levels.}
    \label{fig:Spop}
\end{figure}

An important gauge for the validity of simulations of closed systems is whether they conserve energy or not. We use a symplectic integrator in the classical system (velocity Verlet), and in the exact diagonalization case, we see energy conservation for up to 50,000 timesteps. As can be seen in Figure~\ref{fig:Senergy} the TDVQP algorithm does not conserve energy. This is because the populations are not preserved in the diagonal basis in the p-VQD step, as the optimization is limited to a finite number of iterations and the ansatz is system agnostic. The effect of this can be seen clearly in Figure~\ref{fig:Spop}, where it can be seen that the population in higher states increases much faster than in the ideal case. Although it is not shown, starting the exact evolution from the VQE state does begin with some population spread, but this does not change as the evolution progresses. The population plot is shown at the infinite shot limit for clarity, but the finite shot cases can be found in \ref{app:shots}.

As a consequence of the higher energy levels being increasingly populated as the evolution progresses, it is the case that the fidelity decreases gradually. This is indeed the case and can be seen in \rf{fig:Mfidelity}. This general degradation of quality is not optimal and strategies could be employed in the optimization to mitigate this, such as measuring the energy and allowing the cost function to penalize when the system is not conserving energy. This would require measuring the expectation value of the system Hamiltonian which would increase the cost of this algorithm.  

Despite the problem with energy conservation, using such an algorithm to measure an observable such as the force exerted on the nucleus by the electron (\(F_{el}\)) can still lead to reasonable results. \rf{fig:Sforce} shows the mean of the electron force measurements from TDVQP compared to the ideal measurements at different per-circuit shot counts. It is clear that the mean value slowly deviates from the ideal evolution in even the infinite shot limit, and that you require \(10^5\) shots per circuit to reach qualitatively relevant results at longer times. Efficiently estimating energy gradients is a huge undertaking, and this work does not implement some of the NISQ-friendly techniques that have been developed \cite{azadQuantumChemistryCalculations2022,ceroniTailgatingQuantumCircuits2022}, but it is expected to be a problem even in the fault-tolerant regime \cite{obrienEfficientQuantumComputation2022}.

\begin{figure}[h]
    \centering
    \includegraphics[width=1\columnwidth]{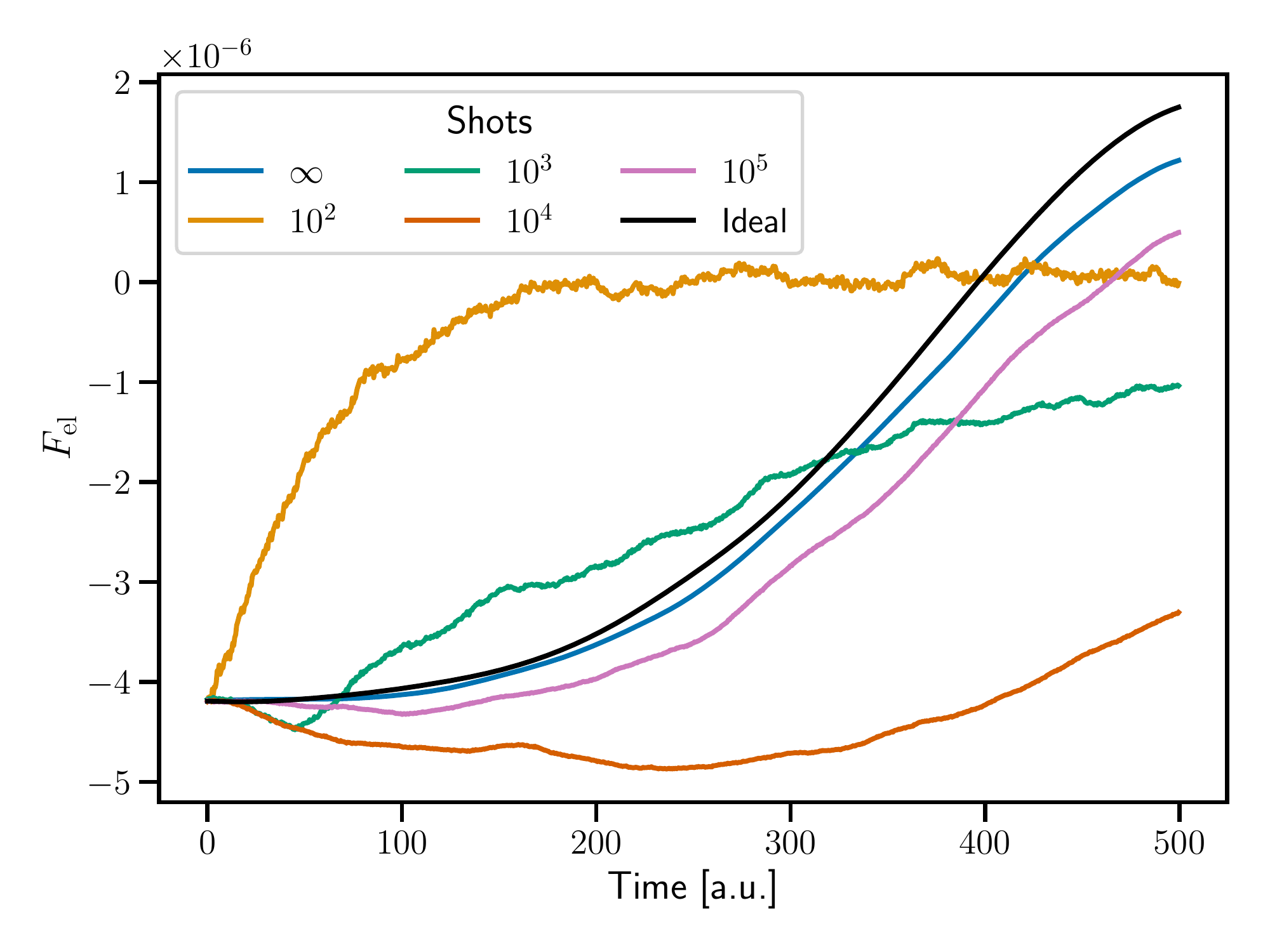}
    \caption{\textbf{Electron force observable for the single initialization}. The ideal simulation (black) and 100 instances of TDVQP evolutions (coloured) over 1000 timesteps of $0.5$ a.u. at varying shot counts per circuit. We observe qualitative agreement of TDVQP with the ideal case at $10^5$ shots and the infinite shot limit, with a shift downwards due to leakage to higher energy levels, which in this case biases the force in this direction.} 
    \label{fig:Sforce}
\end{figure}

\begin{figure}
    \centering
    \includegraphics[width=1\columnwidth]{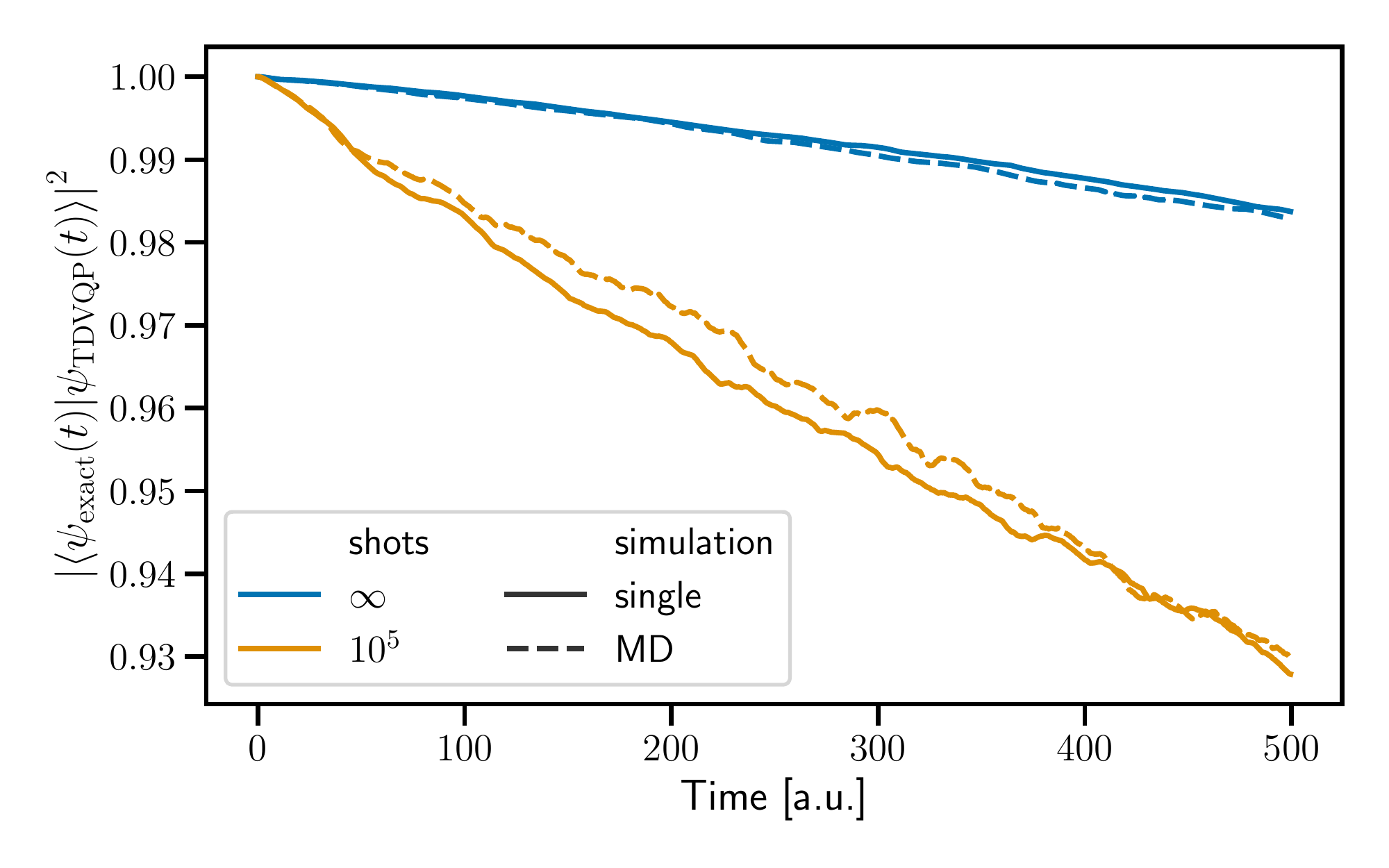}
    \caption{\textbf{Fidelity for the MD simulation} showing the 100 instances of TDVQP evolution (orange) over 1000 timesteps of $0.5$ a.u. for 100 different 'MD' trajectories and 'single' simulations. The MD simulation is of slightly higher fidelity compared to single trajectory simulations at high but finite shot counts. At lower shot counts the fidelity decreases too quickly in all cases.}
    \label{fig:Mforce}
\end{figure}

 We see in \rf{fig:Mfidelity} that the fidelity decays in all cases over time and that for long time evolutions, one requires more than $10^5$ shots when not using any additional techniques to better measure the force or better preserve the populations when not undergoing a transition. At lower shot counts the fidelity falls quickly, following eq.~\ref{eq:fidfall} until the equal superposition is approached, which sets a higher floor than zero for the decay of the fidelity. The potential effects of other noise sources are described in more detail in \ref{app:errors}. 
 \begin{figure}
    \centering
    \includegraphics[width=1\columnwidth]{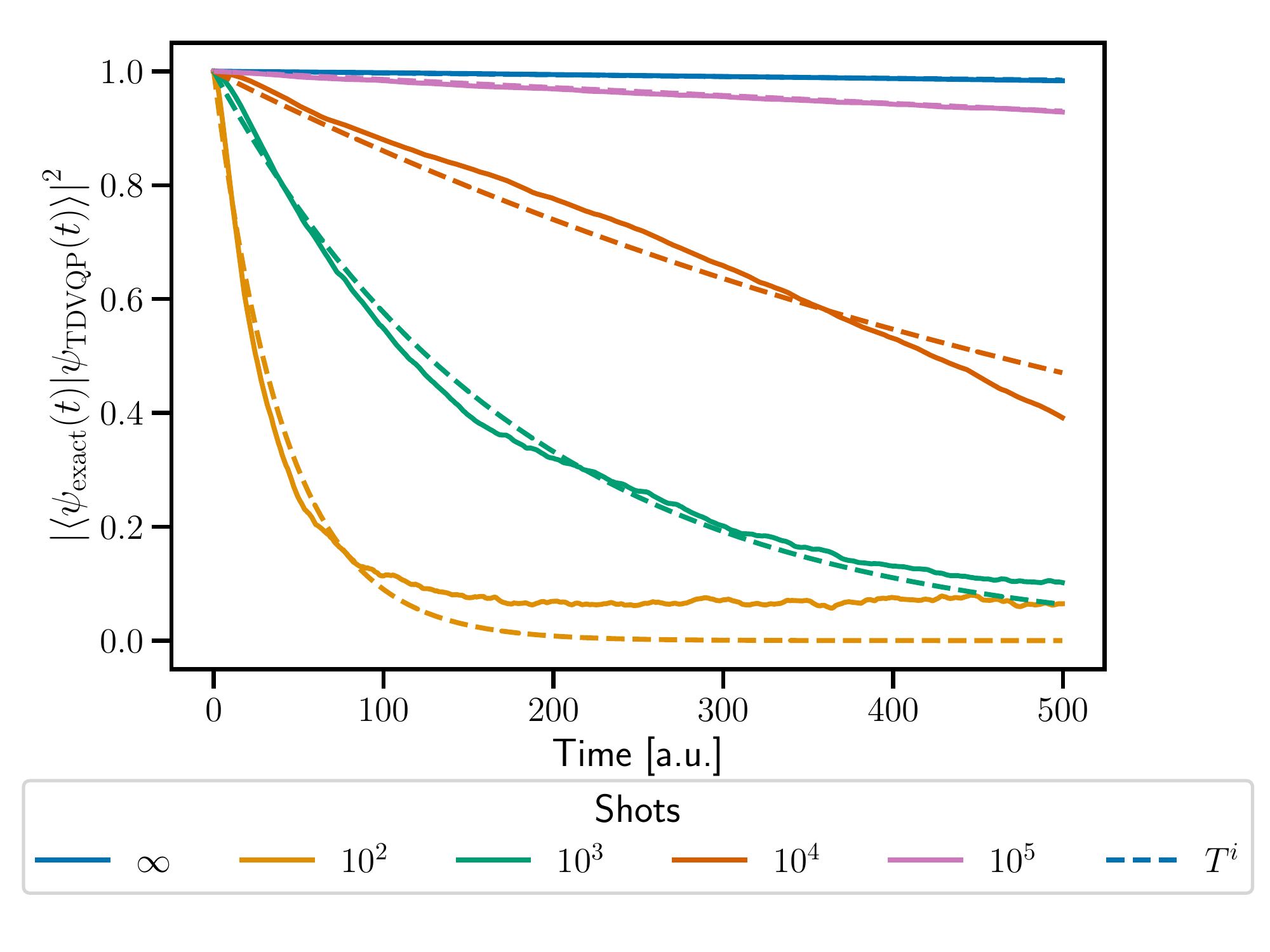}
    \caption{\textbf{TDVQP Fidelity over time} compared to the exact diagonalization evolution of the VQE initialized state over 1000 timesteps of $0.5$ a.u. for 100 single initializations (solid, coloured) at different shot counts. The best-fit approximation of eq.~\ref{eq:fidfall} (dashed) for each line is shown.}
    \label{fig:Mfidelity}
\end{figure}
 Figure~\ref{fig:Mforce} zooms into the two more reasonable fidelity lines, those of the infinite and $10^5$ shot simulations. Here we can see that using the multiple trajectories in an MD sense somewhat improves the simulation fidelity compared to the single trajectory case, and more quantum-tailored algorithms like \cite{kuroiwaQuantumCarParrinelloMolecular2022} may improve this further. In the infinite shot case, the difference is minimal - but due to the larger variance of the MD simulations compared to the ideal trajectory, the performance tends to be minimally worse.
 
\begin{figure}
    \centering
    \includegraphics[width=1\columnwidth]{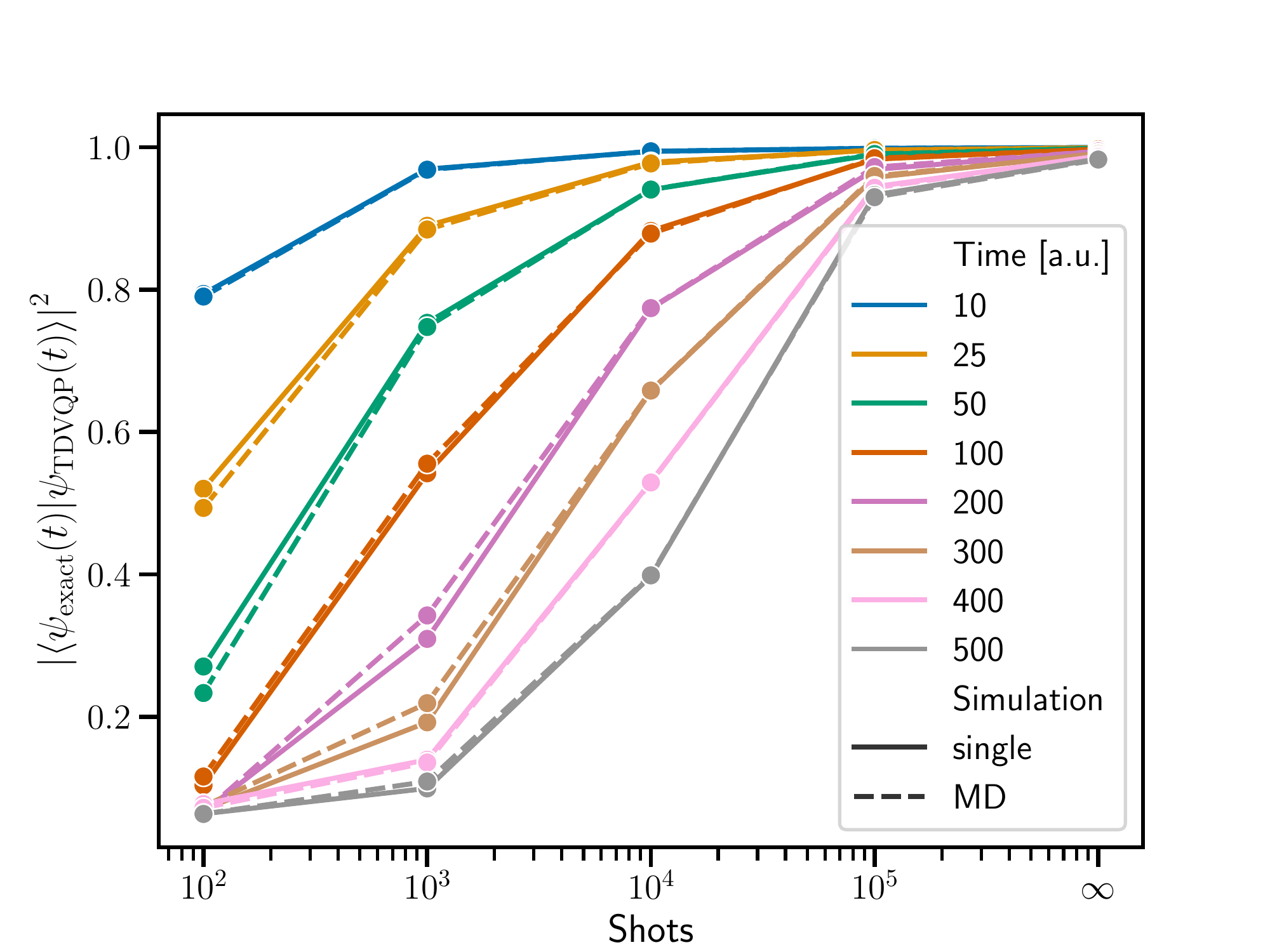}
    \caption{\textbf{TDVQP Fidelity over shots} of the exact diagonalization evolution of the VQE initialized state and the TDVQP evolution over 1000 timesteps of $0.5$ a.u. for 100 different MD-like trajectories (dashed, coloured) and 100 single initializations (solid, coloured) at different shot counts.}
    \label{fig:xshot}
\end{figure}

The relationship between shots and fidelity is also illustrated in Figure~\ref{fig:xshot}, where one can more clearly see that the MD simulation slightly improves the simulation at longer time evolutions when using finite shots. However, this improvement is not massive. It also highlights the large jump in fidelity gained when using higher shot counts. The p-VQD result \cite{barisonEfficientQuantumAlgorithm2021} on which we base our time evolution evolves its system for 40 iterations (20 a.u. here), where we see very high compression fidelities beyond $10^4$ shots per circuit evaluation.  

\begin{figure}
    \centering
    \includegraphics[width=1\columnwidth]{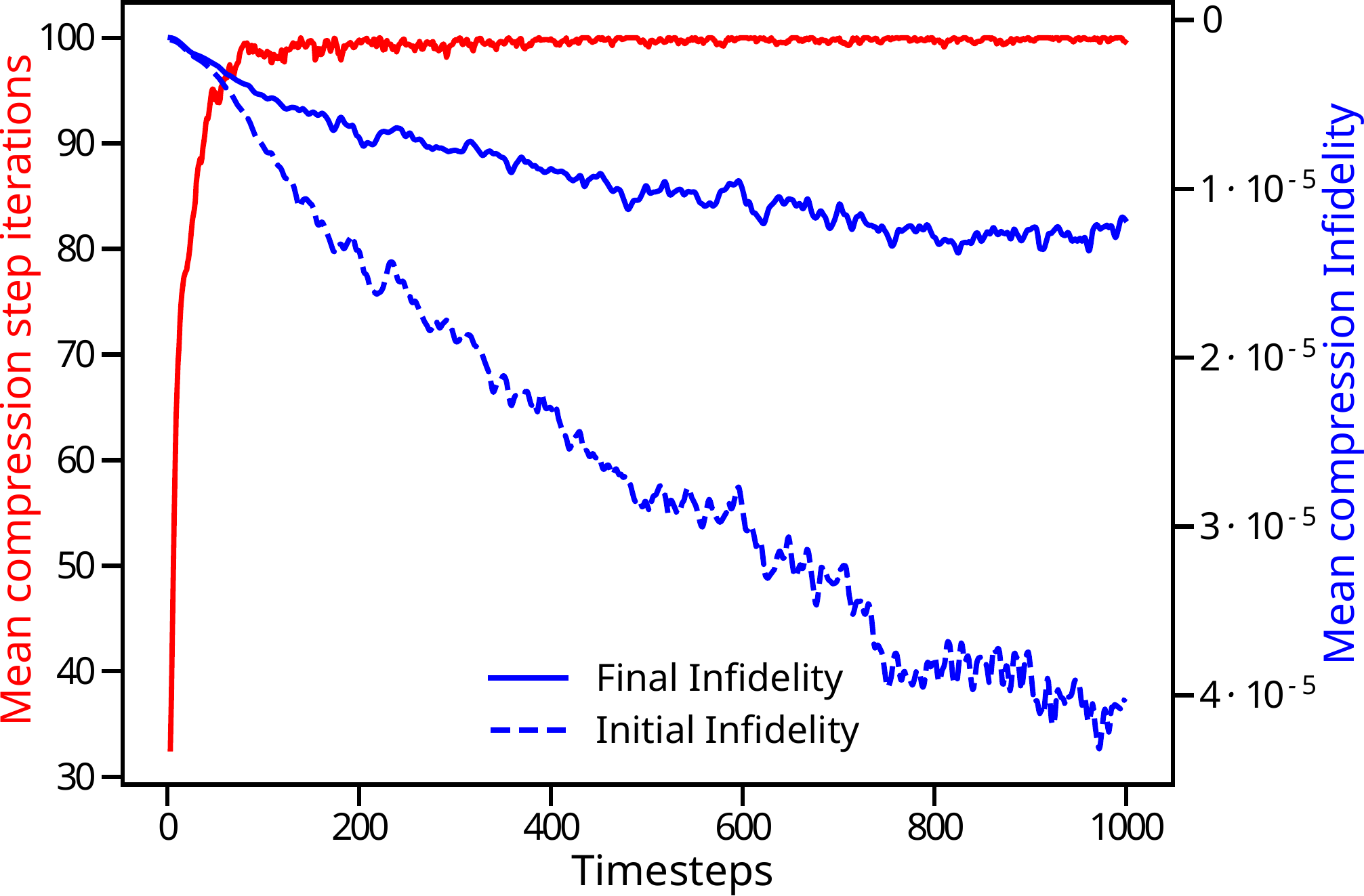}
    \caption{\textbf{Compression infidelity and number of iterations per step} for the single TDVQP setup at the infinite shot limit, showing the mean number of iterations in red with scale to the left. The mean initial infidelity (dashed, blue) and final infidelity (solid, blue) with the scale on the right, where infidelity ($I$) is $I=1-\text{Fidelity}$. The number of iterations taken by the optimizer quickly reaches the limit of 100, while the final and initial infidelities gradually fall with the number of timesteps into the simulation.}
    \label{fig:compfid}
\end{figure}

Finally, Figure~\ref{fig:compfid} illustrates that in the simulation the maximum number of iterations (100) is quickly reached before the 200\textsuperscript{th} iteration at the infinite shot limit. The overall mean final infidelity is $10^{-5}$, although the fitted threshold of eq.~\ref{eq:fidfall} for the overall algorithm is slightly lower at $3\cdot10^5$. The infidelity is $I=1-\mathcal{F}$, where $\mathcal{F}$ is the fidelity. This implies there is an additional error, likely due to the drift of the exact simulation of the system from the TDVQP simulation.  This is consistent with a $10^{-6}$ to $10^{-7}$ shift in the force as described in the numerical simulations in \ref{app:errors}, which is also roughly the difference in the force observable seen in Figure \ref{fig:Sforce}. 

Overall the results show some interesting behaviour. The number of timesteps modelled in this work is very high, and this results in only qualitative agreement in the region of interest where there is a significant population transfer. Furthermore in this simple model, the energy levels are well separated and we begin in the ground state. This results in a strong unidirectional contribution from population leakage to higher energy levels. In a more complex molecule, one would begin, for example, from a thermal ensemble of not only velocities but also states. In turn, the leakage would then result in deviations from both higher and lower energy levels and may be less detrimental to the ensemble average than here. 

\section{discussion}

Time-dependent evolution is an exceptionally interesting problem that can be well explored through quantum computers. Many techniques can be used for full quantum systems \cite{cirstoiuVariationalFastForwarding2020a,leeVariationalQuantumSimulation2022,lowHamiltonianSimulationQubitization2019,yaoAdaptiveVariationalQuantum2021,berthusenQuantumDynamicsSimulations2022,barisonEfficientQuantumAlgorithm2021} which are suitable to both near term and fault-tolerant machines. 

Algorithms that are suitable for MQC dynamics do require efficient and accurate full quantum dynamics, but the interplay between the classical and quantum systems brings a new spate of challenges. To exchange information, one must measure observables from the quantum system, which is expensive and destroys the state, requiring at minimum an efficient way to measure energy gradients, which is an area of active research \cite{azadQuantumChemistryCalculations2022,ceroniTailgatingQuantumCircuits2022,obrienCalculatingEnergyDerivatives2019}. This is a disadvantage, but it also means that one is limited to short-time evolutions between measurements. This makes it possible that a single trotter step is accurate enough \cite{babbushChemicalBasisTrotterSuzuki2015,trotterProductSemiGroupsOperators1959}, which is beneficial to near-term devices. 

Even though larger timesteps may be possible, the longer the time evolution, the longer the optimizer takes to find the time-evolved ansatz parameters. This is because the previous timestep parameters are no longer as close to the evolved ones. At the same time, most classical MQC methods do not update the parameters that govern the classical system's evolution at the small time intervals we use \cite{curchodInitioNonadiabaticQuantum2018}. It would be advantageous to use the largest possible classical timestep for a given integrator. To do this, one could do multiple compression steps with short-time Trotterizations using a constant Hamiltonian and only measuring the desired observables after the quantum system has evolved for the standard timestep of your classical problem, performing updates after this point. This will leverage the underlying compression algorithm to its fullest and reduce the overall number of measurements required.

The algorithm we present takes advantage of the above facts and is highly modular. Although the results are shown using an algorithm like p-VQD \cite{barisonEfficientQuantumAlgorithm2021} with Trotterization of the operator, there is no reason that other efficient time evolution algorithms couldn't be used. This is especially true if the time evolution operator could be efficiently represented by techniques other than the Trotterization of the Hamiltonian. The update step used here measures the Pauli string decomposition of the \(\frac{dH}{dR}\) matrix to compute forces, but other techniques exist in the fault-tolerant regime \cite{obrienCalculatingEnergyDerivatives2019,obrienEfficientQuantumComputation2022}, as well as in the NISQ regime \cite{azadQuantumChemistryCalculations2022,ceroniTailgatingQuantumCircuits2022}.
The main constraint with TDVQP is the fact that throughout the time evolution, there are inevitable inaccuracies in optimization, due to the compression step not preserving the populations in the diagonal basis as would have been expected as shown in \rf{fig:Spop}. This has the direct consequence that energy is not conserved, even though in the ideal simulation this is the case as shown in \rf{fig:Senergy}. Due to this accumulated error, fidelity falls consistently, and the effect is compounded when quantum resources are finite. 

These problems might be tackled by either increasing the threshold of the compression step or by measuring the energy and penalizing the optimizer when energy is not conserved. Another option that may be possible is designing an ansatz with problem-specific constraints \cite{gardEfficientSymmetrypreservingState2020}. Such an ansatz considers properties such as particle preservation within their structure, which may remove the need for expensive additional iteration steps or measurements. Furthermore, it may be possible to replace the p-VQD propagation with other compression methods \cite{berthusenQuantumDynamicsSimulations2022}. Since we are working in the first quantization representation, it may be difficult to find what properties to conserve in the wavefunction, but with that disadvantage, we gain an advantage in not needing to measure non-adiabatic couplings. 

We have also found that the error mostly comes from the compression step or due to finite sampling effects more than from the coupling to the classical system due to the small classical timestep.  Although it is always interesting to see how an algorithm behaves under noisy conditions, the performance of p-VQD under noise has been explored for full quantum dynamics in \cite{berthusenQuantumDynamicsSimulations2022}. This work focuses on the interplay between the scheme under the effect of a Hamiltonian which depends on the measured observables.

Overall we have introduced the TDVQP algorithm for MQC dynamics with the quantum subsystem computed on a quantum computer and have explored it on the Shin-Metiu model as an example of Ehrenfest dynamics in first quantization. However, it is not limited to this setting. It reproduces the expected observables and state evolution qualitatively. The algorithm is modular and refinements to it may be tackled in future research. Inaccuracies of the quantum computer can also be mitigated when computing ensemble averages of the classical properties. This work shows that MQC simulations may be practically feasible on noisy quantum computers if it is proven that variational quantum algorithms can have an advantage in chemical problems. 

\section{Method}

\subsection{Numerical simulations}
\label{sec:simparam}
To gauge the performance of the scheme we implement the Shin-Metiu model as described in Section~\ref{ssec:SM}. In the BOPES, we see an avoided crossing at around \(R=-1.9\) a.u. We initialize the system with the nucleus at an initial position of \(R=-2\) a.u. and an initial velocity of \(v_0=1.14\cdot 10^{-3}\) a.u., the average nuclear velocity from the Boltzmann distribution at 300K. The electronic system is initialized through the VQE with a random set of parameters and is allowed 300 iterations to approximate the ground state. The system is then evolved through the TDVQP algorithm with a timestep of \(\Delta t=0.5\) a.u. Each quantum time evolution step attempts to reach a fidelity threshold of $1-10^{-5}$ or up to 100 iterations of stochastic gradient descent \cite{robbinsStochasticApproximationMethod1951}. to find the optimal circuit parameters to approximate the previous time evolved state. Gradients were computed through the parameter shift rule \cite{wierichsGeneralParametershiftRules2022}. All simulations are done on 16 grid points that can be represented by 4 qubits. 

We examine two different situations. First, keep the initial conditions constant but sample different VQE ground state approximations, which we call the "Single Initial Condition" case. In the second case, we examine the MD-type approach in depth, where we sample a normal distribution of initial conditions for the initial velocity of the nucleus and allow one TDVQP evolution per sample. The velocity distribution is sampled from the Boltzmann distribution, only keeping positive velocities so that the nuclei approach the avoided crossing. The results shown are 100 samples that are evolved for 1000 timesteps which bring the classical trajectory beyond the avoided crossing point. 

Additional examples are provided for longer-time evolutions as well as for non-ground state evolution in Appendix~\ref{app:longtime}. Excited states and superpositions are prepared by using the uncomputation step of the TDVQP, but instead of starting the state with a known circuit from the VQE, the simulator is simply initialized to a desired arbitrary state, and the optimizer attempts to uncompute it with the ansatz and then those parameters are used as the initial step in place of the VQE. Various techniques to prepare excited states exist \cite{gochoExcitedStateCalculations2023,mccleanHybridQuantumclassicalHierarchy2017a}, but are not the focus of this work. 

We use two different metrics to establish the accuracy of the TDVQP algorithm: the so-called "Ideal" evolution begins at the desired state to numerical precision and is evolved by exact diagonalization. But, precise state preparation is another area of intense study \cite{aulicinoStatePreparationEvolution2022}. To better gauge the performance of the TDVQP in isolation, we also perform an "Exact" evolution, which uses the VQE-optimized initial state for evolution via exact diagonalization. This allows us to remove any bias from a poorly optimized ground state. 

\begin{figure}
    \centering
\includegraphics{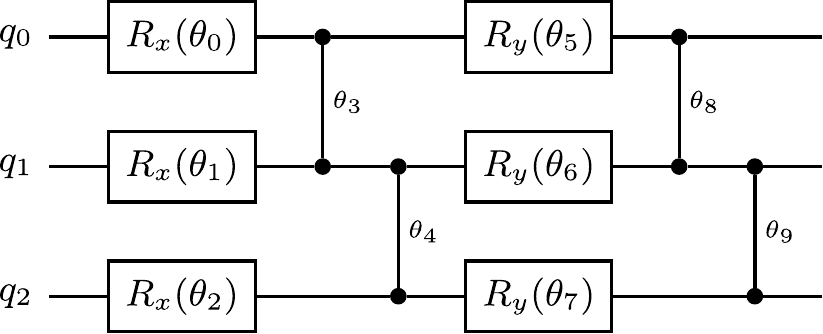}
    \caption{\textbf{One layer of the ansatz used for the VQE and TDVQP for three qubits.} If multiple layers are used then the above circuit is repeated. If more qubits are used then the vertical motif is continued. In both cases, more parameters can be added as needed and \(\vec \theta\) refers to the list of all parameters. The ansatz features $x$ rotations and the parameterized ZZ rotation.}
    \label{fig:ansatz}

\end{figure}

The VQE uses an ansatz of the form shown in Figure~\ref{fig:ansatz}, which was heuristically chosen as it can achieve ground state infidelities of up to \(10^{-5} \) on this system with 4 layers. We use the same ansatz as in \cite{barisonEfficientQuantumAlgorithm2021}, but various ansatze can be used, and for first quantization problems, in particular, there are some examples of how several different heuristic ansatze perform in \cite{ollitraultQuantumAlgorithmsGridbased2022}. The number of repetitions of the Trotterization layer is another important parameter, but as the decomposition of the Trotterized operator into native gates is deep, we limit ourselves to one. Although for full quantum dynamics, this would be very inaccurate for larger timesteps, the interaction with a classical system necessitates that we use short time steps, so that the Hamiltonian of the system is kept up to date with the classical state of the system. This means that a single trotter step is all that is needed, and the number of layers of the ansatz can compensate as shown in \ref{app:trotlay}.

The simulations were run on the Qiskit state vector simulator (version 0.28) \cite{aleksandrowiczQiskitOpensourceFramework2019} using the parameter-shift rule \cite{crooksGradientsParameterizedQuantum2019, wierichsGeneralParametershiftRules2022} to determine the analytic gradients required for gradient-descent based optimization.  Numpy \cite{harrisArrayProgrammingNumPy2020} was used for the exact numerical simulations, to prepare the Hamiltonian and to compute the velocity Verlet steps.

\subsection{Grid-based mapping to qubits}

We treat the Shin-Metiu system on the quantum computer in first quantization. We use a finite difference method on an equidistant grid. For low-dimensional problems, this is an appropriate approximation, but in general discrete variable representations (DVR) are a better choice for problems in higher dimensions. In quantum computing DVRs have been used to explore first quantization simulations in \cite{leeVariationalQuantumSimulation2022} using the Colbert and Miller DVR \cite{colbertNovelDiscreteVariable1992}. Issues exist with using DVRs on quantum computers as they generally require a full matrix Hamiltonian which is costly to measure and implement on quantum computers, and alternatives have been proposed \cite{ollitraultQuantumAlgorithmsGridbased2022}. 

Quantum computing in first quantization has the advantage that $n_g$ grid points can be represented by $N=\log_2(n_g)$ qubits. We choose $n_g=2^N$ to maximize the use of the $N$ available qubits. In the simplest finite differences method each position is an integer multiple $\nicefrac{L}{n_g}$. The grid point $g$ is by the quantum state $\ket{g}$ and mapped as
\begin{equation}
    \ket{g}=\ket{j_0}\otimes\dots\otimes\ket{j_k}\otimes\dots\otimes\ket{j_N},
\end{equation}
where $j_k$ is the k\textsuperscript{th} bit value of the binary representation of $g$. The potential operator \(\hat{V} \) is diagonal in this representation, simply sampling the potential at each grid point. The kinetic energy Hamiltonian is not diagonal in the position representation, and although one could use the split operator method \cite{hermannSplitoperatorSpectralMethod1988} to make it diagonal in momentum space would require a quantum Fourier transform implementation, which, as far as we know, cannot be effectively implemented on existing quantum devices. 

We sidestep all of the issues by using the finite differences method, in which the one-dimensional potential and the Laplacian form of the kinetic energy can be written as
    \begin{align}
        V_{j,j^\prime } & = V(x_j)\delta_{j,j^\prime} \\ 
        T_{j,j^\prime } &= \frac{-\partial ^2 }{\partial r^2} \frac{1}{2m_{el}}f,\ f 
        \begin{dcases}
            -2, &\text{ if } j=j^\prime  ;\\
            1, &\text{ if }  j=j^\prime \pm 1.
        \end{dcases} \\
    \end{align}

The tridiagonal matrix that results has the same value on the off-diagonal terms, which allows them to be decomposed into fewer Pauli strings than a full matrix via an elegant recursive form that is described in \ref{app:tridiagonal} following from \cite{guhneToolboxEntanglementDetection2007}. This is important because the number of non-zero entries is related to the number of terms in the Pauli decomposition, which should be kept minimal to reduce the length of the Trotterized time evolution operator and observable measurements. The finite difference method does require high grid densities to be accurate (although this depends on how oscillatory the system in question is), but this requirement will likely be met by the doubling of grid points per additional qubit.  

Although not implemented in this paper, some algorithms can make the time evolution of Hamiltonians with this form more efficient on near-term devices \cite{kalevQuantumAlgorithmSimulating2021}. Depending on the particular Hamiltonian one chooses to study in this way, different efficient algorithms exist to lessen the cost of the time evolution such as variational fast forwarding and qubitization \cite{cirstoiuVariationalFastForwarding2020,lowHamiltonianSimulationQubitization2019}. It is also possible to efficiently solve for the eigenstates of tridiagonal matrices on quantum computers \cite{wangQuantumEigensolverSymmetric2019}.

\subsection{Circuit compression}

A key building block of the presented algorithm and a fundamental aspect of quantum circuit optimization is the concept of circuit compression \cite{rakytaEfficientQuantumGate2022}. Any operation on a quantum computer must be a unitary operation \(\hat{U} \), but the quantum computer only has a finite set of few-qubit gates. An arbitrary \(\hat{U} \) must be expressed, or compiled, into a set of native gates \cite{nielsenQuantumComputationQuantum2010}, this can always be done, but it is an NP-hard problem. If you allow yourself to implement an approximation of \(\hat{U} \) within some threshold, you may find \(\tilde{U}\) which might have a shorter circuit length than even an optimal decomposition of \(\hat{U} \). This latter definition is what is generally known as circuit compression, although the term is sometimes used to refer to more optimal perfect decompositions \cite{kokcuAlgebraicCompressionQuantum2022}.

With an error-corrected quantum computer, one could use arbitrary circuit depths, but NISQ hardware benefits from using short circuits to minimize errors from occurring. But designing hardware efficient ansatze for VQAs is an unsolved problem \cite{fedorovVQEMethodShort2022}. This means that for most purposes we use a heuristic ansatz \(\hat{C} \) parameterized by some vector \(\theta \) brings the initial computational state \(\ket{0} \) to a desired state \(\ket{\psi } \) via \(\hat{C} (\theta )\ket{0} =\ket{\psi } \). The defining property of unitary matrices, namely

\begin{equation}\label{eq:unitary}
    UU^\dagger=U^\dagger U=I,
\end{equation}  
ensures that $C(\theta)^\dagger$  reverses the action of $C(\theta)$. If we add a unitary \(\hat{U} \), then it may be possible to find some parameters $\theta'$ such that \(\hat{U} \hat{C} (\theta )\ket{0} \approx\hat{C} (\theta ^\prime)\ket{0}\), thus compressing the action of the unitary back into the same quantum circuit, at least approximately. which is what is exploited by \cite{linRealImaginaryTimeEvolution2021,barisonEfficientQuantumAlgorithm2021, berthusenQuantumDynamicsSimulations2022}.

Another approach is to approximate an initial state with such an ansatz and then perform a short-time evolution via Trotterization of the time-evolved operator as is done in \cite{gardEfficientSymmetrypreservingState2020,flickCavityBornOppenheimer2017,swopeComputerSimulationMethod1982}. The adjoint of the ansatz is appended to the circuit and its parameters varied such that the machine state is 'uncomputed' to its initial state. If one must assume that the chosen circuit ansatz is expressive enough to capture the entire state's time evolution, then such an approach is guaranteed to work. One then simply needs this new set of parameters and the original ansatz to express the new timestep without the time evolution operator, hence compressing the circuit. This idea has been implemented almost concurrently by Lin et al. \cite{linRealImaginaryTimeEvolution2021}, and by Barison et al.'s  "projected variational quantum dynamics" (p-VQD) algorithm \cite{barisonEfficientQuantumAlgorithm2021}. Subsequent works, for example, \cite{berthusenQuantumDynamicsSimulations2022}, have built on the circuit compression idea.

\section*{Code availability}
The code required for simulating and plotting all the figures can be found at https://zenodo.org/record/8238985.

\begin{acknowledgments}
This work was supported by the European Union’s Horizon 2020 research and innovation programme under the Marie Sk\l{}odowska-Curie grant agreement No. 955479. Computing resources were provided by the state of Baden-Württemberg through bwHPC and the German Research Foundation (DFG) through grant INST 35/1597-1 FUGG.
\end{acknowledgments}
\bibliography{Library.bib}

\begin{thebibliography}{54}%
\makeatletter
\providecommand \@ifxundefined [1]{%
 \@ifx{#1\undefined}
}%
\providecommand \@ifnum [1]{%
 \ifnum #1\expandafter \@firstoftwo
 \else \expandafter \@secondoftwo
 \fi
}%
\providecommand \@ifx [1]{%
 \ifx #1\expandafter \@firstoftwo
 \else \expandafter \@secondoftwo
 \fi
}%
\providecommand \natexlab [1]{#1}%
\providecommand \enquote  [1]{``#1''}%
\providecommand \bibnamefont  [1]{#1}%
\providecommand \bibfnamefont [1]{#1}%
\providecommand \citenamefont [1]{#1}%
\providecommand \href@noop [0]{\@secondoftwo}%
\providecommand \href [0]{\begingroup \@sanitize@url \@href}%
\providecommand \@href[1]{\@@startlink{#1}\@@href}%
\providecommand \@@href[1]{\endgroup#1\@@endlink}%
\providecommand \@sanitize@url [0]{\catcode `\\12\catcode `\$12\catcode `\&12\catcode `\#12\catcode `\^12\catcode `\_12\catcode `\%12\relax}%
\providecommand \@@startlink[1]{}%
\providecommand \@@endlink[0]{}%
\providecommand \url  [0]{\begingroup\@sanitize@url \@url }%
\providecommand \@url [1]{\endgroup\@href {#1}{\urlprefix }}%
\providecommand \urlprefix  [0]{URL }%
\providecommand \Eprint [0]{\href }%
\providecommand \doibase [0]{https://doi.org/}%
\providecommand \selectlanguage [0]{\@gobble}%
\providecommand \bibinfo  [0]{\@secondoftwo}%
\providecommand \bibfield  [0]{\@secondoftwo}%
\providecommand \translation [1]{[#1]}%
\providecommand \BibitemOpen [0]{}%
\providecommand \bibitemStop [0]{}%
\providecommand \bibitemNoStop [0]{.\EOS\space}%
\providecommand \EOS [0]{\spacefactor3000\relax}%
\providecommand \BibitemShut  [1]{\csname bibitem#1\endcsname}%
\let\auto@bib@innerbib\@empty
\bibitem [{\citenamefont {Peruzzo}\ \emph {et~al.}(2014)\citenamefont {Peruzzo}, \citenamefont {McClean}, \citenamefont {Shadbolt}, \citenamefont {Yung}, \citenamefont {Zhou}, \citenamefont {Love}, \citenamefont {{Aspuru-Guzik}},\ and\ \citenamefont {O'Brien}}]{peruzzoVariationalEigenvalueSolver2014}%
  \BibitemOpen
  \bibfield  {author} {\bibinfo {author} {\bibfnamefont {A.}~\bibnamefont {Peruzzo}}, \bibinfo {author} {\bibfnamefont {J.}~\bibnamefont {McClean}}, \bibinfo {author} {\bibfnamefont {P.}~\bibnamefont {Shadbolt}}, \bibinfo {author} {\bibfnamefont {M.-H.}\ \bibnamefont {Yung}}, \bibinfo {author} {\bibfnamefont {X.-Q.}\ \bibnamefont {Zhou}}, \bibinfo {author} {\bibfnamefont {P.~J.}\ \bibnamefont {Love}}, \bibinfo {author} {\bibfnamefont {A.}~\bibnamefont {{Aspuru-Guzik}}},\ and\ \bibinfo {author} {\bibfnamefont {J.~L.}\ \bibnamefont {O'Brien}},\ }\bibfield  {title} {\bibinfo {title} {A variational eigenvalue solver on a photonic quantum processor},\ }\href {https://doi.org/10.1038/ncomms5213} {\bibfield  {journal} {\bibinfo  {journal} {Nature Communications}\ }\textbf {\bibinfo {volume} {5}},\ \bibinfo {pages} {4213} (\bibinfo {year} {2014})}\BibitemShut {NoStop}%
\bibitem [{\citenamefont {Cerezo}\ \emph {et~al.}(2020)\citenamefont {Cerezo}, \citenamefont {Arrasmith}, \citenamefont {Babbush}, \citenamefont {Benjamin}, \citenamefont {Endo}, \citenamefont {Fujii}, \citenamefont {McClean}, \citenamefont {Mitarai}, \citenamefont {Yuan}, \citenamefont {Cincio},\ and\ \citenamefont {Coles}}]{cerezoVariationalQuantumAlgorithms2020}%
  \BibitemOpen
  \bibfield  {author} {\bibinfo {author} {\bibfnamefont {M.}~\bibnamefont {Cerezo}}, \bibinfo {author} {\bibfnamefont {A.}~\bibnamefont {Arrasmith}}, \bibinfo {author} {\bibfnamefont {R.}~\bibnamefont {Babbush}}, \bibinfo {author} {\bibfnamefont {S.~C.}\ \bibnamefont {Benjamin}}, \bibinfo {author} {\bibfnamefont {S.}~\bibnamefont {Endo}}, \bibinfo {author} {\bibfnamefont {K.}~\bibnamefont {Fujii}}, \bibinfo {author} {\bibfnamefont {J.~R.}\ \bibnamefont {McClean}}, \bibinfo {author} {\bibfnamefont {K.}~\bibnamefont {Mitarai}}, \bibinfo {author} {\bibfnamefont {X.}~\bibnamefont {Yuan}}, \bibinfo {author} {\bibfnamefont {L.}~\bibnamefont {Cincio}},\ and\ \bibinfo {author} {\bibfnamefont {P.~J.}\ \bibnamefont {Coles}},\ }\bibfield  {title} {\bibinfo {title} {Variational {{Quantum Algorithms}}},\ }\href@noop {} {\bibfield  {journal} {\bibinfo  {journal} {arXiv:2012.09265 [quant-ph, stat]}\ } (\bibinfo {year} {2020})},\ \Eprint {https://arxiv.org/abs/2012.09265} {arxiv:2012.09265 [quant-ph, stat]} \BibitemShut {NoStop}%
\bibitem [{\citenamefont {Ollitrault}\ \emph {et~al.}(2021)\citenamefont {Ollitrault}, \citenamefont {Miessen},\ and\ \citenamefont {Tavernelli}}]{ollitraultMolecularQuantumDynamics2021}%
  \BibitemOpen
  \bibfield  {author} {\bibinfo {author} {\bibfnamefont {P.~J.}\ \bibnamefont {Ollitrault}}, \bibinfo {author} {\bibfnamefont {A.}~\bibnamefont {Miessen}},\ and\ \bibinfo {author} {\bibfnamefont {I.}~\bibnamefont {Tavernelli}},\ }\bibfield  {title} {\bibinfo {title} {Molecular {{Quantum Dynamics}}: {{A Quantum Computing Perspective}}},\ }\bibfield  {journal} {\bibinfo  {journal} {Accounts of Chemical Research}\ }\href {https://doi.org/10.1021/acs.accounts.1c00514} {10.1021/acs.accounts.1c00514} (\bibinfo {year} {2021})\BibitemShut {NoStop}%
\bibitem [{\citenamefont {Curchod}\ and\ \citenamefont {Mart{\'i}nez}(2018)}]{curchodInitioNonadiabaticQuantum2018}%
  \BibitemOpen
  \bibfield  {author} {\bibinfo {author} {\bibfnamefont {B.~F.~E.}\ \bibnamefont {Curchod}}\ and\ \bibinfo {author} {\bibfnamefont {T.~J.}\ \bibnamefont {Mart{\'i}nez}},\ }\bibfield  {title} {\bibinfo {title} {Ab {{Initio Nonadiabatic Quantum Molecular Dynamics}}},\ }\href {https://doi.org/10.1021/acs.chemrev.7b00423} {\bibfield  {journal} {\bibinfo  {journal} {Chemical Reviews}\ }\textbf {\bibinfo {volume} {118}},\ \bibinfo {pages} {3305} (\bibinfo {year} {2018})}\BibitemShut {NoStop}%
\bibitem [{\citenamefont {Kirrander}\ and\ \citenamefont {Vacher}(2020)}]{kirranderEhrenfestMethodsElectron2020}%
  \BibitemOpen
  \bibfield  {author} {\bibinfo {author} {\bibfnamefont {A.}~\bibnamefont {Kirrander}}\ and\ \bibinfo {author} {\bibfnamefont {M.}~\bibnamefont {Vacher}},\ }\bibfield  {title} {\bibinfo {title} {Ehrenfest {{Methods}} for {{Electron}} and {{Nuclear Dynamics}}},\ }in\ \href {https://doi.org/10.1002/9781119417774.ch15} {\emph {\bibinfo {booktitle} {Quantum {{Chemistry}} and {{Dynamics}} of {{Excited States}}}}}\ (\bibinfo  {publisher} {{John Wiley \& Sons, Ltd}},\ \bibinfo {year} {2020})\ Chap.~\bibinfo {chapter} {15}, pp.\ \bibinfo {pages} {469--497}\BibitemShut {NoStop}%
\bibitem [{\citenamefont {Ollitrault}\ \emph {et~al.}(2020)\citenamefont {Ollitrault}, \citenamefont {Mazzola},\ and\ \citenamefont {Tavernelli}}]{ollitraultNonadiabaticMolecularQuantum2020}%
  \BibitemOpen
  \bibfield  {author} {\bibinfo {author} {\bibfnamefont {P.~J.}\ \bibnamefont {Ollitrault}}, \bibinfo {author} {\bibfnamefont {G.}~\bibnamefont {Mazzola}},\ and\ \bibinfo {author} {\bibfnamefont {I.}~\bibnamefont {Tavernelli}},\ }\bibfield  {title} {\bibinfo {title} {Nonadiabatic {{Molecular Quantum Dynamics}} with {{Quantum Computers}}},\ }\href {https://doi.org/10.1103/PhysRevLett.125.260511} {\bibfield  {journal} {\bibinfo  {journal} {Physical Review Letters}\ }\textbf {\bibinfo {volume} {125}},\ \bibinfo {pages} {260511} (\bibinfo {year} {2020})}\BibitemShut {NoStop}%
\bibitem [{\citenamefont {Sokolov}\ \emph {et~al.}(2021)\citenamefont {Sokolov}, \citenamefont {Barkoutsos}, \citenamefont {Moeller}, \citenamefont {Suchsland}, \citenamefont {Mazzola},\ and\ \citenamefont {Tavernelli}}]{sokolovMicrocanonicalFinitetemperatureInitio2021}%
  \BibitemOpen
  \bibfield  {author} {\bibinfo {author} {\bibfnamefont {I.~O.}\ \bibnamefont {Sokolov}}, \bibinfo {author} {\bibfnamefont {P.~K.}\ \bibnamefont {Barkoutsos}}, \bibinfo {author} {\bibfnamefont {L.}~\bibnamefont {Moeller}}, \bibinfo {author} {\bibfnamefont {P.}~\bibnamefont {Suchsland}}, \bibinfo {author} {\bibfnamefont {G.}~\bibnamefont {Mazzola}},\ and\ \bibinfo {author} {\bibfnamefont {I.}~\bibnamefont {Tavernelli}},\ }\bibfield  {title} {\bibinfo {title} {Microcanonical and finite-temperature ab initio molecular dynamics simulations on quantum computers},\ }\href {https://doi.org/10.1103/PhysRevResearch.3.013125} {\bibfield  {journal} {\bibinfo  {journal} {Physical Review Research}\ }\textbf {\bibinfo {volume} {3}},\ \bibinfo {pages} {013125} (\bibinfo {year} {2021})}\BibitemShut {NoStop}%
\bibitem [{\citenamefont {Rossmannek}\ \emph {et~al.}(2020)\citenamefont {Rossmannek}, \citenamefont {Barkoutsos}, \citenamefont {Ollitrault},\ and\ \citenamefont {Tavernelli}}]{rossmannekQuantumHFDFTEmbedding2020}%
  \BibitemOpen
  \bibfield  {author} {\bibinfo {author} {\bibfnamefont {M.}~\bibnamefont {Rossmannek}}, \bibinfo {author} {\bibfnamefont {P.~K.}\ \bibnamefont {Barkoutsos}}, \bibinfo {author} {\bibfnamefont {P.~J.}\ \bibnamefont {Ollitrault}},\ and\ \bibinfo {author} {\bibfnamefont {I.}~\bibnamefont {Tavernelli}},\ }\bibfield  {title} {\bibinfo {title} {Quantum {{HF}}/{{DFT-Embedding Algorithms}} for {{Electronic Structure Calculations}}: {{Scaling}} up to {{Complex Molecular Systems}}},\ }\href@noop {} {\bibfield  {journal} {\bibinfo  {journal} {arXiv:2009.01872 [physics, physics:quant-ph]}\ } (\bibinfo {year} {2020})},\ \Eprint {https://arxiv.org/abs/2009.01872} {arxiv:2009.01872 [physics, physics:quant-ph]} \BibitemShut {NoStop}%
\bibitem [{\citenamefont {Levine}\ \emph {et~al.}(2020)\citenamefont {Levine}, \citenamefont {Hait}, \citenamefont {Tubman}, \citenamefont {Lehtola}, \citenamefont {Whaley},\ and\ \citenamefont {{Head-Gordon}}}]{levineCASSCFExtremelyLarge2020}%
  \BibitemOpen
  \bibfield  {author} {\bibinfo {author} {\bibfnamefont {D.~S.}\ \bibnamefont {Levine}}, \bibinfo {author} {\bibfnamefont {D.}~\bibnamefont {Hait}}, \bibinfo {author} {\bibfnamefont {N.~M.}\ \bibnamefont {Tubman}}, \bibinfo {author} {\bibfnamefont {S.}~\bibnamefont {Lehtola}}, \bibinfo {author} {\bibfnamefont {K.~B.}\ \bibnamefont {Whaley}},\ and\ \bibinfo {author} {\bibfnamefont {M.}~\bibnamefont {{Head-Gordon}}},\ }\bibfield  {title} {\bibinfo {title} {{{CASSCF}} with {{Extremely Large Active Spaces Using}} the {{Adaptive Sampling Configuration Interaction Method}}},\ }\href {https://doi.org/10.1021/acs.jctc.9b01255} {\bibfield  {journal} {\bibinfo  {journal} {Journal of Chemical Theory and Computation}\ }\textbf {\bibinfo {volume} {16}},\ \bibinfo {pages} {2340} (\bibinfo {year} {2020})}\BibitemShut {NoStop}%
\bibitem [{\citenamefont {Lin}\ \emph {et~al.}(2021)\citenamefont {Lin}, \citenamefont {Dilip}, \citenamefont {Green}, \citenamefont {Smith},\ and\ \citenamefont {Pollmann}}]{linRealImaginaryTimeEvolution2021}%
  \BibitemOpen
  \bibfield  {author} {\bibinfo {author} {\bibfnamefont {S.-H.}\ \bibnamefont {Lin}}, \bibinfo {author} {\bibfnamefont {R.}~\bibnamefont {Dilip}}, \bibinfo {author} {\bibfnamefont {A.~G.}\ \bibnamefont {Green}}, \bibinfo {author} {\bibfnamefont {A.}~\bibnamefont {Smith}},\ and\ \bibinfo {author} {\bibfnamefont {F.}~\bibnamefont {Pollmann}},\ }\bibfield  {title} {\bibinfo {title} {Real- and {{Imaginary-Time Evolution}} with {{Compressed Quantum Circuits}}},\ }\href {https://doi.org/10.1103/PRXQuantum.2.010342} {\bibfield  {journal} {\bibinfo  {journal} {PRX Quantum}\ }\textbf {\bibinfo {volume} {2}},\ \bibinfo {pages} {010342} (\bibinfo {year} {2021})}\BibitemShut {NoStop}%
\bibitem [{\citenamefont {Barison}\ \emph {et~al.}(2021)\citenamefont {Barison}, \citenamefont {Vicentini},\ and\ \citenamefont {Carleo}}]{barisonEfficientQuantumAlgorithm2021}%
  \BibitemOpen
  \bibfield  {author} {\bibinfo {author} {\bibfnamefont {S.}~\bibnamefont {Barison}}, \bibinfo {author} {\bibfnamefont {F.}~\bibnamefont {Vicentini}},\ and\ \bibinfo {author} {\bibfnamefont {G.}~\bibnamefont {Carleo}},\ }\bibfield  {title} {\bibinfo {title} {An efficient quantum algorithm for the time evolution of parameterized circuits},\ }\href {https://doi.org/10.22331/q-2021-07-28-512} {\bibfield  {journal} {\bibinfo  {journal} {Quantum}\ }\textbf {\bibinfo {volume} {5}},\ \bibinfo {pages} {512} (\bibinfo {year} {2021})}\BibitemShut {NoStop}%
\bibitem [{\citenamefont {Berthusen}\ \emph {et~al.}(2022)\citenamefont {Berthusen}, \citenamefont {Trevisan}, \citenamefont {Iadecola},\ and\ \citenamefont {Orth}}]{berthusenQuantumDynamicsSimulations2022}%
  \BibitemOpen
  \bibfield  {author} {\bibinfo {author} {\bibfnamefont {N.~F.}\ \bibnamefont {Berthusen}}, \bibinfo {author} {\bibfnamefont {T.~V.}\ \bibnamefont {Trevisan}}, \bibinfo {author} {\bibfnamefont {T.}~\bibnamefont {Iadecola}},\ and\ \bibinfo {author} {\bibfnamefont {P.~P.}\ \bibnamefont {Orth}},\ }\bibfield  {title} {\bibinfo {title} {Quantum dynamics simulations beyond the coherence time on noisy intermediate-scale quantum hardware by variational {{Trotter}} compression},\ }\href {https://doi.org/10.1103/PhysRevResearch.4.023097} {\bibfield  {journal} {\bibinfo  {journal} {Physical Review Research}\ }\textbf {\bibinfo {volume} {4}},\ \bibinfo {pages} {023097} (\bibinfo {year} {2022})}\BibitemShut {NoStop}%
\bibitem [{\citenamefont {Shin}\ and\ \citenamefont {Metiu}(1995)}]{shinNonadiabaticEffectsCharge1995}%
  \BibitemOpen
  \bibfield  {author} {\bibinfo {author} {\bibfnamefont {S.}~\bibnamefont {Shin}}\ and\ \bibinfo {author} {\bibfnamefont {H.}~\bibnamefont {Metiu}},\ }\bibfield  {title} {\bibinfo {title} {Nonadiabatic effects on the charge transfer rate constant: {{A}} numerical study of a simple model system},\ }\href {https://doi.org/10.1063/1.468795} {\bibfield  {journal} {\bibinfo  {journal} {The Journal of Chemical Physics}\ }\textbf {\bibinfo {volume} {102}},\ \bibinfo {pages} {9285} (\bibinfo {year} {1995})}\BibitemShut {NoStop}%
\bibitem [{\citenamefont {Albareda}\ \emph {et~al.}(2016)\citenamefont {Albareda}, \citenamefont {Abedi}, \citenamefont {Tavernelli},\ and\ \citenamefont {Rubio}}]{albaredaUniversalStepsQuantum2016}%
  \BibitemOpen
  \bibfield  {author} {\bibinfo {author} {\bibfnamefont {G.}~\bibnamefont {Albareda}}, \bibinfo {author} {\bibfnamefont {A.}~\bibnamefont {Abedi}}, \bibinfo {author} {\bibfnamefont {I.}~\bibnamefont {Tavernelli}},\ and\ \bibinfo {author} {\bibfnamefont {A.}~\bibnamefont {Rubio}},\ }\bibfield  {title} {\bibinfo {title} {Universal steps in quantum dynamics with time-dependent potential-energy surfaces: {{Beyond}} the {{Born-Oppenheimer}} picture},\ }\href {https://doi.org/10.1103/PhysRevA.94.062511} {\bibfield  {journal} {\bibinfo  {journal} {Physical Review A}\ }\textbf {\bibinfo {volume} {94}},\ \bibinfo {pages} {062511} (\bibinfo {year} {2016})}\BibitemShut {NoStop}%
\bibitem [{\citenamefont {Erdmann}\ \emph {et~al.}(2003)\citenamefont {Erdmann}, \citenamefont {Marquetand},\ and\ \citenamefont {Engel}}]{erdmannCombinedElectronicNuclear2003}%
  \BibitemOpen
  \bibfield  {author} {\bibinfo {author} {\bibfnamefont {M.}~\bibnamefont {Erdmann}}, \bibinfo {author} {\bibfnamefont {P.}~\bibnamefont {Marquetand}},\ and\ \bibinfo {author} {\bibfnamefont {V.}~\bibnamefont {Engel}},\ }\bibfield  {title} {\bibinfo {title} {Combined electronic and nuclear dynamics in a simple model system},\ }\href {https://doi.org/10.1063/1.1578618} {\bibfield  {journal} {\bibinfo  {journal} {The Journal of Chemical Physics}\ }\textbf {\bibinfo {volume} {119}},\ \bibinfo {pages} {672} (\bibinfo {year} {2003})}\BibitemShut {NoStop}%
\bibitem [{\citenamefont {Falge}\ \emph {et~al.}(2012)\citenamefont {Falge}, \citenamefont {Engel}, \citenamefont {Lein}, \citenamefont {{Vindel-Zandbergen}}, \citenamefont {Chang},\ and\ \citenamefont {Sola}}]{falgeQuantumWavePacketDynamics2012}%
  \BibitemOpen
  \bibfield  {author} {\bibinfo {author} {\bibfnamefont {M.}~\bibnamefont {Falge}}, \bibinfo {author} {\bibfnamefont {V.}~\bibnamefont {Engel}}, \bibinfo {author} {\bibfnamefont {M.}~\bibnamefont {Lein}}, \bibinfo {author} {\bibfnamefont {P.}~\bibnamefont {{Vindel-Zandbergen}}}, \bibinfo {author} {\bibfnamefont {B.~Y.}\ \bibnamefont {Chang}},\ and\ \bibinfo {author} {\bibfnamefont {I.~R.}\ \bibnamefont {Sola}},\ }\bibfield  {title} {\bibinfo {title} {Quantum {{Wave-Packet Dynamics}} in {{Spin-Coupled Vibronic States}}},\ }\href {https://doi.org/10.1021/jp306566x} {\bibfield  {journal} {\bibinfo  {journal} {The Journal of Physical Chemistry A}\ }\textbf {\bibinfo {volume} {116}},\ \bibinfo {pages} {11427} (\bibinfo {year} {2012})}\BibitemShut {NoStop}%
\bibitem [{\citenamefont {Gossel}\ \emph {et~al.}(2019)\citenamefont {Gossel}, \citenamefont {Lacombe},\ and\ \citenamefont {Maitra}}]{gosselNumericalSolutionExact2019}%
  \BibitemOpen
  \bibfield  {author} {\bibinfo {author} {\bibfnamefont {G.~H.}\ \bibnamefont {Gossel}}, \bibinfo {author} {\bibfnamefont {L.}~\bibnamefont {Lacombe}},\ and\ \bibinfo {author} {\bibfnamefont {N.~T.}\ \bibnamefont {Maitra}},\ }\bibfield  {title} {\bibinfo {title} {On the numerical solution of the exact factorization equations},\ }\href {https://doi.org/10.1063/1.5090802} {\bibfield  {journal} {\bibinfo  {journal} {The Journal of Chemical Physics}\ }\textbf {\bibinfo {volume} {150}},\ \bibinfo {pages} {154112} (\bibinfo {year} {2019})}\BibitemShut {NoStop}%
\bibitem [{\citenamefont {Nielsen}\ and\ \citenamefont {Chuang}(2010)}]{nielsenQuantumComputationQuantum2010}%
  \BibitemOpen
  \bibfield  {author} {\bibinfo {author} {\bibfnamefont {M.~A.}\ \bibnamefont {Nielsen}}\ and\ \bibinfo {author} {\bibfnamefont {I.~L.}\ \bibnamefont {Chuang}},\ }\href {https://doi.org/10.1017/CBO9780511976667} {\bibinfo {title} {Quantum {{Computation}} and {{Quantum Information}}: 10th {{Anniversary Edition}}}},\ \bibinfo {howpublished} {https://www.cambridge.org/highereducation/books/quantum-computation-and-quantum-information/01E10196D0A682A6AEFFEA52D53BE9AE} (\bibinfo {year} {2010})\BibitemShut {NoStop}%
\bibitem [{\citenamefont {Bharti}\ \emph {et~al.}(2021)\citenamefont {Bharti}, \citenamefont {{Cervera-Lierta}}, \citenamefont {Kyaw}, \citenamefont {Haug}, \citenamefont {{Alperin-Lea}}, \citenamefont {Anand}, \citenamefont {Degroote}, \citenamefont {Heimonen}, \citenamefont {Kottmann}, \citenamefont {Menke}, \citenamefont {Mok}, \citenamefont {Sim}, \citenamefont {Kwek},\ and\ \citenamefont {{Aspuru-Guzik}}}]{bhartiNoisyIntermediatescaleQuantum2021}%
  \BibitemOpen
  \bibfield  {author} {\bibinfo {author} {\bibfnamefont {K.}~\bibnamefont {Bharti}}, \bibinfo {author} {\bibfnamefont {A.}~\bibnamefont {{Cervera-Lierta}}}, \bibinfo {author} {\bibfnamefont {T.~H.}\ \bibnamefont {Kyaw}}, \bibinfo {author} {\bibfnamefont {T.}~\bibnamefont {Haug}}, \bibinfo {author} {\bibfnamefont {S.}~\bibnamefont {{Alperin-Lea}}}, \bibinfo {author} {\bibfnamefont {A.}~\bibnamefont {Anand}}, \bibinfo {author} {\bibfnamefont {M.}~\bibnamefont {Degroote}}, \bibinfo {author} {\bibfnamefont {H.}~\bibnamefont {Heimonen}}, \bibinfo {author} {\bibfnamefont {J.~S.}\ \bibnamefont {Kottmann}}, \bibinfo {author} {\bibfnamefont {T.}~\bibnamefont {Menke}}, \bibinfo {author} {\bibfnamefont {W.-K.}\ \bibnamefont {Mok}}, \bibinfo {author} {\bibfnamefont {S.}~\bibnamefont {Sim}}, \bibinfo {author} {\bibfnamefont {L.-C.}\ \bibnamefont {Kwek}},\ and\ \bibinfo {author} {\bibfnamefont {A.}~\bibnamefont {{Aspuru-Guzik}}},\ }\bibfield  {title} {\bibinfo {title} {Noisy intermediate-scale quantum ({{NISQ}}) algorithms},\ }\href@noop {} {\bibfield  {journal} {\bibinfo  {journal} {arXiv:2101.08448 [cond-mat, physics:quant-ph]}\ } (\bibinfo {year} {2021})},\ \Eprint {https://arxiv.org/abs/2101.08448} {arxiv:2101.08448 [cond-mat, physics:quant-ph]} \BibitemShut {NoStop}%
\bibitem [{\citenamefont {Low}\ and\ \citenamefont {Chuang}(2019)}]{lowHamiltonianSimulationQubitization2019}%
  \BibitemOpen
  \bibfield  {author} {\bibinfo {author} {\bibfnamefont {G.~H.}\ \bibnamefont {Low}}\ and\ \bibinfo {author} {\bibfnamefont {I.~L.}\ \bibnamefont {Chuang}},\ }\bibfield  {title} {\bibinfo {title} {Hamiltonian {{Simulation}} by {{Qubitization}}},\ }\href {https://doi.org/10.22331/q-2019-07-12-163} {\bibfield  {journal} {\bibinfo  {journal} {Quantum}\ }\textbf {\bibinfo {volume} {3}},\ \bibinfo {pages} {163} (\bibinfo {year} {2019})}\BibitemShut {NoStop}%
\bibitem [{\citenamefont {C{\^i}rstoiu}\ \emph {et~al.}(2020{\natexlab{a}})\citenamefont {C{\^i}rstoiu}, \citenamefont {Holmes}, \citenamefont {Iosue}, \citenamefont {Cincio}, \citenamefont {Coles},\ and\ \citenamefont {Sornborger}}]{cirstoiuVariationalFastForwarding2020}%
  \BibitemOpen
  \bibfield  {author} {\bibinfo {author} {\bibfnamefont {C.}~\bibnamefont {C{\^i}rstoiu}}, \bibinfo {author} {\bibfnamefont {Z.}~\bibnamefont {Holmes}}, \bibinfo {author} {\bibfnamefont {J.}~\bibnamefont {Iosue}}, \bibinfo {author} {\bibfnamefont {L.}~\bibnamefont {Cincio}}, \bibinfo {author} {\bibfnamefont {P.~J.}\ \bibnamefont {Coles}},\ and\ \bibinfo {author} {\bibfnamefont {A.}~\bibnamefont {Sornborger}},\ }\bibfield  {title} {\bibinfo {title} {Variational fast forwarding for quantum simulation beyond the coherence time},\ }\href {https://doi.org/10.1038/s41534-020-00302-0} {\bibfield  {journal} {\bibinfo  {journal} {npj Quantum Information}\ }\textbf {\bibinfo {volume} {6}},\ \bibinfo {pages} {1} (\bibinfo {year} {2020}{\natexlab{a}})}\BibitemShut {NoStop}%
\bibitem [{\citenamefont {Atia}\ and\ \citenamefont {Aharonov}(2017)}]{atiaFastforwardingHamiltoniansExponentially2017}%
  \BibitemOpen
  \bibfield  {author} {\bibinfo {author} {\bibfnamefont {Y.}~\bibnamefont {Atia}}\ and\ \bibinfo {author} {\bibfnamefont {D.}~\bibnamefont {Aharonov}},\ }\bibfield  {title} {\bibinfo {title} {Fast-forwarding of {{Hamiltonians}} and exponentially precise measurements},\ }\href {https://doi.org/10.1038/s41467-017-01637-7} {\bibfield  {journal} {\bibinfo  {journal} {Nature Communications}\ }\textbf {\bibinfo {volume} {8}},\ \bibinfo {pages} {1572} (\bibinfo {year} {2017})}\BibitemShut {NoStop}%
\bibitem [{\citenamefont {Berry}\ \emph {et~al.}(2007)\citenamefont {Berry}, \citenamefont {Ahokas}, \citenamefont {Cleve},\ and\ \citenamefont {Sanders}}]{berryEfficientQuantumAlgorithms2007}%
  \BibitemOpen
  \bibfield  {author} {\bibinfo {author} {\bibfnamefont {D.~W.}\ \bibnamefont {Berry}}, \bibinfo {author} {\bibfnamefont {G.}~\bibnamefont {Ahokas}}, \bibinfo {author} {\bibfnamefont {R.}~\bibnamefont {Cleve}},\ and\ \bibinfo {author} {\bibfnamefont {B.~C.}\ \bibnamefont {Sanders}},\ }\bibfield  {title} {\bibinfo {title} {Efficient {{Quantum Algorithms}} for {{Simulating Sparse Hamiltonians}}},\ }\href {https://doi.org/10.1007/s00220-006-0150-x} {\bibfield  {journal} {\bibinfo  {journal} {Communications in Mathematical Physics}\ }\textbf {\bibinfo {volume} {270}},\ \bibinfo {pages} {359} (\bibinfo {year} {2007})}\BibitemShut {NoStop}%
\bibitem [{\citenamefont {Childs}\ and\ \citenamefont {Kothari}()}]{childsLimitationsSimulationNonsparse}%
  \BibitemOpen
  \bibfield  {author} {\bibinfo {author} {\bibfnamefont {A.~M.}\ \bibnamefont {Childs}}\ and\ \bibinfo {author} {\bibfnamefont {R.}~\bibnamefont {Kothari}},\ }\bibfield  {title} {\bibinfo {title} {Limitations on the simulation of non-sparse {{Hamiltonians}}},\ }\bibfield  {journal} {\bibinfo  {journal} {Quantum Information and Computation}\ }\textbf {\bibinfo {volume} {10}},\ \href {https://doi.org/10.26421/QIC10.7-8} {10.26421/QIC10.7-8},\ \Eprint {https://arxiv.org/abs/0908.4398} {arxiv:0908.4398 [quant-ph]} \BibitemShut {NoStop}%
\bibitem [{\citenamefont {Flick}\ \emph {et~al.}(2017)\citenamefont {Flick}, \citenamefont {Appel}, \citenamefont {Ruggenthaler},\ and\ \citenamefont {Rubio}}]{flickCavityBornOppenheimer2017}%
  \BibitemOpen
  \bibfield  {author} {\bibinfo {author} {\bibfnamefont {J.}~\bibnamefont {Flick}}, \bibinfo {author} {\bibfnamefont {H.}~\bibnamefont {Appel}}, \bibinfo {author} {\bibfnamefont {M.}~\bibnamefont {Ruggenthaler}},\ and\ \bibinfo {author} {\bibfnamefont {A.}~\bibnamefont {Rubio}},\ }\bibfield  {title} {\bibinfo {title} {Cavity {{Born}}\textendash{{Oppenheimer Approximation}} for {{Correlated Electron}}\textendash{{Nuclear-Photon Systems}}},\ }\href {https://doi.org/10.1021/acs.jctc.6b01126} {\bibfield  {journal} {\bibinfo  {journal} {Journal of Chemical Theory and Computation}\ }\textbf {\bibinfo {volume} {13}},\ \bibinfo {pages} {1616} (\bibinfo {year} {2017})}\BibitemShut {NoStop}%
\bibitem [{\citenamefont {Swope}\ \emph {et~al.}(1982)\citenamefont {Swope}, \citenamefont {Andersen}, \citenamefont {Berens},\ and\ \citenamefont {Wilson}}]{swopeComputerSimulationMethod1982}%
  \BibitemOpen
  \bibfield  {author} {\bibinfo {author} {\bibfnamefont {W.~C.}\ \bibnamefont {Swope}}, \bibinfo {author} {\bibfnamefont {H.~C.}\ \bibnamefont {Andersen}}, \bibinfo {author} {\bibfnamefont {P.~H.}\ \bibnamefont {Berens}},\ and\ \bibinfo {author} {\bibfnamefont {K.~R.}\ \bibnamefont {Wilson}},\ }\bibfield  {title} {\bibinfo {title} {A computer simulation method for the calculation of equilibrium constants for the formation of physical clusters of molecules: {{Application}} to small water clusters},\ }\href {https://doi.org/10.1063/1.442716} {\bibfield  {journal} {\bibinfo  {journal} {The Journal of Chemical Physics}\ }\textbf {\bibinfo {volume} {76}},\ \bibinfo {pages} {637} (\bibinfo {year} {1982})}\BibitemShut {NoStop}%
\bibitem [{\citenamefont {Kuroiwa}\ \emph {et~al.}(2022)\citenamefont {Kuroiwa}, \citenamefont {Ohkuma}, \citenamefont {Sato},\ and\ \citenamefont {Imai}}]{kuroiwaQuantumCarParrinelloMolecular2022}%
  \BibitemOpen
  \bibfield  {author} {\bibinfo {author} {\bibfnamefont {K.}~\bibnamefont {Kuroiwa}}, \bibinfo {author} {\bibfnamefont {T.}~\bibnamefont {Ohkuma}}, \bibinfo {author} {\bibfnamefont {H.}~\bibnamefont {Sato}},\ and\ \bibinfo {author} {\bibfnamefont {R.}~\bibnamefont {Imai}},\ }\href {https://doi.org/10.48550/arXiv.2212.11921} {\bibinfo {title} {Quantum {{Car-Parrinello Molecular Dynamics}}: {{A Cost-Efficient Molecular Simulation Method}} on {{Near-Term Quantum Computers}}}} (\bibinfo {year} {2022}),\ \Eprint {https://arxiv.org/abs/2212.11921} {arxiv:2212.11921 [quant-ph]} \BibitemShut {NoStop}%
\bibitem [{\citenamefont {Azad}\ and\ \citenamefont {Singh}(2022)}]{azadQuantumChemistryCalculations2022}%
  \BibitemOpen
  \bibfield  {author} {\bibinfo {author} {\bibfnamefont {U.}~\bibnamefont {Azad}}\ and\ \bibinfo {author} {\bibfnamefont {H.}~\bibnamefont {Singh}},\ }\bibfield  {title} {\bibinfo {title} {Quantum chemistry calculations using energy derivatives on quantum computers},\ }\href {https://doi.org/10.1016/j.chemphys.2022.111506} {\bibfield  {journal} {\bibinfo  {journal} {Chemical Physics}\ }\textbf {\bibinfo {volume} {558}},\ \bibinfo {pages} {111506} (\bibinfo {year} {2022})}\BibitemShut {NoStop}%
\bibitem [{\citenamefont {Ceroni}\ \emph {et~al.}(2022)\citenamefont {Ceroni}, \citenamefont {Delgado}, \citenamefont {Jahangiri},\ and\ \citenamefont {Arrazola}}]{ceroniTailgatingQuantumCircuits2022}%
  \BibitemOpen
  \bibfield  {author} {\bibinfo {author} {\bibfnamefont {J.}~\bibnamefont {Ceroni}}, \bibinfo {author} {\bibfnamefont {A.}~\bibnamefont {Delgado}}, \bibinfo {author} {\bibfnamefont {S.}~\bibnamefont {Jahangiri}},\ and\ \bibinfo {author} {\bibfnamefont {J.~M.}\ \bibnamefont {Arrazola}},\ }\href {https://doi.org/10.48550/arXiv.2207.11274} {\bibinfo {title} {Tailgating quantum circuits for high-order energy derivatives}} (\bibinfo {year} {2022}),\ \Eprint {https://arxiv.org/abs/2207.11274} {arxiv:2207.11274 [quant-ph]} \BibitemShut {NoStop}%
\bibitem [{\citenamefont {O'Brien}\ \emph {et~al.}(2022)\citenamefont {O'Brien}, \citenamefont {Streif}, \citenamefont {Rubin}, \citenamefont {Santagati}, \citenamefont {Su}, \citenamefont {Huggins}, \citenamefont {Goings}, \citenamefont {Moll}, \citenamefont {Kyoseva}, \citenamefont {Degroote}, \citenamefont {Tautermann}, \citenamefont {Lee}, \citenamefont {Berry}, \citenamefont {Wiebe},\ and\ \citenamefont {Babbush}}]{obrienEfficientQuantumComputation2022}%
  \BibitemOpen
  \bibfield  {author} {\bibinfo {author} {\bibfnamefont {T.~E.}\ \bibnamefont {O'Brien}}, \bibinfo {author} {\bibfnamefont {M.}~\bibnamefont {Streif}}, \bibinfo {author} {\bibfnamefont {N.~C.}\ \bibnamefont {Rubin}}, \bibinfo {author} {\bibfnamefont {R.}~\bibnamefont {Santagati}}, \bibinfo {author} {\bibfnamefont {Y.}~\bibnamefont {Su}}, \bibinfo {author} {\bibfnamefont {W.~J.}\ \bibnamefont {Huggins}}, \bibinfo {author} {\bibfnamefont {J.~J.}\ \bibnamefont {Goings}}, \bibinfo {author} {\bibfnamefont {N.}~\bibnamefont {Moll}}, \bibinfo {author} {\bibfnamefont {E.}~\bibnamefont {Kyoseva}}, \bibinfo {author} {\bibfnamefont {M.}~\bibnamefont {Degroote}}, \bibinfo {author} {\bibfnamefont {C.~S.}\ \bibnamefont {Tautermann}}, \bibinfo {author} {\bibfnamefont {J.}~\bibnamefont {Lee}}, \bibinfo {author} {\bibfnamefont {D.~W.}\ \bibnamefont {Berry}}, \bibinfo {author} {\bibfnamefont {N.}~\bibnamefont {Wiebe}},\ and\ \bibinfo {author} {\bibfnamefont {R.}~\bibnamefont {Babbush}},\ }\bibfield  {title} {\bibinfo {title} {Efficient quantum computation of molecular forces and other energy gradients},\ }\href {https://doi.org/10.1103/PhysRevResearch.4.043210} {\bibfield  {journal} {\bibinfo  {journal} {Physical Review Research}\ }\textbf {\bibinfo {volume} {4}},\ \bibinfo {pages} {043210} (\bibinfo {year} {2022})}\BibitemShut {NoStop}%
\bibitem [{\citenamefont {C{\^i}rstoiu}\ \emph {et~al.}(2020{\natexlab{b}})\citenamefont {C{\^i}rstoiu}, \citenamefont {Holmes}, \citenamefont {Iosue}, \citenamefont {Cincio}, \citenamefont {Coles},\ and\ \citenamefont {Sornborger}}]{cirstoiuVariationalFastForwarding2020a}%
  \BibitemOpen
  \bibfield  {author} {\bibinfo {author} {\bibfnamefont {C.}~\bibnamefont {C{\^i}rstoiu}}, \bibinfo {author} {\bibfnamefont {Z.}~\bibnamefont {Holmes}}, \bibinfo {author} {\bibfnamefont {J.}~\bibnamefont {Iosue}}, \bibinfo {author} {\bibfnamefont {L.}~\bibnamefont {Cincio}}, \bibinfo {author} {\bibfnamefont {P.~J.}\ \bibnamefont {Coles}},\ and\ \bibinfo {author} {\bibfnamefont {A.}~\bibnamefont {Sornborger}},\ }\bibfield  {title} {\bibinfo {title} {Variational fast forwarding for quantum simulation beyond the coherence time},\ }\href {https://doi.org/10.1038/s41534-020-00302-0} {\bibfield  {journal} {\bibinfo  {journal} {npj Quantum Information}\ }\textbf {\bibinfo {volume} {6}},\ \bibinfo {pages} {1} (\bibinfo {year} {2020}{\natexlab{b}})}\BibitemShut {NoStop}%
\bibitem [{\citenamefont {Lee}\ \emph {et~al.}(2022)\citenamefont {Lee}, \citenamefont {Hsieh}, \citenamefont {Zhang},\ and\ \citenamefont {Shi}}]{leeVariationalQuantumSimulation2022}%
  \BibitemOpen
  \bibfield  {author} {\bibinfo {author} {\bibfnamefont {C.-K.}\ \bibnamefont {Lee}}, \bibinfo {author} {\bibfnamefont {C.-Y.}\ \bibnamefont {Hsieh}}, \bibinfo {author} {\bibfnamefont {S.}~\bibnamefont {Zhang}},\ and\ \bibinfo {author} {\bibfnamefont {L.}~\bibnamefont {Shi}},\ }\bibfield  {title} {\bibinfo {title} {Variational {{Quantum Simulation}} of {{Chemical Dynamics}} with {{Quantum Computers}}},\ }\bibfield  {journal} {\bibinfo  {journal} {Journal of Chemical Theory and Computation}\ }\href {https://doi.org/10.1021/acs.jctc.1c01176} {10.1021/acs.jctc.1c01176} (\bibinfo {year} {2022})\BibitemShut {NoStop}%
\bibitem [{\citenamefont {Yao}\ \emph {et~al.}(2021)\citenamefont {Yao}, \citenamefont {Gomes}, \citenamefont {Zhang}, \citenamefont {Wang}, \citenamefont {Ho}, \citenamefont {Iadecola},\ and\ \citenamefont {Orth}}]{yaoAdaptiveVariationalQuantum2021}%
  \BibitemOpen
  \bibfield  {author} {\bibinfo {author} {\bibfnamefont {Y.-X.}\ \bibnamefont {Yao}}, \bibinfo {author} {\bibfnamefont {N.}~\bibnamefont {Gomes}}, \bibinfo {author} {\bibfnamefont {F.}~\bibnamefont {Zhang}}, \bibinfo {author} {\bibfnamefont {C.-Z.}\ \bibnamefont {Wang}}, \bibinfo {author} {\bibfnamefont {K.-M.}\ \bibnamefont {Ho}}, \bibinfo {author} {\bibfnamefont {T.}~\bibnamefont {Iadecola}},\ and\ \bibinfo {author} {\bibfnamefont {P.~P.}\ \bibnamefont {Orth}},\ }\bibfield  {title} {\bibinfo {title} {Adaptive {{Variational Quantum Dynamics Simulations}}},\ }\href {https://doi.org/10.1103/PRXQuantum.2.030307} {\bibfield  {journal} {\bibinfo  {journal} {PRX Quantum}\ }\textbf {\bibinfo {volume} {2}},\ \bibinfo {pages} {030307} (\bibinfo {year} {2021})}\BibitemShut {NoStop}%
\bibitem [{\citenamefont {O'Brien}\ \emph {et~al.}(2019)\citenamefont {O'Brien}, \citenamefont {Senjean}, \citenamefont {Sagastizabal}, \citenamefont {{Bonet-Monroig}}, \citenamefont {Dutkiewicz}, \citenamefont {Buda}, \citenamefont {DiCarlo},\ and\ \citenamefont {Visscher}}]{obrienCalculatingEnergyDerivatives2019}%
  \BibitemOpen
  \bibfield  {author} {\bibinfo {author} {\bibfnamefont {T.~E.}\ \bibnamefont {O'Brien}}, \bibinfo {author} {\bibfnamefont {B.}~\bibnamefont {Senjean}}, \bibinfo {author} {\bibfnamefont {R.}~\bibnamefont {Sagastizabal}}, \bibinfo {author} {\bibfnamefont {X.}~\bibnamefont {{Bonet-Monroig}}}, \bibinfo {author} {\bibfnamefont {A.}~\bibnamefont {Dutkiewicz}}, \bibinfo {author} {\bibfnamefont {F.}~\bibnamefont {Buda}}, \bibinfo {author} {\bibfnamefont {L.}~\bibnamefont {DiCarlo}},\ and\ \bibinfo {author} {\bibfnamefont {L.}~\bibnamefont {Visscher}},\ }\bibfield  {title} {\bibinfo {title} {Calculating energy derivatives for quantum chemistry on a quantum computer},\ }\href {https://doi.org/10.1038/s41534-019-0213-4} {\bibfield  {journal} {\bibinfo  {journal} {npj Quantum Information}\ }\textbf {\bibinfo {volume} {5}},\ \bibinfo {pages} {1} (\bibinfo {year} {2019})}\BibitemShut {NoStop}%
\bibitem [{\citenamefont {Babbush}\ \emph {et~al.}(2015)\citenamefont {Babbush}, \citenamefont {McClean}, \citenamefont {Wecker}, \citenamefont {{Aspuru-Guzik}},\ and\ \citenamefont {Wiebe}}]{babbushChemicalBasisTrotterSuzuki2015}%
  \BibitemOpen
  \bibfield  {author} {\bibinfo {author} {\bibfnamefont {R.}~\bibnamefont {Babbush}}, \bibinfo {author} {\bibfnamefont {J.}~\bibnamefont {McClean}}, \bibinfo {author} {\bibfnamefont {D.}~\bibnamefont {Wecker}}, \bibinfo {author} {\bibfnamefont {A.}~\bibnamefont {{Aspuru-Guzik}}},\ and\ \bibinfo {author} {\bibfnamefont {N.}~\bibnamefont {Wiebe}},\ }\bibfield  {title} {\bibinfo {title} {Chemical basis of {{Trotter-Suzuki}} errors in quantum chemistry simulation},\ }\href {https://doi.org/10.1103/PhysRevA.91.022311} {\bibfield  {journal} {\bibinfo  {journal} {Physical Review A}\ }\textbf {\bibinfo {volume} {91}},\ \bibinfo {pages} {022311} (\bibinfo {year} {2015})}\BibitemShut {NoStop}%
\bibitem [{\citenamefont {Trotter}(1959)}]{trotterProductSemiGroupsOperators1959}%
  \BibitemOpen
  \bibfield  {author} {\bibinfo {author} {\bibfnamefont {H.~F.}\ \bibnamefont {Trotter}},\ }\bibfield  {title} {\bibinfo {title} {On the {{Product}} of {{Semi-Groups}} of {{Operators}}},\ }\href {https://doi.org/10.2307/2033649} {\bibfield  {journal} {\bibinfo  {journal} {Proceedings of the American Mathematical Society}\ }\textbf {\bibinfo {volume} {10}},\ \bibinfo {pages} {545} (\bibinfo {year} {1959})},\ \Eprint {https://arxiv.org/abs/2033649} {2033649} \BibitemShut {NoStop}%
\bibitem [{\citenamefont {Gard}\ \emph {et~al.}(2020)\citenamefont {Gard}, \citenamefont {Zhu}, \citenamefont {Barron}, \citenamefont {Mayhall}, \citenamefont {Economou},\ and\ \citenamefont {Barnes}}]{gardEfficientSymmetrypreservingState2020}%
  \BibitemOpen
  \bibfield  {author} {\bibinfo {author} {\bibfnamefont {B.~T.}\ \bibnamefont {Gard}}, \bibinfo {author} {\bibfnamefont {L.}~\bibnamefont {Zhu}}, \bibinfo {author} {\bibfnamefont {G.~S.}\ \bibnamefont {Barron}}, \bibinfo {author} {\bibfnamefont {N.~J.}\ \bibnamefont {Mayhall}}, \bibinfo {author} {\bibfnamefont {S.~E.}\ \bibnamefont {Economou}},\ and\ \bibinfo {author} {\bibfnamefont {E.}~\bibnamefont {Barnes}},\ }\bibfield  {title} {\bibinfo {title} {Efficient symmetry-preserving state preparation circuits for the variational quantum eigensolver algorithm},\ }\href {https://doi.org/10.1038/s41534-019-0240-1} {\bibfield  {journal} {\bibinfo  {journal} {npj Quantum Information}\ }\textbf {\bibinfo {volume} {6}},\ \bibinfo {pages} {1} (\bibinfo {year} {2020})}\BibitemShut {NoStop}%
\bibitem [{\citenamefont {Robbins}\ and\ \citenamefont {Monro}(1951)}]{robbinsStochasticApproximationMethod1951}%
  \BibitemOpen
  \bibfield  {author} {\bibinfo {author} {\bibfnamefont {H.}~\bibnamefont {Robbins}}\ and\ \bibinfo {author} {\bibfnamefont {S.}~\bibnamefont {Monro}},\ }\bibfield  {title} {\bibinfo {title} {A {{Stochastic Approximation Method}}},\ }\href {https://doi.org/10.1214/aoms/1177729586} {\bibfield  {journal} {\bibinfo  {journal} {The Annals of Mathematical Statistics}\ }\textbf {\bibinfo {volume} {22}},\ \bibinfo {pages} {400} (\bibinfo {year} {1951})}\BibitemShut {NoStop}%
\bibitem [{\citenamefont {Wierichs}\ \emph {et~al.}(2022)\citenamefont {Wierichs}, \citenamefont {Izaac}, \citenamefont {Wang},\ and\ \citenamefont {Lin}}]{wierichsGeneralParametershiftRules2022}%
  \BibitemOpen
  \bibfield  {author} {\bibinfo {author} {\bibfnamefont {D.}~\bibnamefont {Wierichs}}, \bibinfo {author} {\bibfnamefont {J.}~\bibnamefont {Izaac}}, \bibinfo {author} {\bibfnamefont {C.}~\bibnamefont {Wang}},\ and\ \bibinfo {author} {\bibfnamefont {C.~Y.-Y.}\ \bibnamefont {Lin}},\ }\bibfield  {title} {\bibinfo {title} {General parameter-shift rules for quantum gradients},\ }\href {https://doi.org/10.22331/q-2022-03-30-677} {\bibfield  {journal} {\bibinfo  {journal} {Quantum}\ }\textbf {\bibinfo {volume} {6}},\ \bibinfo {pages} {677} (\bibinfo {year} {2022})}\BibitemShut {NoStop}%
\bibitem [{\citenamefont {Gocho}\ \emph {et~al.}(2023)\citenamefont {Gocho}, \citenamefont {Nakamura}, \citenamefont {Kanno}, \citenamefont {Gao}, \citenamefont {Kobayashi}, \citenamefont {Inagaki},\ and\ \citenamefont {Hatanaka}}]{gochoExcitedStateCalculations2023}%
  \BibitemOpen
  \bibfield  {author} {\bibinfo {author} {\bibfnamefont {S.}~\bibnamefont {Gocho}}, \bibinfo {author} {\bibfnamefont {H.}~\bibnamefont {Nakamura}}, \bibinfo {author} {\bibfnamefont {S.}~\bibnamefont {Kanno}}, \bibinfo {author} {\bibfnamefont {Q.}~\bibnamefont {Gao}}, \bibinfo {author} {\bibfnamefont {T.}~\bibnamefont {Kobayashi}}, \bibinfo {author} {\bibfnamefont {T.}~\bibnamefont {Inagaki}},\ and\ \bibinfo {author} {\bibfnamefont {M.}~\bibnamefont {Hatanaka}},\ }\bibfield  {title} {\bibinfo {title} {Excited state calculations using variational quantum eigensolver with spin-restricted ans\"atze and automatically-adjusted constraints},\ }\href {https://doi.org/10.1038/s41524-023-00965-1} {\bibfield  {journal} {\bibinfo  {journal} {npj Computational Materials}\ }\textbf {\bibinfo {volume} {9}},\ \bibinfo {pages} {1} (\bibinfo {year} {2023})}\BibitemShut {NoStop}%
\bibitem [{\citenamefont {McClean}\ \emph {et~al.}(2017)\citenamefont {McClean}, \citenamefont {{Kimchi-Schwartz}}, \citenamefont {Carter},\ and\ \citenamefont {{de Jong}}}]{mccleanHybridQuantumclassicalHierarchy2017a}%
  \BibitemOpen
  \bibfield  {author} {\bibinfo {author} {\bibfnamefont {J.~R.}\ \bibnamefont {McClean}}, \bibinfo {author} {\bibfnamefont {M.~E.}\ \bibnamefont {{Kimchi-Schwartz}}}, \bibinfo {author} {\bibfnamefont {J.}~\bibnamefont {Carter}},\ and\ \bibinfo {author} {\bibfnamefont {W.~A.}\ \bibnamefont {{de Jong}}},\ }\bibfield  {title} {\bibinfo {title} {Hybrid quantum-classical hierarchy for mitigation of decoherence and determination of excited states},\ }\href {https://doi.org/10.1103/PhysRevA.95.042308} {\bibfield  {journal} {\bibinfo  {journal} {Physical Review A}\ }\textbf {\bibinfo {volume} {95}},\ \bibinfo {pages} {042308} (\bibinfo {year} {2017})}\BibitemShut {NoStop}%
\bibitem [{\citenamefont {Aulicino}\ \emph {et~al.}(2022)\citenamefont {Aulicino}, \citenamefont {Keen},\ and\ \citenamefont {Peng}}]{aulicinoStatePreparationEvolution2022}%
  \BibitemOpen
  \bibfield  {author} {\bibinfo {author} {\bibfnamefont {J.~C.}\ \bibnamefont {Aulicino}}, \bibinfo {author} {\bibfnamefont {T.}~\bibnamefont {Keen}},\ and\ \bibinfo {author} {\bibfnamefont {B.}~\bibnamefont {Peng}},\ }\bibfield  {title} {\bibinfo {title} {State preparation and evolution in quantum computing: {{A}} perspective from {{Hamiltonian}} moments},\ }\href {https://doi.org/10.1002/qua.26853} {\bibfield  {journal} {\bibinfo  {journal} {International Journal of Quantum Chemistry}\ }\textbf {\bibinfo {volume} {122}},\ \bibinfo {pages} {e26853} (\bibinfo {year} {2022})}\BibitemShut {NoStop}%
\bibitem [{\citenamefont {Ollitrault}\ \emph {et~al.}(2022)\citenamefont {Ollitrault}, \citenamefont {Jandura}, \citenamefont {Miessen}, \citenamefont {Burghardt}, \citenamefont {Martinazzo}, \citenamefont {Tacchino},\ and\ \citenamefont {Tavernelli}}]{ollitraultQuantumAlgorithmsGridbased2022}%
  \BibitemOpen
  \bibfield  {author} {\bibinfo {author} {\bibfnamefont {P.~J.}\ \bibnamefont {Ollitrault}}, \bibinfo {author} {\bibfnamefont {S.}~\bibnamefont {Jandura}}, \bibinfo {author} {\bibfnamefont {A.}~\bibnamefont {Miessen}}, \bibinfo {author} {\bibfnamefont {I.}~\bibnamefont {Burghardt}}, \bibinfo {author} {\bibfnamefont {R.}~\bibnamefont {Martinazzo}}, \bibinfo {author} {\bibfnamefont {F.}~\bibnamefont {Tacchino}},\ and\ \bibinfo {author} {\bibfnamefont {I.}~\bibnamefont {Tavernelli}},\ }\bibfield  {title} {\bibinfo {title} {Quantum algorithms for grid-based variational time evolution},\ }\href@noop {} {\bibfield  {journal} {\bibinfo  {journal} {arXiv:2203.02521 [quant-ph]}\ } (\bibinfo {year} {2022})},\ \Eprint {https://arxiv.org/abs/2203.02521} {arxiv:2203.02521 [quant-ph]} \BibitemShut {NoStop}%
\bibitem [{\citenamefont {Aleksandrowicz}\ \emph {et~al.}(2019)\citenamefont {Aleksandrowicz}, \citenamefont {Alexander}, \citenamefont {Barkoutsos}, \citenamefont {Bello}, \citenamefont {{Ben-Haim}}, \citenamefont {Bucher}, \citenamefont {{Cabrera-Hern{\'a}ndez}}, \citenamefont {{Carballo-Franquis}}, \citenamefont {Chen}, \citenamefont {Chen}, \citenamefont {Chow}, \citenamefont {{C{\'o}rcoles-Gonzales}}, \citenamefont {Cross}, \citenamefont {Cross}, \citenamefont {{Cruz-Benito}}, \citenamefont {Culver}, \citenamefont {Gonz{\'a}lez}, \citenamefont {Torre}, \citenamefont {Ding}, \citenamefont {Dumitrescu}, \citenamefont {Duran}, \citenamefont {Eendebak}, \citenamefont {Everitt}, \citenamefont {Sertage}, \citenamefont {Frisch}, \citenamefont {Fuhrer}, \citenamefont {Gambetta}, \citenamefont {Gago}, \citenamefont {{Gomez-Mosquera}}, \citenamefont {Greenberg}, \citenamefont {Hamamura}, \citenamefont {Havlicek}, \citenamefont {Hellmers}, \citenamefont {Herok}, \citenamefont {Horii}, \citenamefont {Hu}, \citenamefont {Imamichi}, \citenamefont {Itoko}, \citenamefont {{Javadi-Abhari}}, \citenamefont {Kanazawa}, \citenamefont {Karazeev}, \citenamefont {Krsulich}, \citenamefont {Liu}, \citenamefont {Luh}, \citenamefont {Maeng}, \citenamefont {Marques}, \citenamefont {{Mart{\'i}n-Fern{\'a}ndez}}, \citenamefont {McClure}, \citenamefont {McKay}, \citenamefont {Meesala}, \citenamefont {Mezzacapo}, \citenamefont {Moll}, \citenamefont {Rodr{\'i}guez}, \citenamefont {Nannicini}, \citenamefont {Nation}, \citenamefont {Ollitrault}, \citenamefont {O'Riordan}, \citenamefont {Paik}, \citenamefont {P{\'e}rez}, \citenamefont {Phan}, \citenamefont {Pistoia}, \citenamefont {Prutyanov}, \citenamefont {Reuter}, \citenamefont {Rice}, \citenamefont {Davila}, \citenamefont {Rudy}, \citenamefont {Ryu}, \citenamefont {Sathaye}, \citenamefont {Schnabel}, \citenamefont {Schoute}, \citenamefont {Setia}, \citenamefont {Shi}, \citenamefont {Silva}, \citenamefont {Siraichi}, \citenamefont {Sivarajah}, \citenamefont {Smolin}, \citenamefont {Soeken}, \citenamefont {Takahashi}, \citenamefont {Tavernelli}, \citenamefont {Taylor}, \citenamefont {Taylour}, \citenamefont {Trabing}, \citenamefont {Treinish}, \citenamefont {Turner}, \citenamefont {{Vogt-Lee}}, \citenamefont {Vuillot}, \citenamefont {Wildstrom}, \citenamefont {Wilson}, \citenamefont {Winston}, \citenamefont {Wood}, \citenamefont {Wood}, \citenamefont {W{\"o}rner}, \citenamefont {Akhalwaya},\ and\ \citenamefont {Zoufal}}]{aleksandrowiczQiskitOpensourceFramework2019}%
  \BibitemOpen
  \bibfield  {author} {\bibinfo {author} {\bibfnamefont {G.}~\bibnamefont {Aleksandrowicz}}, \bibinfo {author} {\bibfnamefont {T.}~\bibnamefont {Alexander}}, \bibinfo {author} {\bibfnamefont {P.}~\bibnamefont {Barkoutsos}}, \bibinfo {author} {\bibfnamefont {L.}~\bibnamefont {Bello}}, \bibinfo {author} {\bibfnamefont {Y.}~\bibnamefont {{Ben-Haim}}}, \bibinfo {author} {\bibfnamefont {D.}~\bibnamefont {Bucher}}, \bibinfo {author} {\bibfnamefont {F.~J.}\ \bibnamefont {{Cabrera-Hern{\'a}ndez}}}, \bibinfo {author} {\bibfnamefont {J.}~\bibnamefont {{Carballo-Franquis}}}, \bibinfo {author} {\bibfnamefont {A.}~\bibnamefont {Chen}}, \bibinfo {author} {\bibfnamefont {C.-F.}\ \bibnamefont {Chen}}, \bibinfo {author} {\bibfnamefont {J.~M.}\ \bibnamefont {Chow}}, \bibinfo {author} {\bibfnamefont {A.~D.}\ \bibnamefont {{C{\'o}rcoles-Gonzales}}}, \bibinfo {author} {\bibfnamefont {A.~J.}\ \bibnamefont {Cross}}, \bibinfo {author} {\bibfnamefont {A.}~\bibnamefont {Cross}}, \bibinfo {author} {\bibfnamefont {J.}~\bibnamefont {{Cruz-Benito}}}, \bibinfo {author} {\bibfnamefont {C.}~\bibnamefont {Culver}}, \bibinfo {author} {\bibfnamefont {S.~D. L.~P.}\ \bibnamefont {Gonz{\'a}lez}}, \bibinfo {author} {\bibfnamefont {E.~D.~L.}\ \bibnamefont {Torre}}, \bibinfo {author} {\bibfnamefont {D.}~\bibnamefont {Ding}}, \bibinfo {author} {\bibfnamefont {E.}~\bibnamefont {Dumitrescu}}, \bibinfo {author} {\bibfnamefont {I.}~\bibnamefont {Duran}}, \bibinfo {author} {\bibfnamefont {P.}~\bibnamefont {Eendebak}}, \bibinfo {author} {\bibfnamefont {M.}~\bibnamefont {Everitt}}, \bibinfo {author} {\bibfnamefont {I.~F.}\ \bibnamefont {Sertage}}, \bibinfo {author} {\bibfnamefont {A.}~\bibnamefont {Frisch}}, \bibinfo {author} {\bibfnamefont {A.}~\bibnamefont {Fuhrer}}, \bibinfo {author} {\bibfnamefont {J.}~\bibnamefont {Gambetta}}, \bibinfo {author} {\bibfnamefont {B.~G.}\ \bibnamefont {Gago}}, \bibinfo {author} {\bibfnamefont {J.}~\bibnamefont {{Gomez-Mosquera}}}, \bibinfo {author} {\bibfnamefont {D.}~\bibnamefont {Greenberg}}, \bibinfo {author} {\bibfnamefont {I.}~\bibnamefont {Hamamura}}, \bibinfo {author} {\bibfnamefont {V.}~\bibnamefont {Havlicek}}, \bibinfo {author} {\bibfnamefont {J.}~\bibnamefont {Hellmers}}, \bibinfo {author} {\bibfnamefont {{\L}.}~\bibnamefont {Herok}}, \bibinfo {author} {\bibfnamefont {H.}~\bibnamefont {Horii}}, \bibinfo {author} {\bibfnamefont {S.}~\bibnamefont {Hu}}, \bibinfo {author} {\bibfnamefont {T.}~\bibnamefont {Imamichi}}, \bibinfo {author} {\bibfnamefont {T.}~\bibnamefont {Itoko}}, \bibinfo {author} {\bibfnamefont {A.}~\bibnamefont {{Javadi-Abhari}}}, \bibinfo {author} {\bibfnamefont {N.}~\bibnamefont {Kanazawa}}, \bibinfo {author} {\bibfnamefont {A.}~\bibnamefont {Karazeev}}, \bibinfo {author} {\bibfnamefont {K.}~\bibnamefont {Krsulich}}, \bibinfo {author} {\bibfnamefont {P.}~\bibnamefont {Liu}}, \bibinfo {author} {\bibfnamefont {Y.}~\bibnamefont {Luh}}, \bibinfo {author} {\bibfnamefont {Y.}~\bibnamefont {Maeng}}, \bibinfo {author} {\bibfnamefont {M.}~\bibnamefont {Marques}}, \bibinfo {author} {\bibfnamefont {F.~J.}\ \bibnamefont {{Mart{\'i}n-Fern{\'a}ndez}}}, \bibinfo {author} {\bibfnamefont {D.~T.}\ \bibnamefont {McClure}}, \bibinfo {author} {\bibfnamefont {D.}~\bibnamefont {McKay}}, \bibinfo {author} {\bibfnamefont {S.}~\bibnamefont {Meesala}}, \bibinfo {author} {\bibfnamefont {A.}~\bibnamefont {Mezzacapo}}, \bibinfo {author} {\bibfnamefont {N.}~\bibnamefont {Moll}}, \bibinfo {author} {\bibfnamefont {D.~M.}\ \bibnamefont {Rodr{\'i}guez}}, \bibinfo {author} {\bibfnamefont {G.}~\bibnamefont {Nannicini}}, \bibinfo {author} {\bibfnamefont {P.}~\bibnamefont {Nation}}, \bibinfo {author} {\bibfnamefont {P.}~\bibnamefont {Ollitrault}}, \bibinfo {author} {\bibfnamefont {L.~J.}\ \bibnamefont {O'Riordan}}, \bibinfo {author} {\bibfnamefont {H.}~\bibnamefont {Paik}}, \bibinfo {author} {\bibfnamefont {J.}~\bibnamefont {P{\'e}rez}}, \bibinfo {author} {\bibfnamefont {A.}~\bibnamefont {Phan}}, \bibinfo {author} {\bibfnamefont {M.}~\bibnamefont {Pistoia}}, \bibinfo {author} {\bibfnamefont {V.}~\bibnamefont {Prutyanov}}, \bibinfo {author} {\bibfnamefont {M.}~\bibnamefont {Reuter}}, \bibinfo {author} {\bibfnamefont {J.}~\bibnamefont {Rice}}, \bibinfo {author} {\bibfnamefont {A.~R.}\ \bibnamefont {Davila}}, \bibinfo {author} {\bibfnamefont {R.~H.~P.}\ \bibnamefont {Rudy}}, \bibinfo {author} {\bibfnamefont {M.}~\bibnamefont {Ryu}}, \bibinfo {author} {\bibfnamefont {N.}~\bibnamefont {Sathaye}}, \bibinfo {author} {\bibfnamefont {C.}~\bibnamefont {Schnabel}}, \bibinfo {author} {\bibfnamefont {E.}~\bibnamefont {Schoute}}, \bibinfo {author} {\bibfnamefont {K.}~\bibnamefont {Setia}}, \bibinfo {author} {\bibfnamefont {Y.}~\bibnamefont {Shi}}, \bibinfo {author} {\bibfnamefont {A.}~\bibnamefont {Silva}}, \bibinfo {author} {\bibfnamefont {Y.}~\bibnamefont {Siraichi}}, \bibinfo {author} {\bibfnamefont {S.}~\bibnamefont {Sivarajah}}, \bibinfo {author} {\bibfnamefont {J.~A.}\ \bibnamefont {Smolin}}, \bibinfo {author} {\bibfnamefont {M.}~\bibnamefont {Soeken}}, \bibinfo {author} {\bibfnamefont {H.}~\bibnamefont {Takahashi}}, \bibinfo {author} {\bibfnamefont {I.}~\bibnamefont {Tavernelli}}, \bibinfo {author} {\bibfnamefont {C.}~\bibnamefont {Taylor}}, \bibinfo {author} {\bibfnamefont {P.}~\bibnamefont {Taylour}}, \bibinfo {author} {\bibfnamefont {K.}~\bibnamefont {Trabing}}, \bibinfo {author} {\bibfnamefont {M.}~\bibnamefont {Treinish}}, \bibinfo {author} {\bibfnamefont {W.}~\bibnamefont {Turner}}, \bibinfo {author} {\bibfnamefont {D.}~\bibnamefont {{Vogt-Lee}}}, \bibinfo {author} {\bibfnamefont {C.}~\bibnamefont {Vuillot}}, \bibinfo {author} {\bibfnamefont {J.~A.}\ \bibnamefont {Wildstrom}}, \bibinfo {author} {\bibfnamefont {J.}~\bibnamefont {Wilson}}, \bibinfo {author} {\bibfnamefont {E.}~\bibnamefont {Winston}}, \bibinfo {author} {\bibfnamefont {C.}~\bibnamefont {Wood}}, \bibinfo {author} {\bibfnamefont {S.}~\bibnamefont {Wood}}, \bibinfo {author} {\bibfnamefont {S.}~\bibnamefont {W{\"o}rner}}, \bibinfo {author} {\bibfnamefont {I.~Y.}\ \bibnamefont {Akhalwaya}},\ and\ \bibinfo {author} {\bibfnamefont {C.}~\bibnamefont {Zoufal}},\ }\href {https://doi.org/10.5281/zenodo.2562111} {\bibinfo {title} {Qiskit: {{An Open-source Framework}} for {{Quantum Computing}}}},\ \bibinfo {howpublished} {Zenodo} (\bibinfo {year} {2019})\BibitemShut {NoStop}%
\bibitem [{\citenamefont {Crooks}(2019)}]{crooksGradientsParameterizedQuantum2019}%
  \BibitemOpen
  \bibfield  {author} {\bibinfo {author} {\bibfnamefont {G.~E.}\ \bibnamefont {Crooks}},\ }\href {https://doi.org/10.48550/arXiv.1905.13311} {\bibinfo {title} {Gradients of parameterized quantum gates using the parameter-shift rule and gate decomposition}} (\bibinfo {year} {2019}),\ \Eprint {https://arxiv.org/abs/1905.13311} {arxiv:1905.13311 [quant-ph]} \BibitemShut {NoStop}%
\bibitem [{\citenamefont {Harris}\ \emph {et~al.}(2020)\citenamefont {Harris}, \citenamefont {Millman}, \citenamefont {{van der Walt}}, \citenamefont {Gommers}, \citenamefont {Virtanen}, \citenamefont {Cournapeau}, \citenamefont {Wieser}, \citenamefont {Taylor}, \citenamefont {Berg}, \citenamefont {Smith}, \citenamefont {Kern}, \citenamefont {Picus}, \citenamefont {Hoyer}, \citenamefont {{van Kerkwijk}}, \citenamefont {Brett}, \citenamefont {Haldane}, \citenamefont {{del R{\'i}o}}, \citenamefont {Wiebe}, \citenamefont {Peterson}, \citenamefont {{G{\'e}rard-Marchant}}, \citenamefont {Sheppard}, \citenamefont {Reddy}, \citenamefont {Weckesser}, \citenamefont {Abbasi}, \citenamefont {Gohlke},\ and\ \citenamefont {Oliphant}}]{harrisArrayProgrammingNumPy2020}%
  \BibitemOpen
  \bibfield  {author} {\bibinfo {author} {\bibfnamefont {C.~R.}\ \bibnamefont {Harris}}, \bibinfo {author} {\bibfnamefont {K.~J.}\ \bibnamefont {Millman}}, \bibinfo {author} {\bibfnamefont {S.~J.}\ \bibnamefont {{van der Walt}}}, \bibinfo {author} {\bibfnamefont {R.}~\bibnamefont {Gommers}}, \bibinfo {author} {\bibfnamefont {P.}~\bibnamefont {Virtanen}}, \bibinfo {author} {\bibfnamefont {D.}~\bibnamefont {Cournapeau}}, \bibinfo {author} {\bibfnamefont {E.}~\bibnamefont {Wieser}}, \bibinfo {author} {\bibfnamefont {J.}~\bibnamefont {Taylor}}, \bibinfo {author} {\bibfnamefont {S.}~\bibnamefont {Berg}}, \bibinfo {author} {\bibfnamefont {N.~J.}\ \bibnamefont {Smith}}, \bibinfo {author} {\bibfnamefont {R.}~\bibnamefont {Kern}}, \bibinfo {author} {\bibfnamefont {M.}~\bibnamefont {Picus}}, \bibinfo {author} {\bibfnamefont {S.}~\bibnamefont {Hoyer}}, \bibinfo {author} {\bibfnamefont {M.~H.}\ \bibnamefont {{van Kerkwijk}}}, \bibinfo {author} {\bibfnamefont {M.}~\bibnamefont {Brett}}, \bibinfo {author} {\bibfnamefont {A.}~\bibnamefont {Haldane}}, \bibinfo {author} {\bibfnamefont {J.~F.}\ \bibnamefont {{del R{\'i}o}}}, \bibinfo {author} {\bibfnamefont {M.}~\bibnamefont {Wiebe}}, \bibinfo {author} {\bibfnamefont {P.}~\bibnamefont {Peterson}}, \bibinfo {author} {\bibfnamefont {P.}~\bibnamefont {{G{\'e}rard-Marchant}}}, \bibinfo {author} {\bibfnamefont {K.}~\bibnamefont {Sheppard}}, \bibinfo {author} {\bibfnamefont {T.}~\bibnamefont {Reddy}}, \bibinfo {author} {\bibfnamefont {W.}~\bibnamefont {Weckesser}}, \bibinfo {author} {\bibfnamefont {H.}~\bibnamefont {Abbasi}}, \bibinfo {author} {\bibfnamefont {C.}~\bibnamefont {Gohlke}},\ and\ \bibinfo {author} {\bibfnamefont {T.~E.}\ \bibnamefont {Oliphant}},\ }\bibfield  {title} {\bibinfo {title} {Array programming with {{NumPy}}},\ }\href {https://doi.org/10.1038/s41586-020-2649-2} {\bibfield  {journal} {\bibinfo  {journal} {Nature}\ }\textbf {\bibinfo {volume} {585}},\ \bibinfo {pages} {357} (\bibinfo {year} {2020})}\BibitemShut {NoStop}%
\bibitem [{\citenamefont {Colbert}\ and\ \citenamefont {Miller}(1992)}]{colbertNovelDiscreteVariable1992}%
  \BibitemOpen
  \bibfield  {author} {\bibinfo {author} {\bibfnamefont {D.~T.}\ \bibnamefont {Colbert}}\ and\ \bibinfo {author} {\bibfnamefont {W.~H.}\ \bibnamefont {Miller}},\ }\bibfield  {title} {\bibinfo {title} {A novel discrete variable representation for quantum mechanical reactive scattering via the {{S}}-matrix {{Kohn}} method},\ }\href {https://doi.org/10.1063/1.462100} {\bibfield  {journal} {\bibinfo  {journal} {The Journal of Chemical Physics}\ }\textbf {\bibinfo {volume} {96}},\ \bibinfo {pages} {1982} (\bibinfo {year} {1992})}\BibitemShut {NoStop}%
\bibitem [{\citenamefont {Hermann}\ and\ \citenamefont {Fleck}(1988)}]{hermannSplitoperatorSpectralMethod1988}%
  \BibitemOpen
  \bibfield  {author} {\bibinfo {author} {\bibfnamefont {M.~R.}\ \bibnamefont {Hermann}}\ and\ \bibinfo {author} {\bibfnamefont {J.~A.}\ \bibnamefont {Fleck}},\ }\bibfield  {title} {\bibinfo {title} {Split-operator spectral method for solving the time-dependent {{Schr}}\textbackslash "odinger equation in spherical coordinates},\ }\href {https://doi.org/10.1103/PhysRevA.38.6000} {\bibfield  {journal} {\bibinfo  {journal} {Physical Review A}\ }\textbf {\bibinfo {volume} {38}},\ \bibinfo {pages} {6000} (\bibinfo {year} {1988})}\BibitemShut {NoStop}%
\bibitem [{\citenamefont {G{\"u}hne}\ \emph {et~al.}(2007)\citenamefont {G{\"u}hne}, \citenamefont {Lu}, \citenamefont {Gao},\ and\ \citenamefont {Pan}}]{guhneToolboxEntanglementDetection2007}%
  \BibitemOpen
  \bibfield  {author} {\bibinfo {author} {\bibfnamefont {O.}~\bibnamefont {G{\"u}hne}}, \bibinfo {author} {\bibfnamefont {C.-Y.}\ \bibnamefont {Lu}}, \bibinfo {author} {\bibfnamefont {W.-B.}\ \bibnamefont {Gao}},\ and\ \bibinfo {author} {\bibfnamefont {J.-W.}\ \bibnamefont {Pan}},\ }\bibfield  {title} {\bibinfo {title} {Toolbox for entanglement detection and fidelity estimation},\ }\href {https://doi.org/10.1103/PhysRevA.76.030305} {\bibfield  {journal} {\bibinfo  {journal} {Physical Review A}\ }\textbf {\bibinfo {volume} {76}},\ \bibinfo {pages} {030305} (\bibinfo {year} {2007})}\BibitemShut {NoStop}%
\bibitem [{\citenamefont {Kalev}\ and\ \citenamefont {Hen}(2021)}]{kalevQuantumAlgorithmSimulating2021}%
  \BibitemOpen
  \bibfield  {author} {\bibinfo {author} {\bibfnamefont {A.}~\bibnamefont {Kalev}}\ and\ \bibinfo {author} {\bibfnamefont {I.}~\bibnamefont {Hen}},\ }\bibfield  {title} {\bibinfo {title} {Quantum {{Algorithm}} for {{Simulating Hamiltonian Dynamics}} with an {{Off-diagonal Series Expansion}}},\ }\href {https://doi.org/10.22331/q-2021-04-08-426} {\bibfield  {journal} {\bibinfo  {journal} {Quantum}\ }\textbf {\bibinfo {volume} {5}},\ \bibinfo {pages} {426} (\bibinfo {year} {2021})}\BibitemShut {NoStop}%
\bibitem [{\citenamefont {Wang}\ and\ \citenamefont {Xiang}(2019)}]{wangQuantumEigensolverSymmetric2019}%
  \BibitemOpen
  \bibfield  {author} {\bibinfo {author} {\bibfnamefont {H.}~\bibnamefont {Wang}}\ and\ \bibinfo {author} {\bibfnamefont {H.}~\bibnamefont {Xiang}},\ }\bibfield  {title} {\bibinfo {title} {A quantum eigensolver for symmetric tridiagonal matrices},\ }\href {https://doi.org/10.1007/s11128-019-2211-z} {\bibfield  {journal} {\bibinfo  {journal} {Quantum Information Processing}\ }\textbf {\bibinfo {volume} {18}},\ \bibinfo {pages} {93} (\bibinfo {year} {2019})}\BibitemShut {NoStop}%
\bibitem [{\citenamefont {Rakyta}\ and\ \citenamefont {Zimbor{\'a}s}(2022)}]{rakytaEfficientQuantumGate2022}%
  \BibitemOpen
  \bibfield  {author} {\bibinfo {author} {\bibfnamefont {P.}~\bibnamefont {Rakyta}}\ and\ \bibinfo {author} {\bibfnamefont {Z.}~\bibnamefont {Zimbor{\'a}s}},\ }\href {https://doi.org/10.48550/arXiv.2203.04426} {\bibinfo {title} {Efficient quantum gate decomposition via adaptive circuit compression}} (\bibinfo {year} {2022}),\ \Eprint {https://arxiv.org/abs/2203.04426} {arxiv:2203.04426 [quant-ph]} \BibitemShut {NoStop}%
\bibitem [{\citenamefont {K{\"o}kc{\"u}}\ \emph {et~al.}(2022)\citenamefont {K{\"o}kc{\"u}}, \citenamefont {Camps}, \citenamefont {Bassman~Oftelie}, \citenamefont {Freericks}, \citenamefont {{de Jong}}, \citenamefont {Van~Beeumen},\ and\ \citenamefont {Kemper}}]{kokcuAlgebraicCompressionQuantum2022}%
  \BibitemOpen
  \bibfield  {author} {\bibinfo {author} {\bibfnamefont {E.}~\bibnamefont {K{\"o}kc{\"u}}}, \bibinfo {author} {\bibfnamefont {D.}~\bibnamefont {Camps}}, \bibinfo {author} {\bibfnamefont {L.}~\bibnamefont {Bassman~Oftelie}}, \bibinfo {author} {\bibfnamefont {J.~K.}\ \bibnamefont {Freericks}}, \bibinfo {author} {\bibfnamefont {W.~A.}\ \bibnamefont {{de Jong}}}, \bibinfo {author} {\bibfnamefont {R.}~\bibnamefont {Van~Beeumen}},\ and\ \bibinfo {author} {\bibfnamefont {A.~F.}\ \bibnamefont {Kemper}},\ }\bibfield  {title} {\bibinfo {title} {Algebraic compression of quantum circuits for {{Hamiltonian}} evolution},\ }\href {https://doi.org/10.1103/PhysRevA.105.032420} {\bibfield  {journal} {\bibinfo  {journal} {Physical Review A}\ }\textbf {\bibinfo {volume} {105}},\ \bibinfo {pages} {032420} (\bibinfo {year} {2022})}\BibitemShut {NoStop}%
\bibitem [{\citenamefont {Fedorov}\ \emph {et~al.}(2022)\citenamefont {Fedorov}, \citenamefont {Peng}, \citenamefont {Govind},\ and\ \citenamefont {Alexeev}}]{fedorovVQEMethodShort2022}%
  \BibitemOpen
  \bibfield  {author} {\bibinfo {author} {\bibfnamefont {D.~A.}\ \bibnamefont {Fedorov}}, \bibinfo {author} {\bibfnamefont {B.}~\bibnamefont {Peng}}, \bibinfo {author} {\bibfnamefont {N.}~\bibnamefont {Govind}},\ and\ \bibinfo {author} {\bibfnamefont {Y.}~\bibnamefont {Alexeev}},\ }\bibfield  {title} {\bibinfo {title} {{{VQE}} method: A short survey and recent developments},\ }\href {https://doi.org/10.1186/s41313-021-00032-6} {\bibfield  {journal} {\bibinfo  {journal} {Materials Theory}\ }\textbf {\bibinfo {volume} {6}},\ \bibinfo {pages} {2} (\bibinfo {year} {2022})}\BibitemShut {NoStop}%
\end{thebibliography}%
\clearpage
\newpage
\pagebreak
\begin{center}
\textbf{\large Supplemental Materials: Mixed Quantum-Classical Dynamics for Near Term Quantum Computers}
\end{center}

\setcounter{equation}{0}
\setcounter{figure}{0}
\setcounter{table}{0}
\setcounter{page}{1}
\setcounter{section}{0}
\renewcommand{\thesection}{S-\Roman{section}}
\renewcommand{\theequation}{S\arabic{equation}}
\renewcommand{\thefigure}{S\arabic{figure}}
\section{Propagation and sources of errors in the TDVQP algorithm}\label{app:errors}

The TDVQP algorithm inherits all of the errors of its constituent parts. This includes the chosen circuit compression algorithm, time evolution approximation, and in the classical propagator. Nonetheless, it is important to have an intuition of the potential pitfalls of the algorithm. This section will begin with a short derivation of the effect of the magnitude of the infidelity of the wave function on an observable. This is followed by an analysis of the propagation through the velocity Verlet integrator, and finally, numerical experiments comparing the effect of various potential errors on the Shin-Metiu model. 

We can think of an error as a superposition of the desired state $\ket{\psi}$ some combination of undesired orthogonal states $\ket{\phi}$ are in a superposition of $\ket{\tilde\psi}=\sqrt{ 1-I^{2} }\ket{\psi}+I\ket{\phi}$, where $I$ is the infidelity. When we measure a Hermitian observable $\mathcal{O}$ we will get 

\begin{align}
\bra{\tilde \psi} \mathcal{O} \ket {\tilde \psi}      = \left( 1-I^{2} \right)\bra{\psi} \mathcal{O} \ket {\psi} +I^{2}\bra{\phi} \mathcal{O} \ket {\phi}.
\end{align}

The actual measured observable is completely system dependent, so the effect of its magnitude on the rest of the algorithm cannot be estimated at this stage. Nonetheless, we can get an idea of the effect of this on the integrator by assuming that this directly translates to a worst-case error in the force observable, so that for an error of some magnitude we can replace $F_{e}$ in eq.~\ref{eq:verlet_start} with $F_{e}(R_{i},\ket{\psi_{i}})=F_{i}+F_{i\epsilon}$ and propagate the first timestep. 

\begin{align}
    \tilde R_1 &= R_{0}+\dot R_{0} \Delta t +\frac{F_{0}+F_{0\epsilon}}{M}\Delta t^2\\ 
     & = R_{1}+\frac{F_{\epsilon}}{M}\Delta t^{2},\\
\end{align}

This shows us that we are linear in the force error and quadratic in the timestep. This now enters the generation of the new position-dependent Hamiltonian such that we have $H_{el}(R_{1}+\frac{F_{\epsilon}}{M}\Delta t^{2})$ which is used on the subsequent step. The first evolution begins at a known position and the second is already affected by the previous error as
\begin{align}
\ket{\tilde \psi_{1}} &=\exp(-iH_{el}( R_{0})\Delta t)\ket{\tilde \psi_{0}},  \\
\ket{\tilde \psi_{2}} &=\exp(-iH_{el}( \tilde R_{1})\Delta t)\ket{\tilde \psi_{1}}.  
\end{align}

Sadly even in this simple 1-dimensional model, it is difficult to analytically determine the effect on the evolution of the subsequent wavefunction, so we assume this effect is negligible within one timestep. We can now compute the effect on the velocity update, which yields

\begin{align}
\tilde{ \dot{R}}_{1}= &   \dot{R}_{0}+\frac{F_{0}+F_{0\epsilon}+F_{1}+F_{1\epsilon}}{2M}\Delta t
\\
\tilde{ \dot{R}}_{1}= &   \dot{R}_{0}+\frac{F_{0}+F_{1}}{2M}\Delta t+\frac{F_{0\epsilon}+F_{1\epsilon}}{2M}\Delta t
\\
\tilde{ \dot{R}}_{1}= & \dot{R}_{1}+\frac{F_{0\epsilon}+F_{1\epsilon}}{2M}\Delta t
\end{align}

The update can always be separated into the contributions of the ideal integration and the integration of the error. This shows us that the velocity estimation error is linear in the error in the force and linear in time. From the second timestep onwards, the error will accumulate leading to behaviour like

\begin{align}
    \tilde R_i &= \tilde R_{i-1}+\tilde {\dot R}_{i-1} \Delta t +\frac{F_{i}+F_{i\epsilon}}{2M}\Delta t^2.
\end{align}

This expression can then be expressed as the ideal contributions and the contributions from the force error as

\begin{align}
     \tilde R_i & = R_{i}+\dot{R}_{i-1}\Delta t +\sum_{j=0}^{i}\left( \frac{F_{(j-2)\epsilon}+F_{(j-1)\epsilon}}{2M}+\frac{F_{j\epsilon}}{M} \right) \Delta t^2.\\
\end{align}

And if the error in the force is constant the expression simplifies to 
\begin{equation}\label{eq:powlaw}
    \tilde R_{i}=  R_{i}+\dot{R}_{i-1}\Delta t +\frac{(i^{2}+i){F_{\epsilon}}}{2M} \Delta t^{2},
\end{equation}

which is quadratic in the timestep, linear in the magnitude of the error, and quadratic with respect to the number of time steps. The error is also likely to be proportional to $I^{2}$, which will initially be small but may increase unexpectedly as the desired position and subsequent error-prone time evolution will increasingly diverge from the ideal time evolution. 

The above section illustrates that there is an effect due to the inherent interplay between the observables and the classical propagation which increases through time. But The fidelity of the wavefunction is always affected by the set optimizer threshold for fidelity. Assuming this is always met within the maximum allowed number of iterations per timestep, with a threshold of $T$ and assuming no other errors, we expect to see the fidelity with the number of iterations $i$ to fall as 
\begin{equation}
   |\braket{\psi_{\text{exact}}(i)}{\psi_{\text{TDVQP}}(i)}|^2 \approx T^i.
\end{equation}

\begin{figure}
    \centering
    \includegraphics[width=\columnwidth]{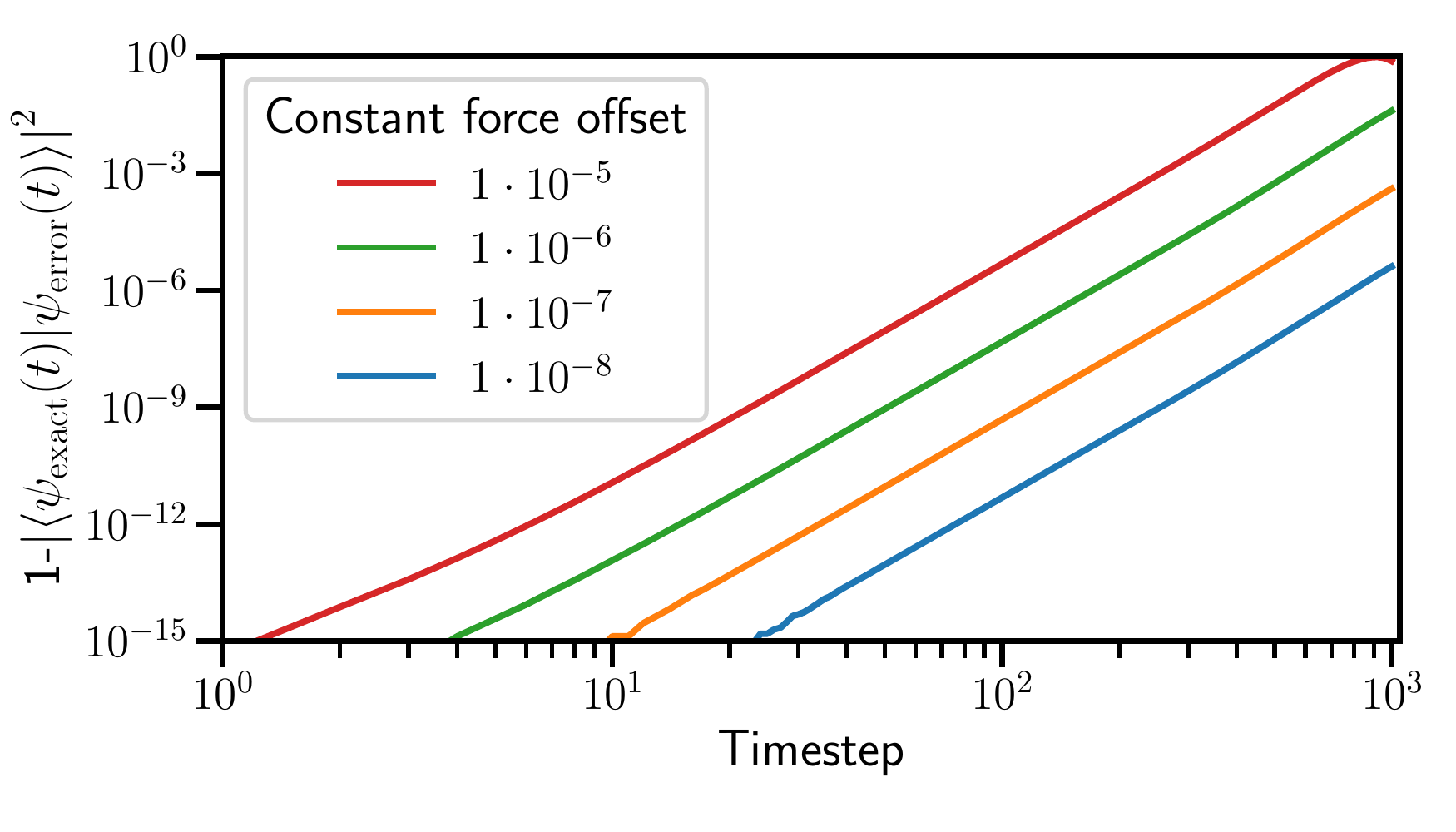}
    \caption{\textbf{Effect of an additive force measurement error on ideal propagation.} The force is a constant addition of $F_\epsilon$ at each step such that $\tilde F_i=F_i+F_\epsilon .$ Different colours represent different error factors.}
    \label{fig:adderr}
\end{figure}

\begin{figure}
    \centering
    \includegraphics[width=\columnwidth]{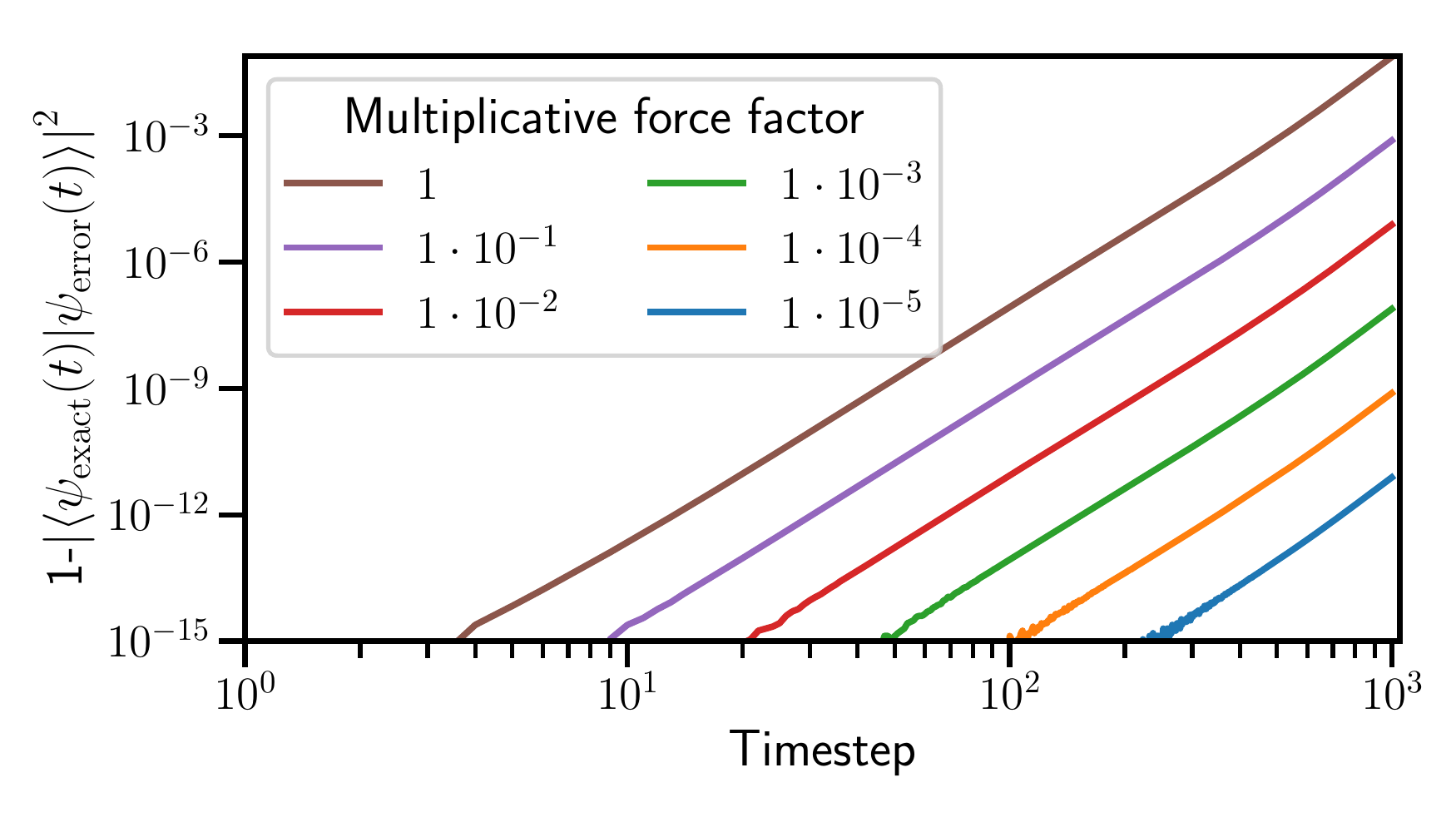}
    \caption{\textbf{Effect of a multiplicative force measurement error on ideal propagation.} Here the force is a constant addition of a magnitude $(\epsilon)$ dependent error such that $\tilde F_i=F_i+F_i\cdot\epsilon.$ Different colours represent different error magnitudes.}
    \label{fig:multerr}
\end{figure}

The overall effect of the threshold error is already illustrated in Figure \ref{fig:Mfidelity}, but the effect of the observable deserves numerical simulations. We have attempted two different types of errors. The first is a constant additive offset \((\tilde F=F+F_{\epsilon}) \) that simply adds a force of the stated magnitude at each timestep and is shown in Figure~\ref{fig:adderr}. The second is a multiplicative factor \(\tilde F=F+F\cdot\epsilon \), which is force-dependent and its effect is shown in Figure~\ref{fig:multerr}. The effect of both types of errors is quantified by comparing the fidelity of the ideal evolution to the one in which this error is injected at each timestep, but otherwise, the quantum evolution and integrator are the same as in the ideal case. We see that the effect in both is a straight line in the log-log plot of $1-\text{fidelity}$ over time, which hints at there being a power law as in eq.~\ref{eq:powlaw}. The gradient of the lines is identical in all cases within one error group (additive or multiplicative), but it is larger than 2. This may be due to the change in the Hamiltonian as the positions diverge which is not taken into account in eq.~\ref{eq:powlaw}.

Another potential error that should be disentangled from the rest is the effect of the trajectory on the TDVQP. To do this, the nucleus was set to follow the path it would have followed on the exact simulation irrespective of the measured observable. In effect, an evolution under an external time-dependent Hamiltonian. \rf{fig:param} shows the effect of the parameterization, which is insignificant at the infinite shot limit, but quite noticeable in the finite shot case. The difference is particularly pronounced near the transition point (which is beginning to be approached around 200 a.u.), where the speed of the nucleus has a large effect on the transition probability, as expected by the Landau-Zener formula. The lower measured force over the simulation as seen in \rf{fig:Sforce} means that the speed of approach differs enough for the transition to cause a divergence in fidelity of the two.

\begin{figure}
    \centering
    \includegraphics[width=\columnwidth]{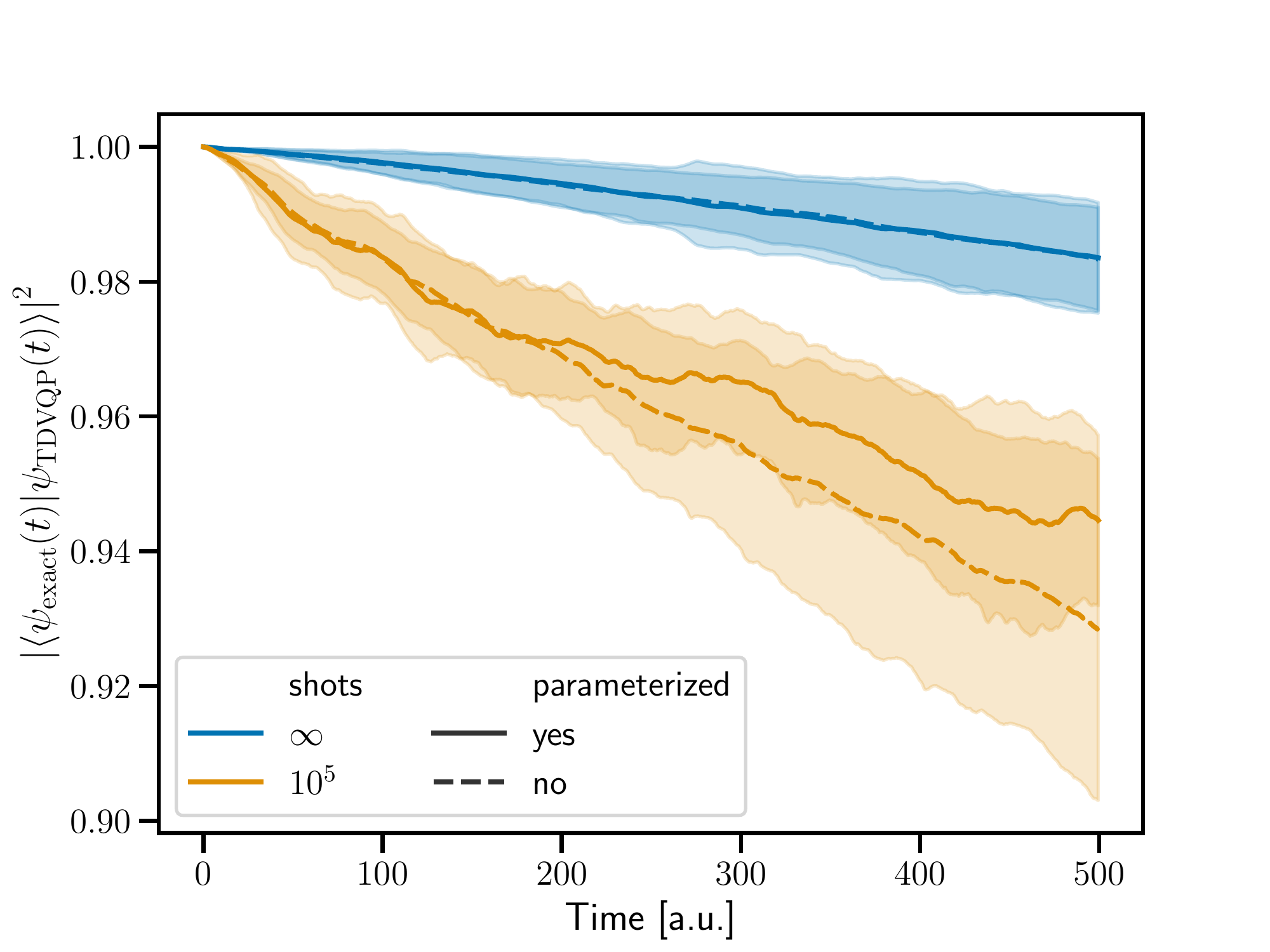}
    \caption{\textbf{Effect of using a parameterized trajectory} on the p-VQD algorithm, showing the infinite shot limit in blue and $10^5$ shots in orange, with highlights showing the standard deviation. Solid lines are unparameterized while dashed lines are. In the infinite shot limit, there is little difference between the two, but in the finite shot case, fidelity does not fall as quickly nearing the transition point, likely due to a difference in the speed of approach of the nucleus.}
    \label{fig:param}
\end{figure}

\subsection{Resource cost of TDVQP}\label{app:resources}

To determine the overall number of circuit evaluations required by the algorithm, the analysis
is split into two parts. The first one-time cost is in finding the initial state circuit parameters. This is the same as in VQE and is $\mathcal{O}(N_{H})$, where $N_{H}$ is the number of Pauli strings required to express the Hamiltonian. The propagation then consists of finding the maximal overlap between the time-evolved state and its approximation, which only requires a single circuit to evaluate for the maximum number of iterations $N_{\text{iter}}$, and is thus $\mathcal{O}(N_{\text{iter}})$. The underlying optimizer will require $\mathcal{O}((N_{\text{param}}^n)$, where $N_{\text{param}}$ is the number of circuit parameters, and $n$ is generally small. The observable measurement requires the most circuit evaluations and is $\mathcal{O}(N_{\text{obs}})$, where $N_{\text{obs}}$ is the number of Pauli strings required to express the observable in question. The whole propagation is linear in the number of timesteps desired $N_{\text{steps}}$. Thus, the circuit evaluation cost is $\mathcal{O}(N_{H}+N_{\text{steps}}(N_{\text{iter}}N_{\text{param}}^n+N_{\text{obs}}))$. In our example, omitting the first step, we have $N_{\text{obs}}=7$ which grows either linearly with grouped observables or exponentially without, then $N_{\text{param}}=70$ and $N_{\text{iter}}=100$, which are set by choice. This gives an overall cost of around $7\cdot10^3$ circuits per timestep.

\subsection{Finite shot effects on the population}
\label{app:shots}

\rf{fig:shotpop} shows the effect of finite sampling error on the state populations. When shot noise is taken into account, the system tends to move towards the equal superposition state. Interestingly, the population transfer between states 0 and 1 is enhanced, likely due to the faster decrease of the 0 state to other states. 

\begin{figure}[H]
    \centering
    \includegraphics[width=\columnwidth]{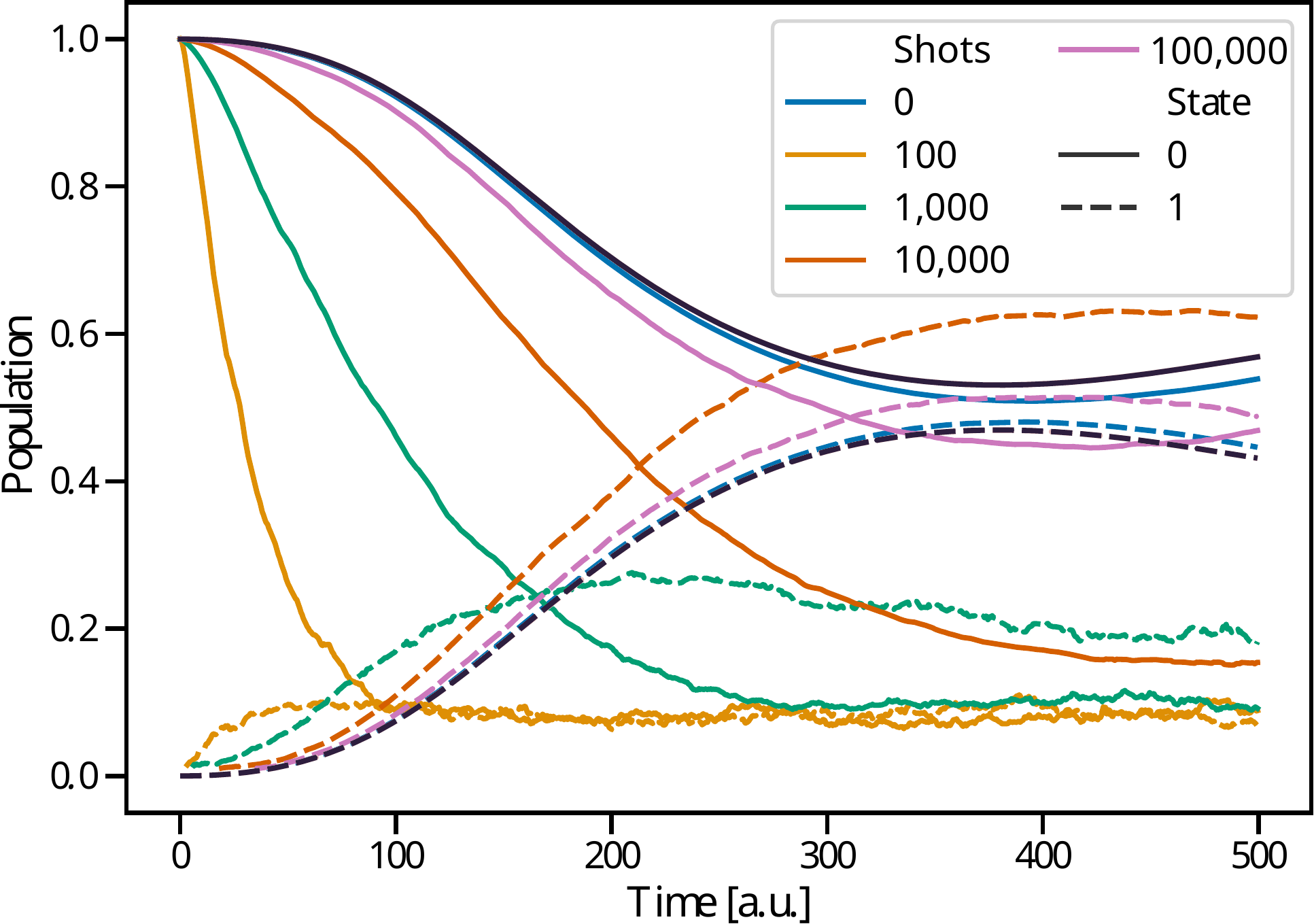}
    \caption{\textbf{Populations with finite sampling effect} for the MD initialization simulation, showing different colours for the various shot counts, 0 being the infinite shot limit. Dashed lines represent the excited state while solid lines represent the ground state.}
    \label{fig:shotpop}
\end{figure}

\section{On Trotterization and ansatz layers}
\label{app:trotlay}
\begin{figure}[h!]
    \centering
    \includegraphics[width=\columnwidth]{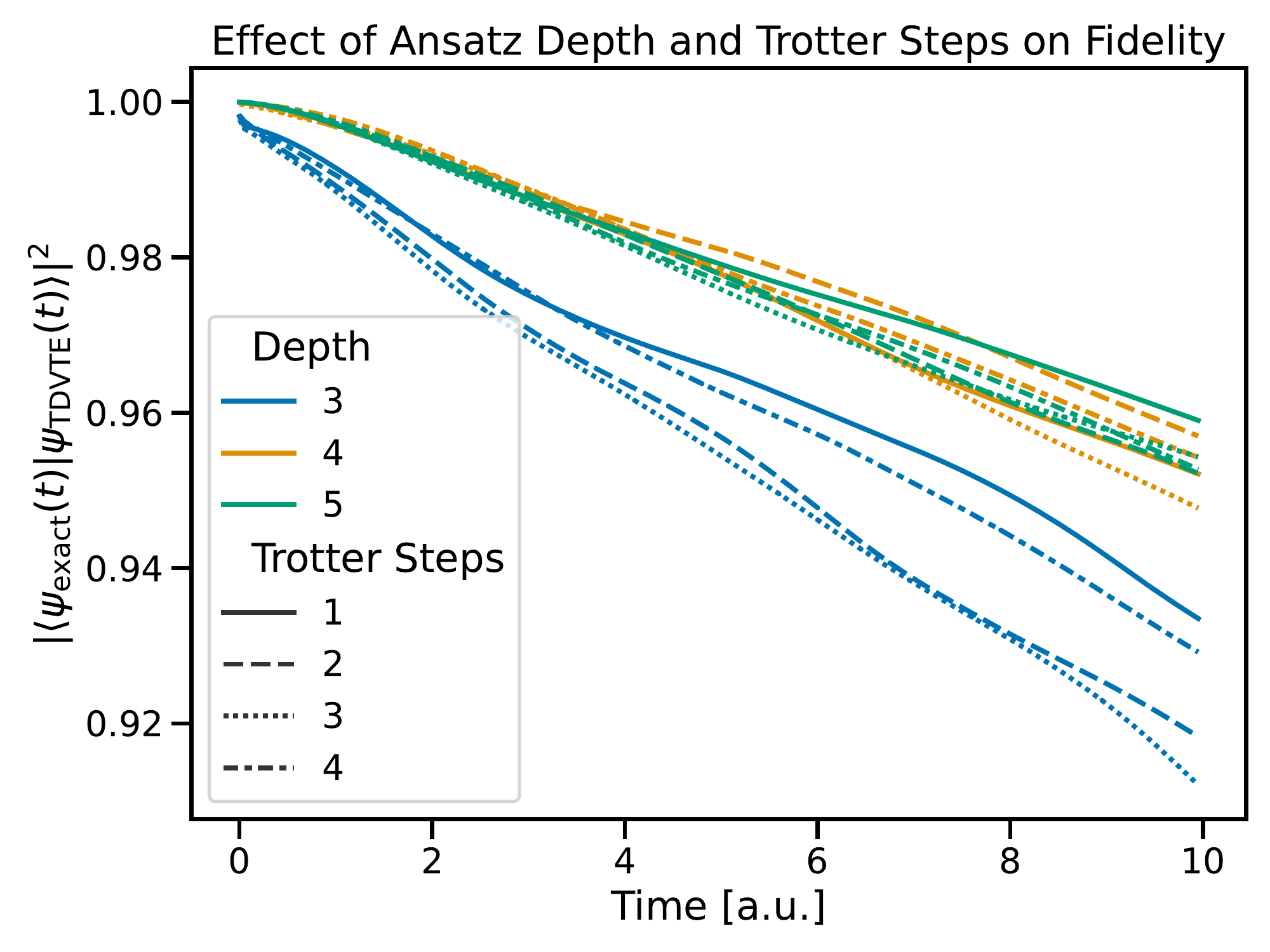}
    \caption{\textbf{Mean fidelity with respect to the depth of the ansatz and the number of trotter steps} in the evolution for 10 samples of 200 timesteps. The lowest depth is 3 (blue), while the next step is of depth 4 (orange), and the deepest is 5 (green). The different Trotter approximations are various dashed or solid lines, but since the time evolution is short, this has little effect compared to the depth of the ansatz.}
    \label{fig:fid_comp}
\end{figure}

Although one can have a near infinite amount of variability in the heuristic form of the ansatz, which we have chosen to be the one shown in \rf{fig:ansatz}. But even for a single ansatz it is important to find what the optimal depth is for a given problem, and in this algorithm, we also want to see the importance of the number of steps in the Trotter approximation. \rf{fig:fid_comp} shows that in the modified Shin-Metiu model we use the timestep of 0.05 a.u. the order of the Trotterization has a small effect compared to the depth of the ansatz. This is not unexpected, as the timestep is very small. For the simulations, we have used a depth of 5 and a single Trotter step for the best compromise between depth and precision. 

\begin{figure}[h!]
    \centering
    \includegraphics[width=\columnwidth]{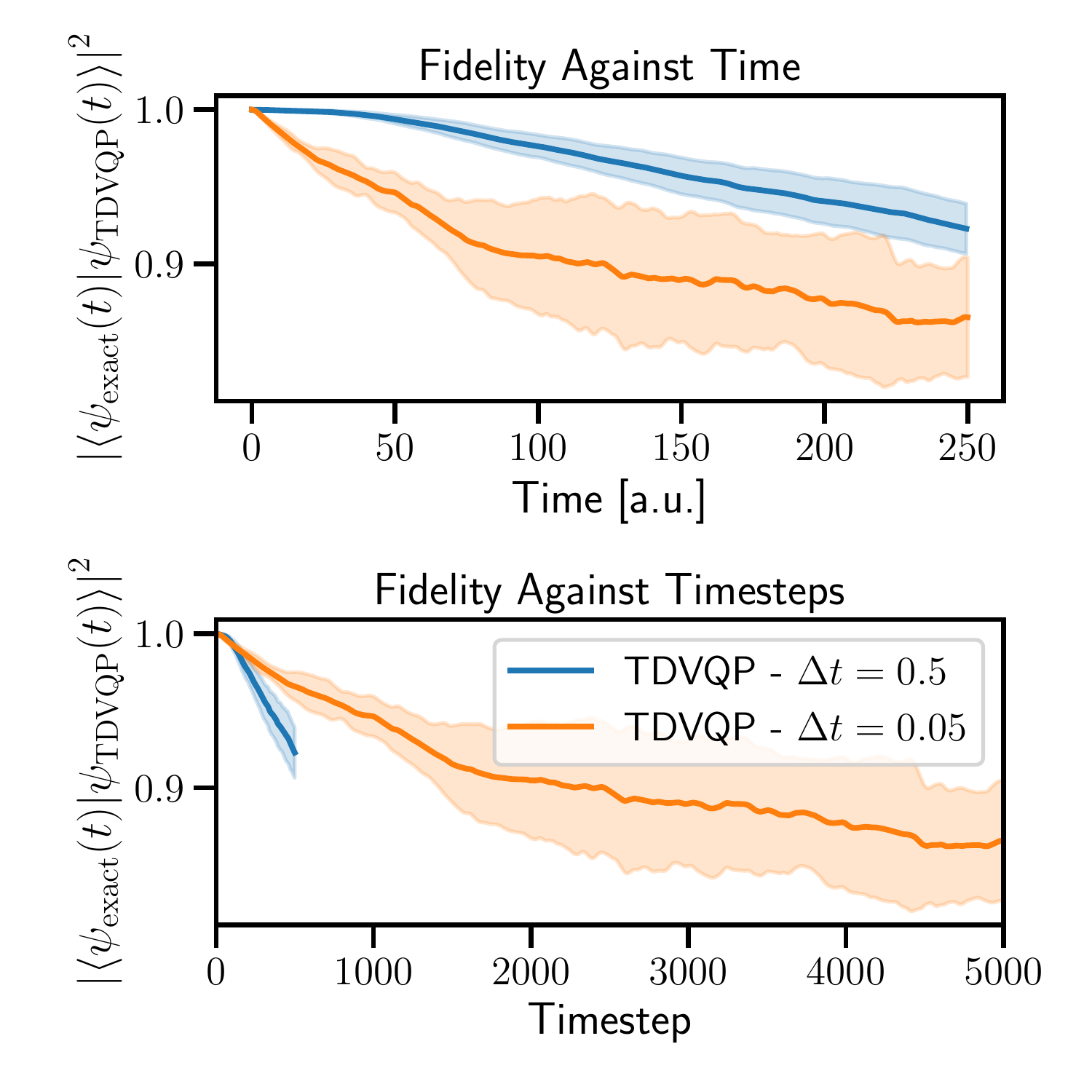}
    \caption{\textbf{Difference between many small timesteps and fewer large timesteps.} Using fewer larger timesteps (blue, 500 steps of 0.5) compared to many smaller timesteps (orange, 5000 steps of 0.05) leads to an interesting observation that smaller time steps lose fidelity slower, but when observing the simulated time, it can be beneficial to use fewer larger timesteps. The respective standard deviations are shown as the highlighted areas.}
    \label{fig:timestep_comp}
\end{figure}

Another important point is whether it is beneficial to use longer timesteps within the limits of the chosen Trotterization depth or to use conservatively short steps. To see the effect of this, we can look at the fidelity of the same simulation as the single case in the main text but with 500 steps of \(\Delta t=0.5\) [a.u.] and 5000 steps of \(\Delta t=0.05\) [a.u.]. Both have a total time of 250 [a.u.], but as should be evident from Figure~\ref{fig:timestep_comp} the two approaches show very different behaviours. When looking at the quality over 'simulated time', choosing larger timesteps is obvious. But if you look at fidelity over the number of timesteps, the shorter timesteps do have an advantage. The reason for this is because the optimizer has a fidelity threshold of \(\phi\) compared to its previous step; as a first approximation, one can assume this fidelity is reached exactly at each timestep \(T\), then after \(T\) steps, we would have a fidelity of \(\phi ^T\) compared to our ideal situation. 

Realistically, the size of the timestep also has a large effect on the iterations the optimizer must take to converge. $1,000$ steps of $\Delta t=0.5$ take longer than $10,000$ steps of $\Delta t=0.05$ given the same treshold. There is also such a thing as a 'golden' initial state, which has the property that many parameters in the ansatz are initialized so that they stay constant or vary smoothly throughout the evolution. Such initializations optimize faster and retain higher fidelities than initial parameters that have much more chaotic 'spiky' evolutions. Although this is hard to quantify, it is something that could be used to filter out badly behaving initializations early on in the evolution.   

\section{Pauli string representation of a Tridiagonal Hermitian matrix}
\label{app:tridiagonal}
The real-space Hamiltonian for the Shin Metiu model is tridiagonal. There is a recursive solution to expressing tridiagonal Hermitian matrices in the Pauli basis. The off-diagonal matrices can be expressed as:

\begin{align*}
A_1 = & X\\
    A_n  =& I_{2} \otimes A_{n-1} +(X \otimes I_{2^{n-1}} )\otimes \\
    & \left[ \frac{1}{2^n}
    \sum_{t=0}^{\lfloor \nicefrac{n}{2}\rfloor} (-1)^t \sum_{\pi}S_\pi\left( X^{\otimes(n-2t)}\otimes Y^{\otimes 2t}\right) \right]\otimes\\
    &(X\otimes I_{2^{n-1}}),
\end{align*}

Where X and Y are the Pauli matrices, and $S_{\pi}$ is the permutation function that returns a unique combination $\pi$ of the Pauli string. That is to say that if we have the string $XYY$, we would get the sum $XYY+YXY+YYX$. This grows exponentially, but with a qubit-wise recursive largest first commutator, it grows as $\mathcal{O}(2^{\nicefrac{n}{2}})$ and if we look at the fully commuting largest first approach, then the number of terms grows as $\mathcal{O}(n)$. The diagonal matrix is then the \(2^{n} \) term weighted linear combination of all possible \(n\) length Pauli strings comprising of \(I\) and \(Z\) matrices. 

Using either qubit-wise commutation relations or Pauli-word wise commutation relations, we can drastically reduce the number of Pauli strings required to implement and measure tridiagonal Hamiltonians. We have analyzed systems of up to 10 qubits via Qiskit and grouped the Pauli string decomposition through the above methods. These groupings can be used to reduce the number of measurements that have to be made, but the results are only presented here for completeness with no comment on how these measurements will be done in practice. \rf{fig:groupin} shows the number of grouped terms using different existing techniques. These are not necessarily realizable on current machines.

\begin{figure}[h!]
    \centering
    \includegraphics[width=\columnwidth]{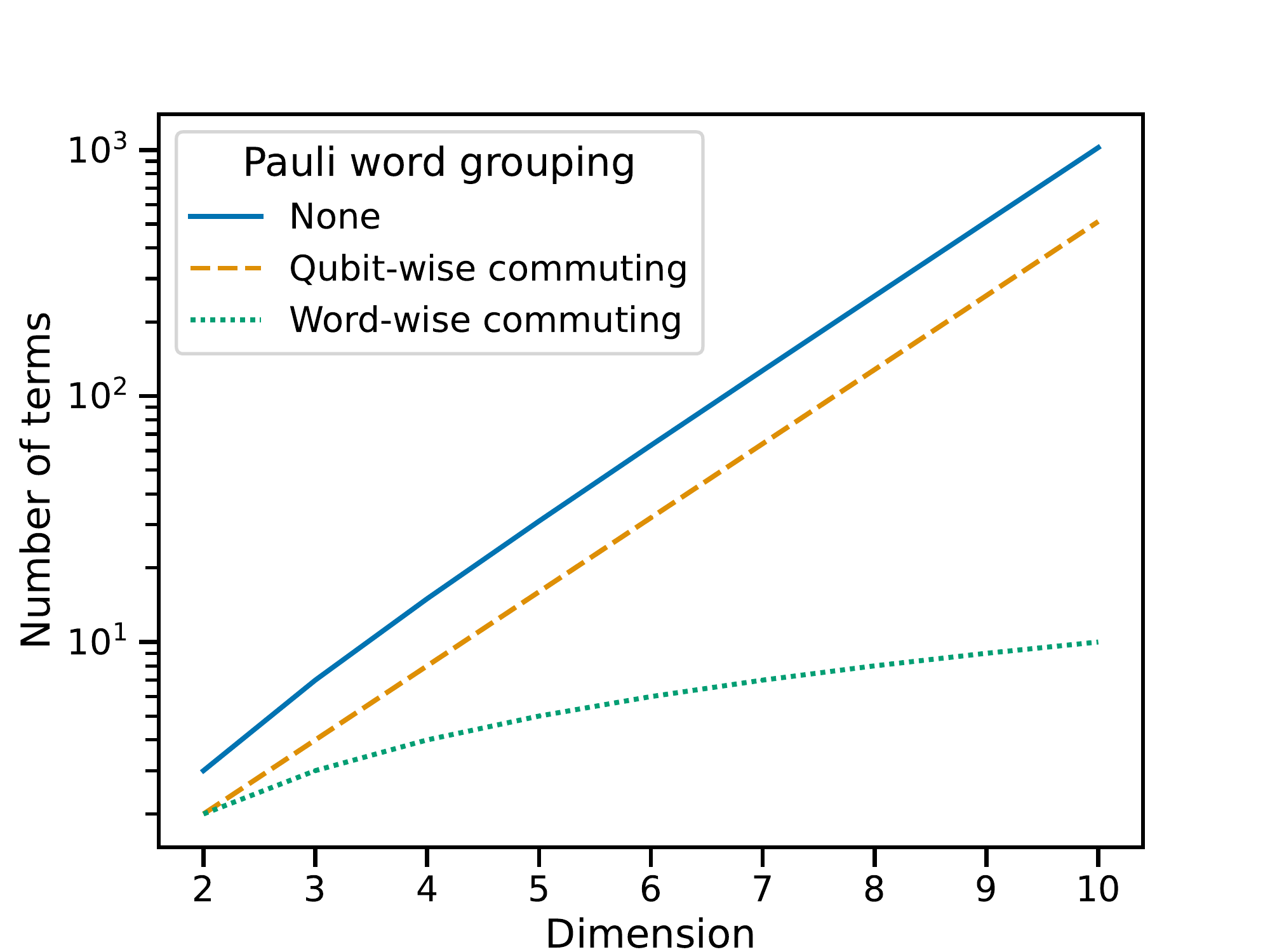}
    \caption{\textbf{The number of grouped terms in a Hermitian tridiagonal matrix} using different grouping strategies. If no strategy is used (solid blue) then the number of terms grows exponentially, which is also the case when using qubit-wise commuting groups (dashed orange). If one uses word-wise grouping then the number of terms grows linearly (dotted green).}
    \label{fig:groupin}
\end{figure}

\section{Additional examples}
\label{app:longtime}

\subsection{Multiple transitions}
\begin{figure}[h!]
    \centering
    \includegraphics[width=\columnwidth]{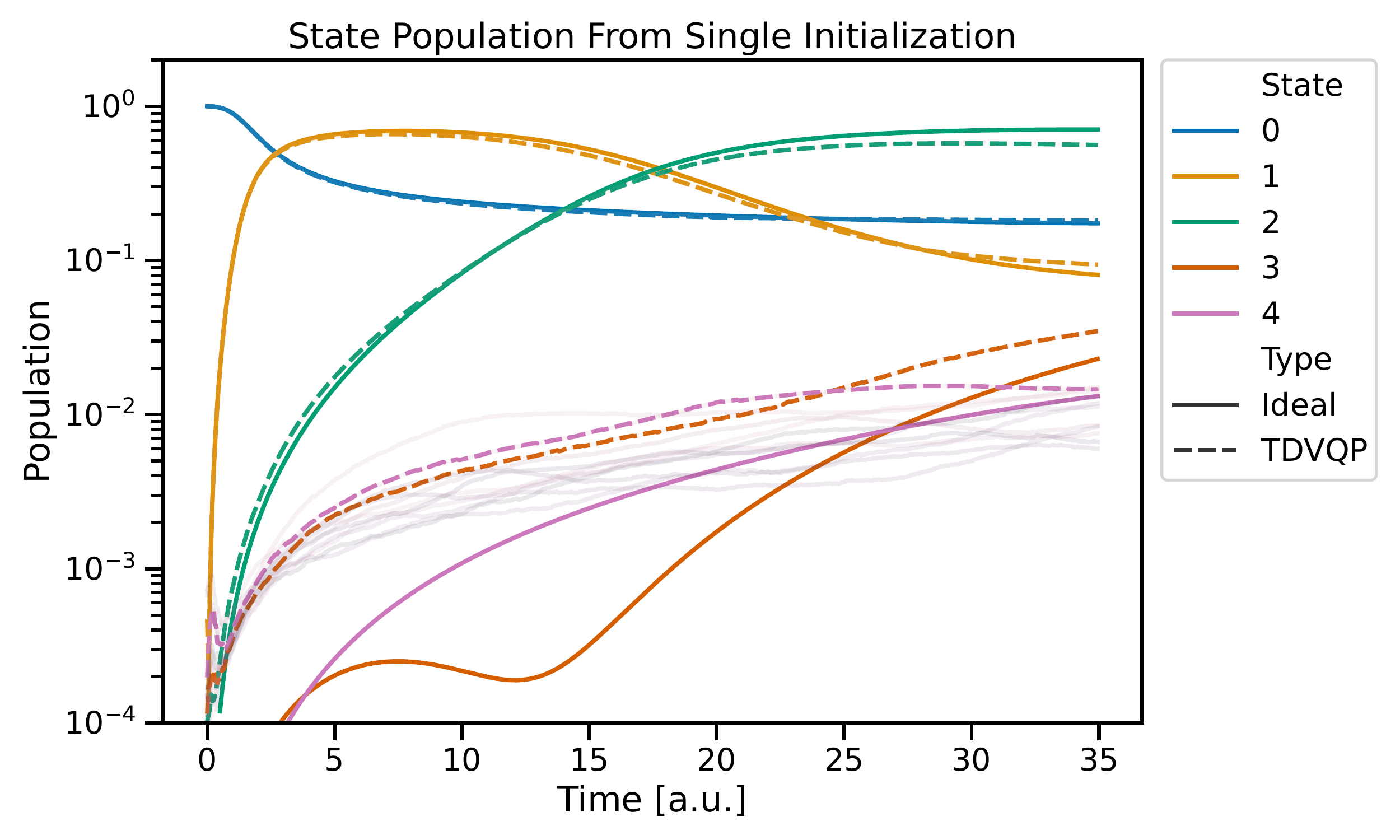}
    \caption{\textbf{State population evolution for multiple transitions.} The simulation of the Shin-Metiu model is initialized in the ground state and evolved for 700 steps of \(\Delta t=0.05\) with a fast-moving proton (\(v=0.2\)). This induces 4 transitions. The lines show the mean BOPES state populations (solid for ideal and dashed for TDVQP) with faint lines representing higher energy levels of the TDVQP.}
    \label{fig:single_pop_long}
\end{figure}

The simulation parameters for these examples are synthetic, with a \(\Delta t=0.05\) and an initial velocity of \(v=0.2\), which allows us to see if the algorithm can deal with more complex dynamics within 700 timesteps. All other parameters are kept as in the main text. Figure~\ref{fig:single_pop_long} shows the dynamics of the populations with 4 population crossings that are well described between states 0,1 and 2. 

\begin{figure}[h!]
    \centering
    \includegraphics[width=\columnwidth]{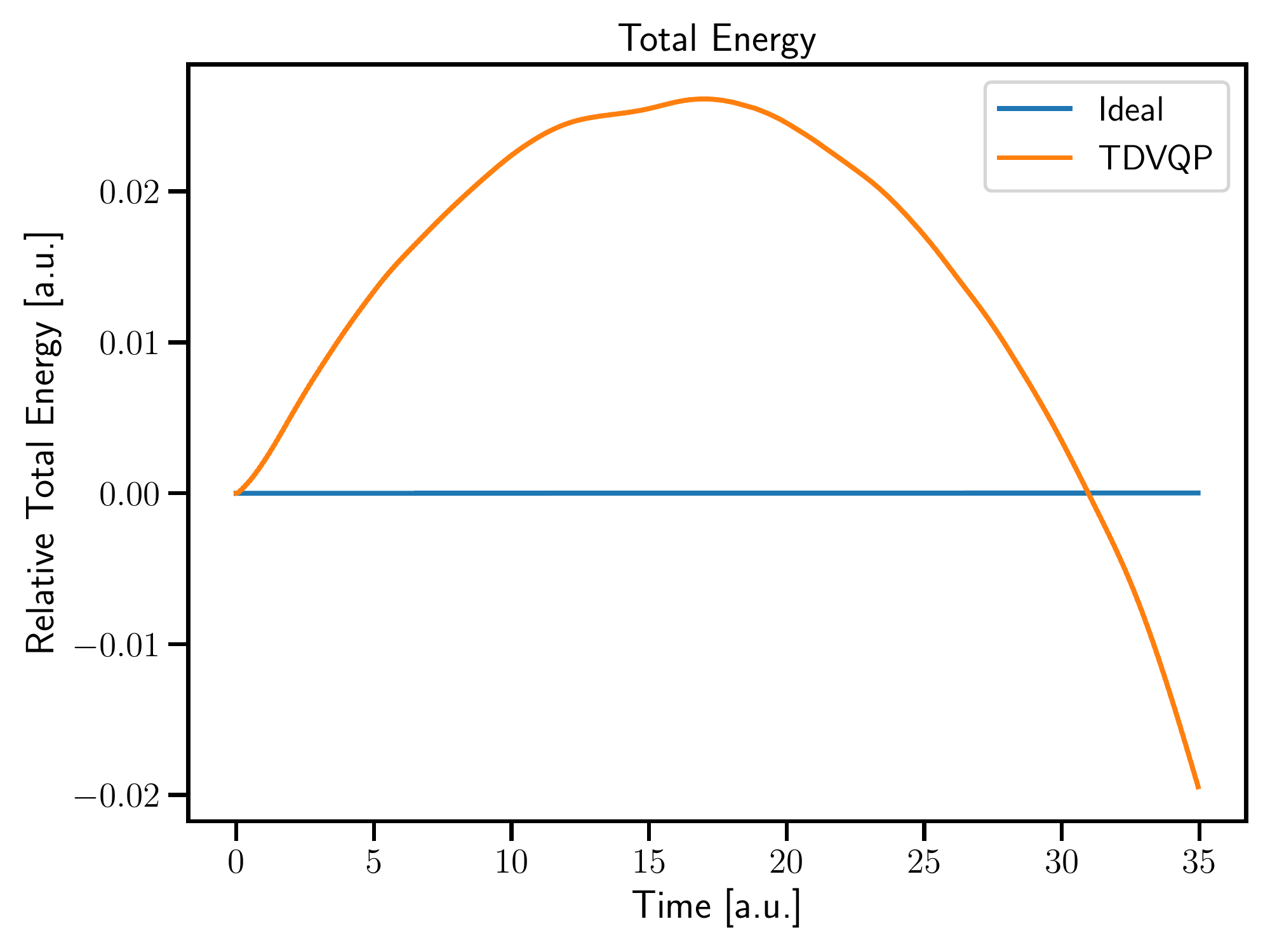}
    \caption{\textbf{Change in energy for the multiple transition simulation.} The plot shows the energy behaviour of the TDVQP algorithm (orange solid line) as it progresses over 700 timesteps of \(\Delta t=0.05\) starting in the ground state, but with a fast-moving proton (with an initial velocity of \(v=0.2\)). The ideal simulation is energy-conserving. The rise in TDVQP energy is due to leakage to higher energy levels, which then transitions to leakage in lower energy levels as the higher energy level is populated.}
    \label{fig:energy_long}
\end{figure}

What is particularly interesting is that around the end of the simulation (\(t>17\)) we see that there is a fall in occupation of the second state and an increase in the first state. This is nicely matched by what we can see in the energy in Figure~\ref{fig:energy_long}. We see that now that the state is highly excited, population loss to lower energy states causes a drop in the energy rather than the increase we have generally seen. 

This fall in energy is not accompanied by an increase in fidelity, and as \rf{fig:fidelity_long} shows, the fidelity keeps decreasing at a steady rate, although likely that at very long times it would begin to oscillate at around 0.5.

\begin{figure}[h!]
    \centering
    \includegraphics[width=\columnwidth]{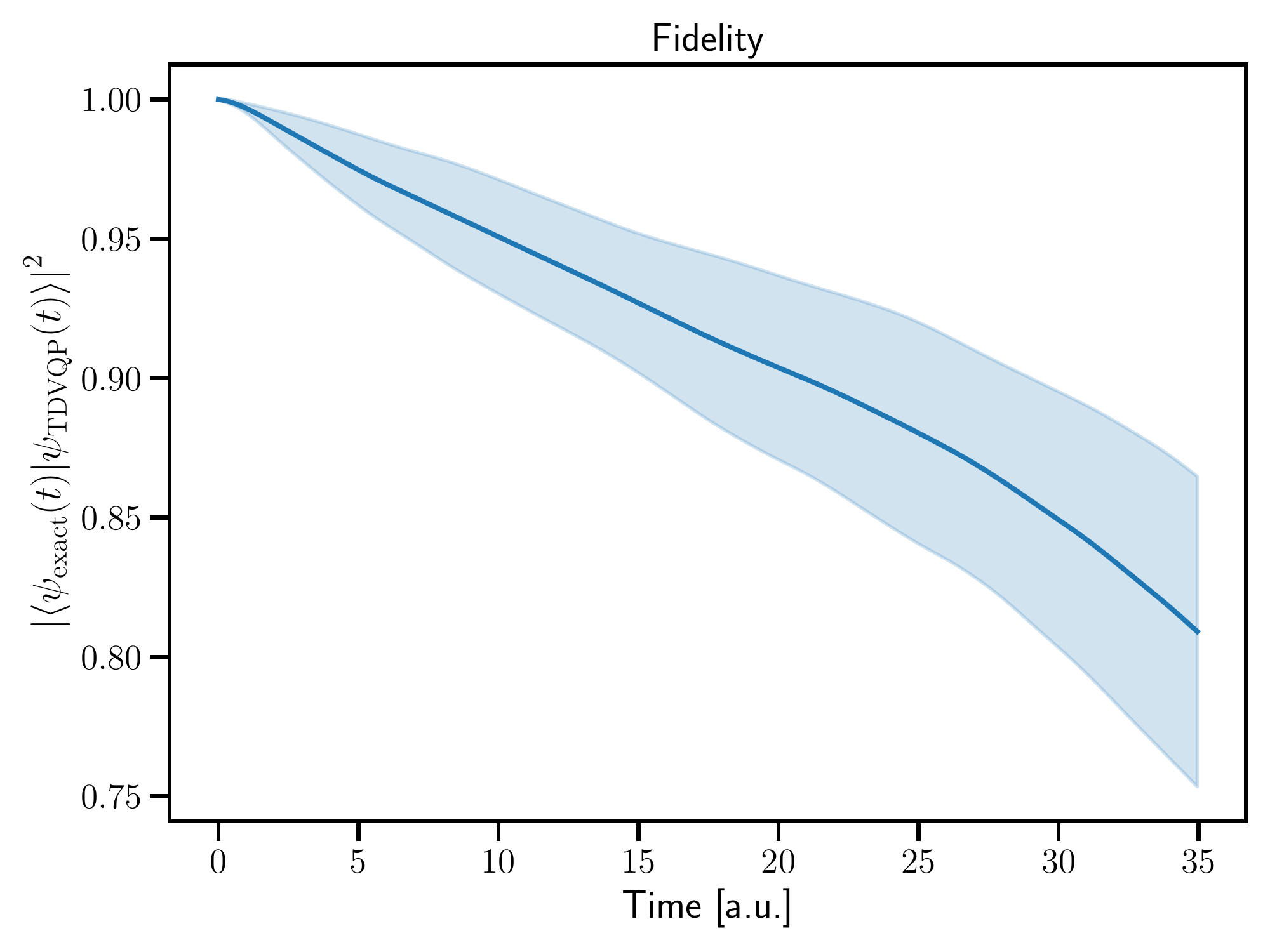}
    \caption{\textbf{Fidelity for the multiple transition simulation.} The fidelity (solid blue line) of the TDVQP wavefunctions compared to the exact evolution of the initial state for 700 timesteps of \(\Delta t=0.05\). It can be seen that it decreases over time more quickly than the ground state evolution.}
    \label{fig:fidelity_long}
\end{figure}

\subsection{Arbitrary state evolution}
To prepare an arbitrary state we use the fact that we expect to be able to find some unitary \(U\) that can operate on a state \(\ket{\psi } \) such that \(U\ket{\psi } =\ket{0} \). If we make this circuit have the form of our desired ansatz and only allow ourselves to vary the parameters \(\theta \). We decide on some threshold or number of iterations and optimize the expression \(\mathop{\min} _\theta  (1-|\bra{\psi }U(\theta )\ket{0}|^{2}) \) to some threshold. This is done by running the circuit in \rf{fig:arbitrary_prep}, where we simply initialize the quantum computer with the excited state in some way. For our simulation, we simply set the starting state to be our desired state, and we show an example for initializing in the first excited state in \rf{fig:excited_pop} and a superposition of the two lowest BOPES in \rf{fig:superposition_pop}. 
\begin{figure}[h]
    \centering
   \begin{quantikz}
       \lstick[]{$\ket{\psi_{\text{desired}  }}$}&
       \gate{U(\theta)^\dagger}\qwbundle[alternate]{}&
       \qwbundle[alternate]{} \rstick[]{$\approx\ket{0}$}\\
   \end{quantikz}
    \caption{\textbf{Computing the ansatz parameters for arbitrary state preparation.} The quantum simulator is initialized in the desired state, and the ansatz \(U\)  parameters \(\theta \) are varied until the final state is close to \(\ket{0} \). }
    \label{fig:arbitrary_prep}
\end{figure}

\begin{figure}[H]
    \centering
    \includegraphics[width=\columnwidth]{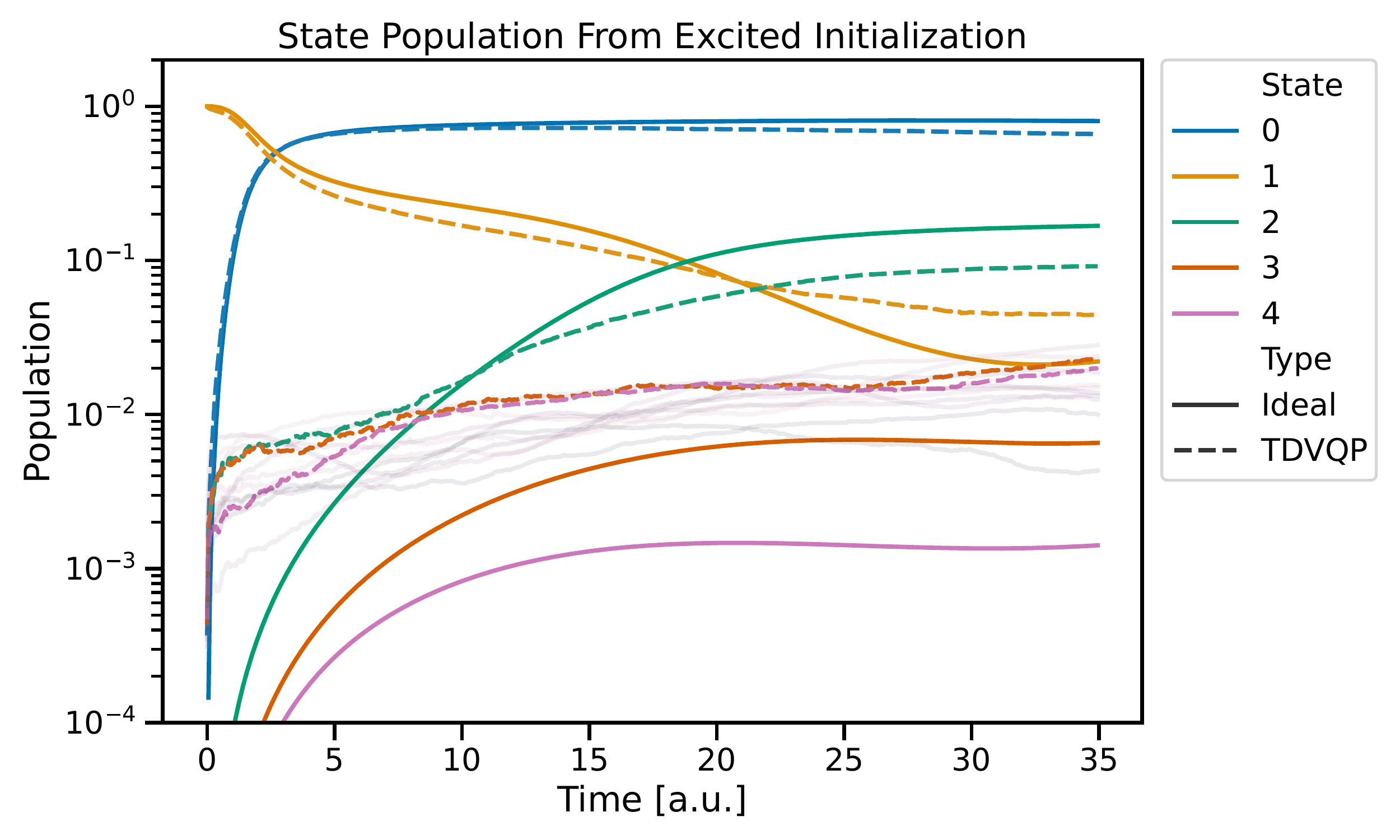}
    \caption{\textbf{State population evolution of the first excited state.} The simulation of the Shin-Metiu model initialized in the first excited state and evolved for 700 steps of \(\Delta t=0.05\). Initial conditions are the same as in \rs{sec:simparam} and the lines show the mean BOPES state populations. The faint lines represent higher energy levels of the TDVQP.}
    \label{fig:excited_pop}
\end{figure}

\begin{figure}[H]
    \centering
    \includegraphics[width=\columnwidth]{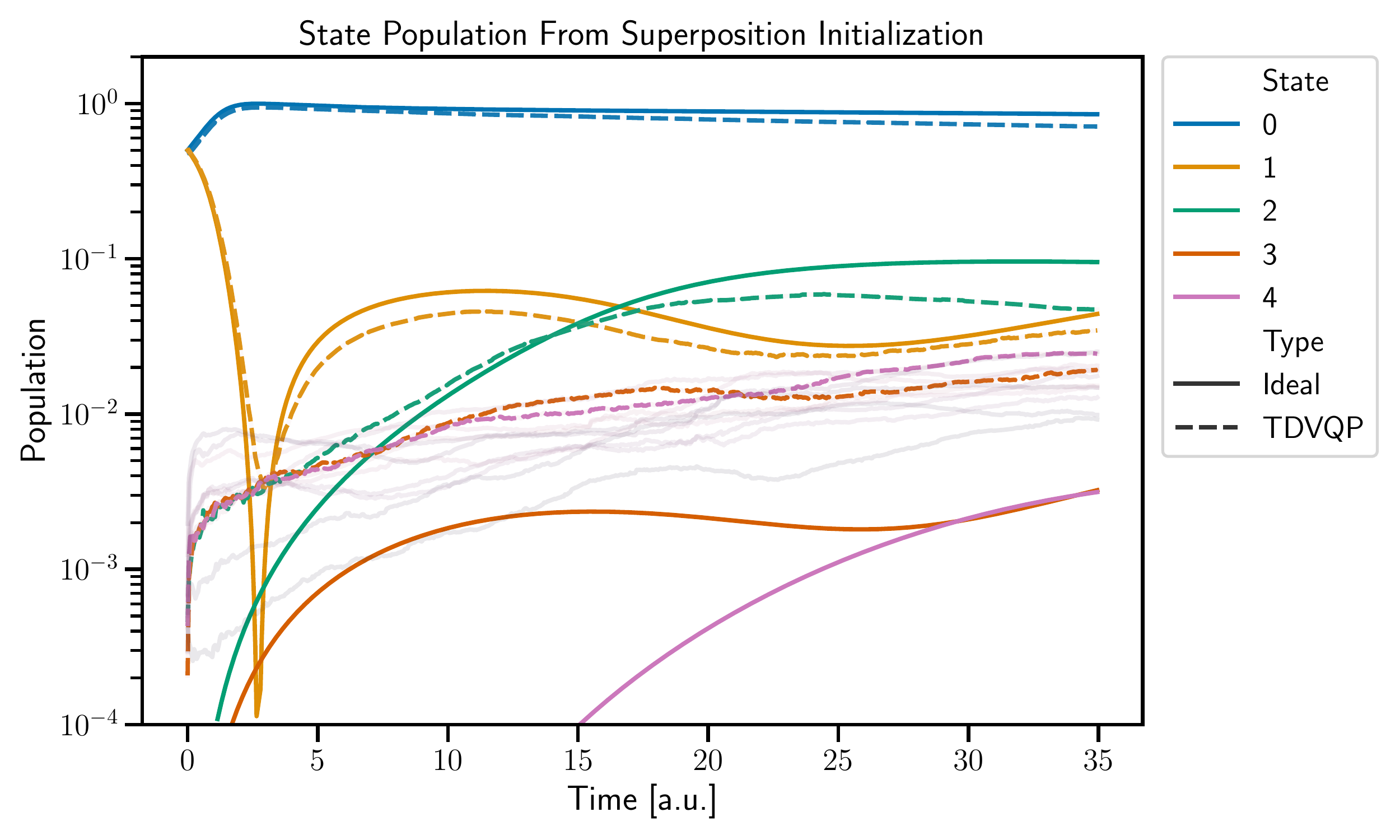}
    \caption{\textbf{State population evolution of the superposition state.} The simulation of the Shin-Metiu model initialized in an equal superposition of the 0 and 1 states showing the mean BOPES state populations and evolved for 700 steps of \(\Delta t=0.05\).  Initial conditions are the same as in \rs{sec:simparam} and the lines show the mean BOPES state populations. The faint lines represent higher energy levels of the TDVQP.}
    \label{fig:superposition_pop}
\end{figure}

\begin{figure}[H]
    \centering
    \includegraphics[width=\columnwidth]{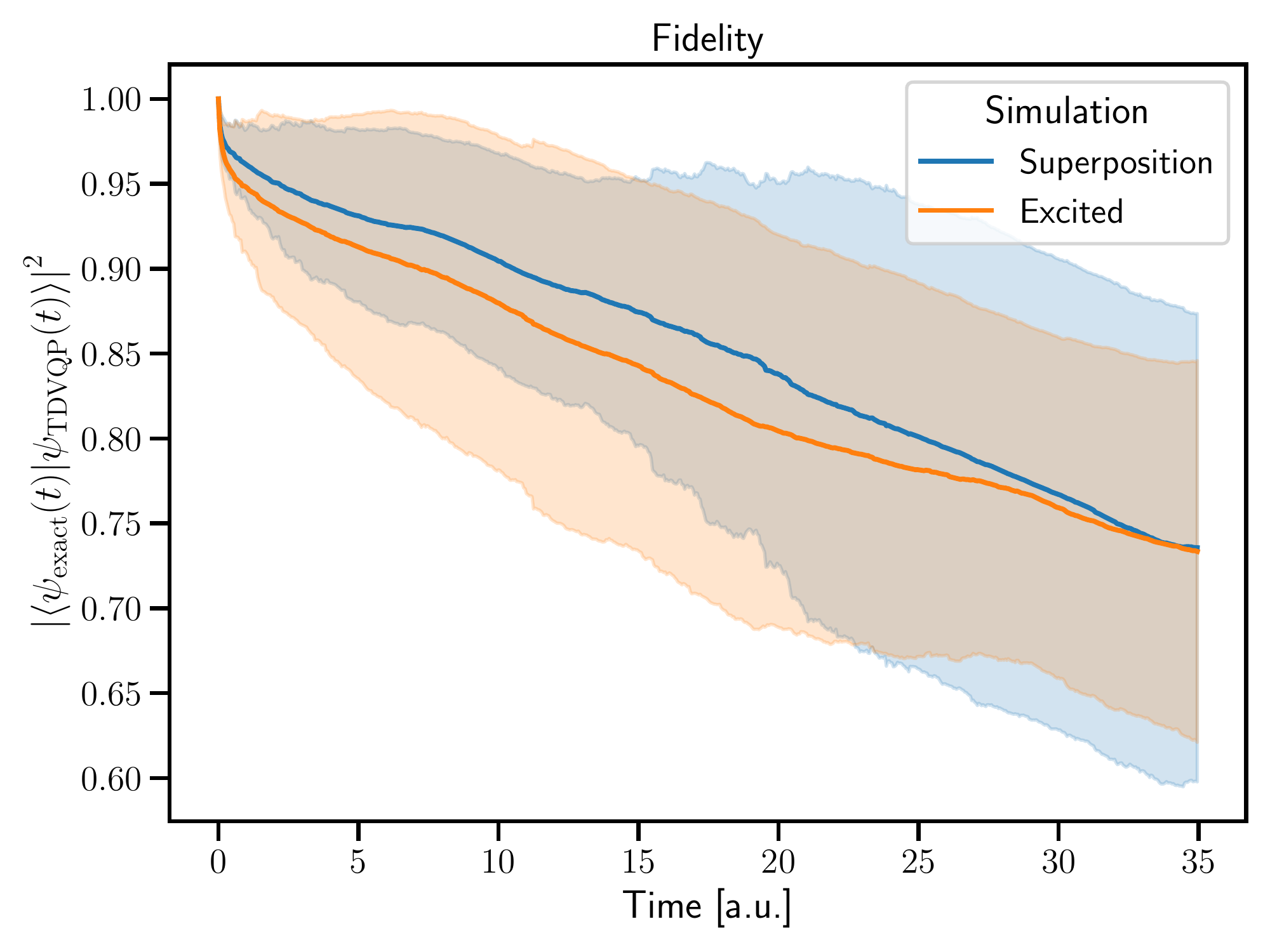}
    \caption{\textbf{Mean fidelity} over the evolution of the equal superposition showing the lowest BOPES states (solid, blue) and first excited state (orange, dashed). The highlighted areas show the standard deviation of their respective values.}
    \label{fig:superposition_en}
\end{figure}

\begin{figure}[H]
    \centering
    \includegraphics[width=\columnwidth]{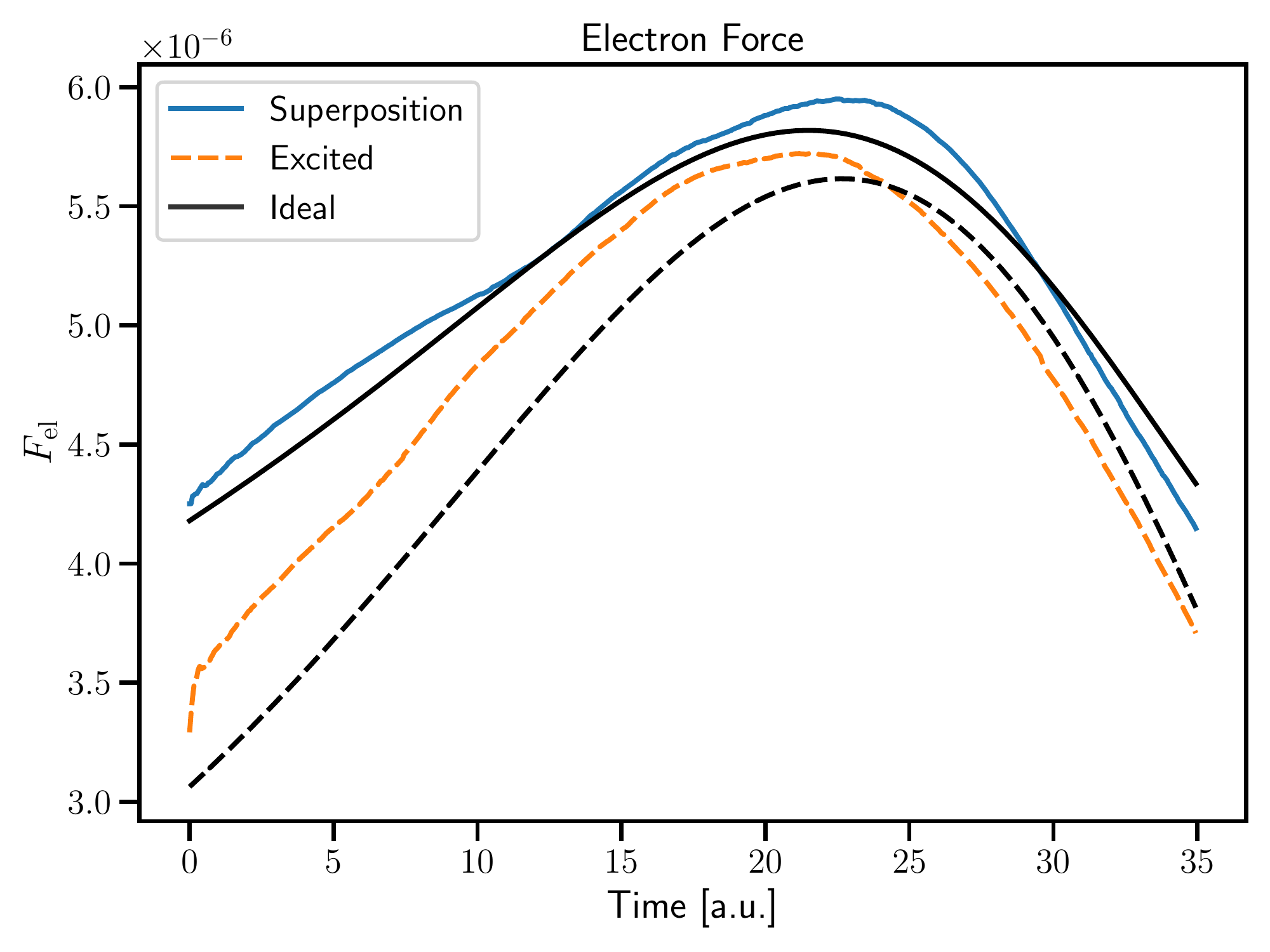}
    \caption{\textbf{Mean force measurements} for an equal superposition of the lowest BOPES states (solid, blue) and first excited state (orange, dashed) compared to their respective ideal simulations (black).}
    \label{fig:superposition_for}
\end{figure}

Figures \ref{fig:arbitrary_prep} and \ref{fig:excited_pop} both show that the states of interest are initially well represented by the algorithm. As the evolution continues the evolution remains qualitatively similar, but degrades, especially when populations approach the 'noise floor' of the algorithm around \(10^{-3}\) where the higher energy levels are populated. \rf{fig:superposition_en} shows a very sharp decrease in fidelity in the first timesteps and quite a large standard deviation compared to ground state results shown in Figure \ref{fig:fidelity_long}. This may either be due to not being able to initialize the ansatz as well in non-ground state settings with our approach here. Forces are still well followed qualitatively as in \rf{fig:superposition_for}.

\end{document}